\documentclass[preprints,review,accept,oneauthor,pdftex]{Definitions/mdpi} 
\usepackage[utf8]{inputenc}
\usepackage{amsmath,amssymb}

\firstpage{1} 
\pubvolume{1}
\issuenum{1}
\articlenumber{0}
\pubyear{2021}
\copyrightyear{2021}
\externaleditor{Academic Editor: Binbin Zhang} 
\datereceived{} 
\dateaccepted{} 
\datepublished{} 
\hreflink{https://doi.org/} 
\pdfoutput=1

\Title{The Low Frequency Perspective on Fast Radio Bursts}

\TitleCitation{The Low Frequency Perspective on Fast Radio Bursts}

\Author{{Maura Pilia} \orcidA{0000-0001-7397-8091}}

\AuthorNames{Maura Pilia}

\AuthorCitation{Pilia, M.}

\address[1]{INAF---Osservatorio Astronomico di Cagliari, Via della Scienza 5, I-09047 Selargius, Italy; maura.pilia@inaf.it}

\abstract{Fast radio bursts (FRBs) represent one of the most exciting astrophysical discoveries of the recent past. The study of their low-frequency emission, which was only effectively picked up  about ten years after their discovery, has helped shape the field thanks to some of the most important detections to date.
Observations between 400 and 800 MHz, carried out by the CHIME/FRB telescope, in particular, have led to the detection of $\sim$500 FRBs in little more than 1 year and, among them, $\sim$20 repeating sources.
Detections at low frequencies have uncovered a nearby population that we can study in detail via continuous monitoring and targeted campaigns. The latest, most important discoveries include: periodicity, both at the days level in repeaters and at the millisecond level in apparently non-repeating sources; the detection of an FRB-like burst from a galactic magnetar; and the localisation of an FRB inside a globular cluster in a nearby galaxy.
The systematic study of the population at low frequencies is important for the characterisation of the environment surrounding the FRBs and, at a global level, to understand the environment of the local universe.
This review is intended to give an overview of the efforts leading to the current rich variety of low-frequency studies and to put into a common context the results achieved in order to trace a possible roadmap for future progress in the field.}


\def\ri{FRB20121102A}
\def\re{FRB20180916B}
\def \pc{pc\,cm$^{-3}$}
\def\sgr{SGR\,1935$+$2154}

\begin{document}

\section{Introduction}
\label{sec:intro}

Fast radio bursts (FRBs) are millisecond-duration bright ($\sim$Jy) signals which have so far only been detected in the radio band. They were discovered in 2007 \cite{lorimer07} through an archival data search of pulsar observations at the Murryiang Parkes radio telescope. All the first detections of FRBs came from Parkes, where these observations were performed at 1.4 GHz. 
FRBs exhibit a dispersion measure (DM) relation that is consistent with the propagation expected through cold plasma \cite{lorimer07}, with values that largely exceed the ones predicted for galactic objects by the ionised interstellar medium (ISM) models toward their direction in the sky \cite{cordeslazio02,ymw17}. 
The first FRB to be detected with the Arecibo radio telescope was \ri\ \cite{spitler14}. \ri\ was a first in many respects.  Its discovery was the first confirmation that FRBs had to be astrophysical objects by removing the direct link of the phenomenon to the Murryiang Parkes telescope. \ri\ was also the first (and only, for a while) FRB whose signal was observed more than once \cite{spitler16}. 
\ri\ allowed the first deeper view into the FRB phenomenon, and it naturally became the archetype for FRB models. 
Before \ri, the locations of the few known FRBs \cite{lorimer07,keane12,thornton13} had been searched for, extensively, both in archival data and with dedicated observations, but no trace of other bursts had been found. For instance, $\sim$80 h of follow up for the Lorimer burst \cite{lorimer07}, $\sim$80 h for FRB 131104 \cite{ravi15} and $\sim$110 h for selected FRB positions \cite{petroff15} at the Parkes radio telescope yielded no repeats. This led to the hypothesis that FRBs must originate from cataclysmic events, such as a merger of two neutron stars (e.g.,~\cite{totani13}), which had no prior and could not have a follow-up signal.
With the discovery of \ri, at least some FRBs could not be explained by one-time events, and new models flourished ranging from SETI activity (\cite{lingam17}) through magnetars to active galactic nuclei (e.g.,~\cite{katz17}). A comprehensive and living review of the models is presented in \cite{platts19}.

The past couple of years have seen the birth of a rapidly growing, statistically significant population of FRBs (with 610 members as of September 2021), among which repeaters are a small but growing fraction (24) \cite{chime_cat1_21}.
The majority of the detections are currently attributed to the Canadian Hydrogen Intensity Mapping Experiment (CHIME, \cite{chime_frb_18}), a new transit instrument which came online in 2018 and was set up for FRB detections in real time in the radio band between 400 and 800 MHz. 
\re\ \cite{chime8_19} is the third known repeater. It was published together with seven other repeating sources which were detected by CHIME during its commissioning phase. 
With 10 detected bursts in little more than 4 months, \re\ was the most active of the lot. Its $DM=349$ \pc\ was not small in absolute terms but, given its direction in the sky, it hinted to a nearby source. 

The DMs of extragalactic objects can be factorised in the following macro components:
\begin{equation}
    DM_{measured}=DM_{MW}+DM_{IGM}+DM_{host}.
\end{equation}

In this empirical formula, the galactic contribution ($DM_{MW}$) can be estimated from the galactic electron density models, such as NE2001 and YMW16 \cite{cordeslazio02,ymw17}; it includes both the plane and the halo contribution, depending on where the galactic plane observations are pointing at. 
Equivalently, the contribution from the host galaxy of the FRB ($DM_{host}$) takes into account the general ISM probed by our observations within the FRB galaxy and its halo. This value, however, also includes the local contribution of the environment surrounding the FRB.
The contribution of the intergalactic medium ($DM_{IGM}$) is dependent on the redshift of the FRB \cite{inoue04,ioka03} and is what was used for the measurements of the missing baryons by \cite{macquart20}.

NE2001 and YMW16 differed in their prediction for the galactic DM in the direction of \re\ by more than 100 \pc, with YMW16 quoting a maximum DM of $\sim$325\,\pc, which made the source location compatible with the Milky Way halo.
The high discrepancy in the models was a consequence of the low galactic latitude of this source (3.7 deg \cite{chime8_19}). Soon after the discovery, this source was indeed localised by the telescopes of the European VLBI Network (EVN) \cite{marcote20} to a massive spiral galaxy at a luminosity distance of 149 Mpc (z= 0.0337).  

Further detections of bursts from \re\ with CHIME led to the first-ever detection of {\bf periodic activity} from an FRB \cite{chime_period_20}.
{ {Rotational periodicity  on the order of seconds or milliseconds}} was expected from FRBs in the context of the magnetar models (see Section~\ref{sec:models}). Searches for periodicity were performed extensively on \ri \cite{spitler14,spitler16,scholz16}.  
Ref. \cite{zhang18} detected 93 pulses over 5 hours of observations of \ri\ in the band 4--8 GHz using machine learning techniques, providing the first and most complete sample of bursts available. Even in this case, the search for periodicity resulted in a robust rejection of periods greater than 10 ms.   
The { {observed periodic activity}} of \re\ was $16.35\pm 0.15$ days. The active cycle lasts 2.7 days, with 50\% of the bursts happening in a $\pm 0.6$\,day interval. 
This unprecedented findi ng has also led to the detection of possible periodicity in \ri. In this case, the source has an active cycle of 40\% within a periodicity window of $157\pm7$ days \cite{rajwade20a, cruces21}.
From an observational point of view, the existence of predictable windows of activity has meant a big shift in paradigm: it is now possible to look at an FRB and know that chances are high to see it active. This has sparked a renewed observational effort not only in diverse radio frequencies, but also in the quest for the so-far elusive multiwavelength counterpart.

\sgr\ is a galactic magnetar with a period of 3.24\,s and period derivative\linebreak $\dot P = 1.43\times 10^{-11}$\,s\,s$^{-1}$, implying a characteristic age of 3.6\,kyr and a surface magnetic field $B=2\times 10^{14}$\,G \cite{israel16}. This source is known from X-ray observations, while no radio emission had been detected in the past \cite{burgay14}.
Starting on April 24th 2020, the source became very active in X-rays and, on April 27, 2020, a ``burst forest'' was detected by several instruments (e.g.,~\cite{palmer20}). On the following day, one single radio burst was detected by CHIME/FRB \cite{chime20sgr}, which was closely followed by the confirmation of the same event by STARE2 \cite{bochenek20}. This same burst was also detected by four X-ray telescopes \cite{mereghetti20,tavani20b,hxmt,ridnaia20}. It was relatively mild and isolated in X-rays compared to the events of the previous days, the only peculiarity being its harder-than-average spectrum. The radio burst, however, was one of a kind, as this was the first time a galactic magnetar emitted a burst which had a fluence $F = 700$\,kJy\,ms in the CHIME band (400--800 MHz) and $F = 1.5$\,MJy\,ms in the STARE2 observation at 1.4 GHz. Its inferred spectral luminosity is about one order of magnitude lower than the lowest spectral luminosity detected for \re\ and higher than the one observed in some bursts of FRB20200120E \cite{bhardwaj21}, but it would have been detectable out to the distance of the closest FRB by a sensitive single dish such as FAST.
If confirmed, this event, which was dubbed FRB 200428, would represent the first FRB to be observed at two different frequency bands. The fluence at 1.5 GHz is double the fluence at 600 MHz. The two points are hardly enough to define a spectral index, but, if we assume a power law spectrum $F({\nu}) \propto \nu^{\alpha}$, the spectral index would be, in this case, $\alpha \sim -1$. 

The technological advances of the latest years and the interest sparkled by the ever-growing exciting science linked to FRBs have paved the way for an exponentially evolving field. Thanks to dedicated survey instruments and large amounts of follow-up time granted by more sensitive telescopes, the last year alone has led to---among other discoveries---the already mentioned detection of periodicity in FRBs, as well as hints for the existence of rare periodicity at the millisecond level (see Section \ref{sec:profile}); the detection of a burst close to FRB energies from a galactic magnetar simultaneous to X-ray emission; the detection of FRB emission down to the lowest radio frequencies (see Section \ref{sec:lowest}); and the discovery and precise localisation of an active repeater from a globular cluster in the nearby galaxy M81 (see Section \ref{sec:chime}). All these findings have been triggered by low-frequency observations.
The motivation for this review analysing the multivariate complexion of FRBs at low frequencies, {  {which will here be considered as frequencies below} $\sim$1\,GHz,} is to provide a comprehensive overview of what aspects have been important to characterise the low-frequency population and, complementarily, to derive some general constraints that have emerged from low-frequency observations that can help us shape future progress in the field.

\section{Observational Properties of FRBs}
\label{sec:properties}

The main observational characteristics of FRBs at first sight were their very short duration ($\sim$ ms to tens of ms), their high fluence and their DM in excess of the expected values for galactic sources (up to $\sim3000$ \pc; see the Transient Name Server, {TNS}~\url{https://www.wis-tns.org/} ({accessed on 10 October 2021}), for a list of known FRBs and their published properties).
The first observations (things have changed thanks to the wider use of interferometers for discoveries) of non-repeating FRBs could only go as far as giving a loose estimate of the source's distance and of its energetics. 
The lack of an established model for FRBs validated the use of FRB observables, i.e.,~fluence and DM, to establish a link with the intrinsic properties of FRBs in a model-independent manner. Flux density is affected both by the detector's temporal resolution and by the temporal smearing of the pulse due to multipath propagation, and this is why fluence measurements were soon preferred over flux density ones, as well as, e.g.,~to incorporate in the source counts formalism.
In addition to these properties, polarisation information was available in some cases, which made it possible to try to characterise the environment around the source.

Repeating bursts from the same source, on the other hand, provide a much larger wealth of information: spectral properties, burst distribution and repetition rate, and the evolution of DM and of rotation measure (RM) with time.
It is not easy to evaluate whether repeaters and apparent non-repeaters have the same behaviour given the fact that we have a distribution of parameters in the case of the former and only one instance for the latter, albeit from a much wider sample. However, comparisons have been made both for all bursts by taking into account only the first detected bursts of repeaters, which are not biased by the detection
threshold for subsequent bursts being lower, and by adding up a disproportionate number of bursts from the more prolific repeater sources (see, e.g.,~Section \ref{sec:chime}).

\subsection{FRB Distributions}
An estimate of the fluence distribution of FRBs is complicated by the fact that different telescopes observe at different frequencies and with different limiting sensitivities. Fluence, however, is the best suited parameter for consistent measurements in a case like that of FRBs, where the temporal duration of the burst would make flux density a biased indicator of the burst characteristics.
The study of source counts, which can be combined with distance estimates, is a very important indicator to define the properties of a new population of astrophysical objects. 
Fluence distributions can provide indications on the cosmological properties of a population, but they may not be enough to exploit the potential of the class as a cosmological probe. 
Fluence itself, however, is a biased measurement, and one should not extrapolate group considerations only based on fluence unless the completeness of each instrument can be taken into account. 
The DM distribution provides a complementing tool which could, for example, help quantify the extent of the missing baryon problem \cite{macquart20}.
The fluence and DM distributions combined have provided a $\log N - \log F$ distribution (where $N$ is the number of sources and $F$ is the fluence), indicating that FRBs are cosmological sources which seem to follow the canonical star formation rate distribution. 
%
According to~\cite{meI18,meII18}, FRBs can be seen at large distances.  Their analysis on fluence, redshift and DM distributions indicated the need to perform FRB surveys using large field-of-view (FoV) telescopes to characterise the high fluence tail of the FRB distribution which contains information on the cosmological evolution of the population. For transient events, this tail is not probed uniquely by the initial observations. 

\vspace{10pt} 

\begin{figure}[H]
   \hspace{-0.2cm}  \includegraphics[width=0.6\textwidth]{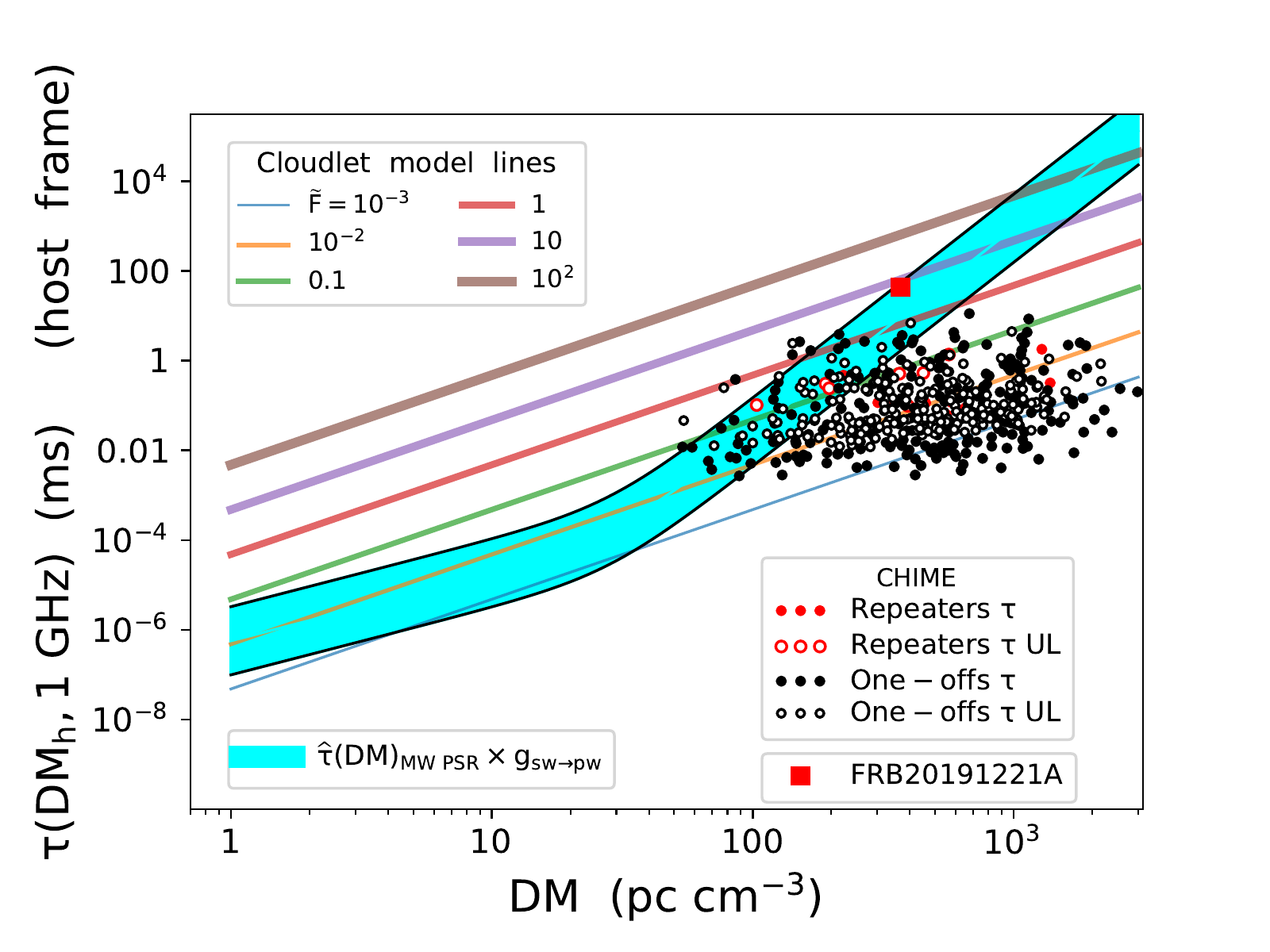}
    \caption{{Figure 5 from} \cite{cordes21}, ``Redshift Estimation and Constraints on Intergalactic and Interstellar Media from Dispersion and Scattering of Fast Radio Bursts'', reproduced with permission from Jim Cordes, ; published by ApJ (IOP Publishing), 2021. It shows the expected scattering time vs DM from the galactic relation derived for pulsars (cyan shaded region, from \cite{bhat04}). The dots represent all published FRBs from CHIME \cite{chime_cat1_21}: black dots represent one-off FRBs, red dots represent repeaters and the empty dots in both colors represent upper limits on scattering. The red square represents the extremely scattered FRB20191221A \cite{chime_ms_period}.}
    \label{fig:lorimer}
\end{figure}

Scattering is an effect of propagation through the ISM with a sharp inverse dependence on frequency ($\propto \nu ^{-4}$). This effect is particularly evident in galactic pulsars at low frequencies. Scattering produces a characteristic asymmetric tail at the trailing edge of the pulsed profile. 
The observed width of a pulse or a burst is given by the convolution
of its intrinsic width with the minimum scattering time.
If, at some frequency, the scattering time is larger than the pulsed width, it can wrap around within the profile and will make the detection impossible. In the case of FRBs, their high DM made it likely that they might also experience strong scattering. 
Scattering, however, can be highly influenced by localised overdensities more than by the total amount of  medium transversed, as is the case for DM. This could either be due to extended emission away from the source or from a dense region in the vicinity of it.
Another factor that comes into play specifically for signals from cosmological distances is cosmological time dilation, which stretches the pulse by a factor of $1 + z$.
Signals with high DMs are expected to be highly scattered at low frequencies according to the observed relation between DM and scattering delay ($DM-\tau_{sc}$, \cite{bhat04}). This relation, however, was determined for the ISM within our own galaxy, and the ionized inter-galactic medium (IGM) might not be distributed in a similar fashion. Furthermore, it has a large scatter, with at least an order of magnitude uncertainty.
For sources outside the Milky Way such as FRBs, the total scattering is made up of two contributions: interstellar scattering in the host galaxy and our own galaxy, and intergalactic scattering caused by the intervening IGM. 
The first observations of FRBs laid significantly below the bulk of pulsars in the $DM-\tau_{sc}$ curve from \cite{bhat04}, suggesting that, for a particular DM, there is less scattering for an extragalactic source than would be expected from the galactic relation.
Ref. \cite{lorimer13} assumed a simple cosmological model, considering the sources as standard candles with a constant number density per unit comoving volume. 
They further assumed a spectral index $\alpha = -1.4$ in an analogy with the radio pulsar population \cite{bates13}. 
At lower frequencies, these assumptions would lead to peak flux densities of the order of tens of Jy, with the event rates significantly dependent on the assumed spectral shape.
Their model implied that, for a given survey at some frequency $\nu$, there is a unique correspondence between the observed flux density and the redshift probed.
Ref. \cite{mk13} confirmed that the scattering due to the IGM per unit of dispersion measure would be several orders of magnitude lower than that found in the ISM, with the constant of proportionality between DM and $\tau_{sc}$ changing
with redshift as $(1+z)^3$.
The two works demonstrated that the scattering effects at lower radio frequencies were less than thought, and that the bursts could be detectable at redshifts out to about $z = 0.5$ in surveys below 1 GHz.
{  M{ore recent studies including the many more bursts that have been detected since then and that have scattering measurements confirm this trend (see Figure} \ref{fig:lorimer} taken from \cite{cordes21}). }
These estimates of the scattering measure at $z \leq 3$ suggested that temporal smearing may be less
than $\sim 1$\,ms for observations at frequencies above 300 MHz.
This implied that surveys with the existing facilities would not necessarily be sensitivity limited and could be carried out with small arrays to maximize the sky coverage.

\subsection{FRB Profiles}
\label{sec:profile}
\begin{figure}[H]
    \includegraphics[width=0.75\textwidth]{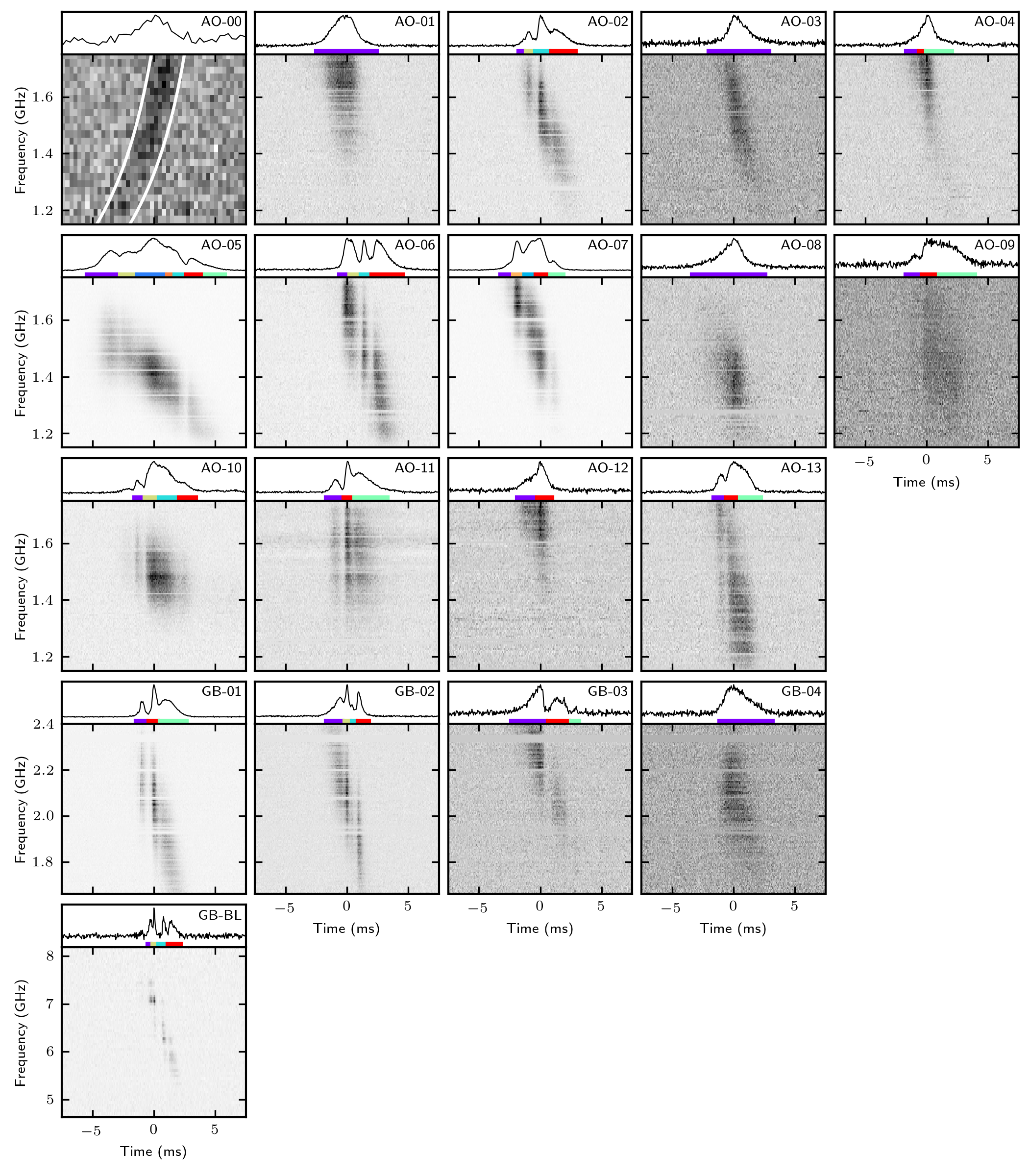}
    \caption{{Figure 1 from} \cite{hessels19} ``FRB 121102 Bursts Show Complex Time-Frequency Structure'', reproduced with permission from Jason Hessels; published by ApJL (IOP Publishing), 2019. 
  It shows the complex morphology of different bursts from \ri\ as observed by GBT and Arecibo. The top of each plot represents the profile, and the bottom represents the dynamic spectrum dedispersed at $DM=560.5$\,\pc. The presence of a sub-burst structure is evident in these bursts, and it is highlighted by the colored bars below each profile. The sub-bursts show a downward drifting trend with the frequency which has been dubbed ``sad trombone'' (see Section \ref{sec:profile}).}
    \label{fig:hessels}
\end{figure}

As more FRBs fill up the sample, the panorama of bursts has become quite diversified. {  Figure \ref{fig:hessels} {shows an example of the variety of burst flavours. They belong to the same source,} \ri\ {in this case, but a similar trend has been seen in other sources as well, see, e.g.,} \cite{day20,pleunis21b}}. Some bursts have bright single-peaked profiles. Other show a wider profile which, when seen in the dynamic spectrum, shows components which appear and disappear at subsequent times from different spectral regions.
 \ri\ has been known to exhibit variable spectral behaviour where the emission drifts across the
frequency band. 
The observed structure of the subpulses typically follows a trend which has been dubbed ``sad trombone'' (evident in Figure \ref{fig:hessels}). In the sad trombone effect, different subpulses, which constitute a large impulse, occupy adjacent spectral bands at contiguous times. 
Ref. \cite{hessels19} conducted a multifrequency study of this behaviour and showed that the drift rate increases at lower frequencies, suggesting that drifting sub-pulses across the frequency band can cause the resulting pulse to be wider.
{  {The detection of bursts comprised of downward-drifiting sub-bursts is not optimised by the typical matched filtering techniques} \cite{cm03} used for FRB detection, as they assume a $\nu^{-2}$ pulse.}
DM optimisation is important when structure is present in the bursts: the optimisation of DM only based on the signal-to-noise ratio (S/N) could mean that the effect of drifting subpulses might imply extra corrections on the DM value. Structure, however, could be disguised below the sensitivity of the observation and could be misinterpreted as scattering. 
It is important, in order to fully account for it, to work on the way the dedispersion is performed. The best DM is typically selected by the detection algorithms as the one which maximises the S/N of the impulse. However, this value is not representative of the correct DM if a sad trombone sub-burst structure is present, mimicking a DM trend. In case of substructures, the best DM should then be chosen as the one which better resolves the substructure. Slightly different approaches  to produce a structure-maximising DM are described in, for example, \cite{hessels19,gajjar18}, and the {\tt DM\_phase} code described in \cite{seymour19} is publicly {available}~\url{https://github.com/danielemichilli/DM\_phase}, {(accessed on 10 October 2021}). 
This behaviour has not been seen previously in the much better studied galactic pulsars, even though new observations at low frequencies show that a similar trend (but also its inverse: a ``happy'' trombone) can be observed with more sensitive, single pulse analyses of known nearby pulsars \cite{bilous21}.

 Burst substructure has now been observed down to $60$\,ns \cite{majid21,nimmo21} when the time resolution of the observations allows it to be detected. 
Notably, the analysis of sub-burst structures has led to the first detection of millisecond periodicity by  CHIME/FRB \cite{chime_ms_period}. FRB20191221A 
shows at least (scattering affects the measurement) nine overlapping components, regularly spaced, spanning a total burst time of 3\,s. A timing analysis of the sub-burst components hinted at a periodicity of 216.8\,ms at a significance level of $6.5\sigma$. 
Such a large number of components is not usual in FRBs, and indeed, only two more have been detected by CHIME with, respectively, six and five components, where a possible hint of ms periodicity has been found at a $<$2$\sigma$-level. 
Strictly speaking, these sources are one-off FRBs, as no repetition has been observed outside the one within the one burst envelope. Indeed, no downward drifting is observed in this case. The ms periodicity may suggest a misclassified galactic pulsar origin, but the DM of the source strongly points towards an extragalactic source. One intriguing explanation for this behaviour in the context of non-repeating FRBs is the generation of periodic pulses from merging neutron stars, with the production of FRBs originating from the interaction from their magnetospheres long before the merger \cite{zhang20ApJL}. 

\subsection{FRB Spectra}

The first band at which FRBs have been exploited is the so-called L-band, at frequencies around 1.4 GHz.
Observations of FRBs were initially carried out mainly at 1.4 GHz thanks to the wealth of available facilities that can observe at these frequencies and the wealth of available archival data. 
This band has historically been an optimal compromise for pulsar observations between their steep spectrum ($\alpha \propto \nu^{-1.4} - \nu^{-1.6}$) and the steep dependence of the ISM contributions to the degradation/distortion of the signal with the inverse of the frequency ($\tau_{DM} \propto \nu^{-2}$, $\tau_{sc} \propto \nu^{-4}$).
\ri\ was first detected at 1.5 GHz but, later on, repetition was also searched for, and found, at 6~GHz~\cite{michilli18}. Currently, \ri\ is the FRB which has been observed at higher frequencies, with detections up to 8~GHz~\cite{gajjar18}.
No systematic searches for FRBs have been conducted extensively above 1.5 GHz, which is also due to the fact that single dishes are not extremely well suited for surveys of the entire sky, as would be needed for FRBs. At lower frequencies, on the other hand, a population of interferometers which could observe a large portion of the sky for a long time exist (or were refurbished having FRBs in mind), and extensive surveys have been performed since soon after FRBs were established as a population.
With the detection of FRBs at high frequencies, and the theories leading to the magnetar models (see Section \ref{sec:models}), their spectra were initially thought to be flat. 
The initial non-detections at low frequencies, also compatible with the magnetar model, were attributed to a cutoff in the emission which must happen below 1\,GHz either due to the internal emission mechanism or absorption by the opaque surrounding medium. 
The first discoveries by CHIME \cite{chime_1r_19,chime8frb} extended the range of the observed emission frequencies down to 400 MHz, and they made it clear that emissions at low frequencies were not exceptions.
A cutoff, if present, must be at lower frequencies and, in that case, ISM effects could not be excluded as the cause of non-detection, especially in the case of very high DM FRBs.

The study of repeaters, and of \ri\ and \re\ in particular because of their frequent and predictable activity, led to a more comprehensive view of the matter.
Observations have shown that \ri\ is very active at intermediate to high frequencies:  \cite{zhang18} published 93 bursts detected with convolutional neural networks in the time span of 5 h in the same dataset of \cite{gajjar18} at a 6 GHz central frequency. More recently, FAST published the detection of 1652 bursts in 59.5 hours spanning 47 days,  with a peak burst rate of 122~hr$^{-1}$ \cite{li21}. In contrast to these, only one single pulse has ever been detected from \ri\ by CHIME below 800 MHz \cite{josephy19} despite a constant on-sky time. 
The burst was detected down to at least 600 MHz, the lowest frequency detection of this source yet, with complex time and frequency morphology. The 34 ms width is larger than the typical width of \ri's bursts, and a sub-burst structure was evident with a downward drift
in frequency at a rate of $-3.9$\,MHz\,ms$^{-1}$, to be
compared to $200$\,MHz\,ms$^{-1}$ reported at 1.4~GHz~\cite{hessels19}. This means that pulses at even lower frequencies might get completely smeared out by the sad trombone effect and that this phenomenon {  {possibly affects the detection of a fraction of the sources.}}
\re, on the other hand, is seen very regularly in CHIME's band, is seen much less at 1.5 GHZ (although a systematic study with APERTIF showed 54 bursts spread along different cycles \cite{pastor21}) and has never been observed at frequencies above 1.5 GHz \cite{pearlman20}. 
It has been proposed that the difference between the two spectra can be ascribed to the external environment and, consequently, on the age of the source (e.g.,~\cite{katz21}). DM and RM measurements and variations \cite{michilli18}, the evidence of a persistent radio source spatially coincident with the repeater \cite{marcote17} and, not least, the paucity of low-frequency emissions \cite{josephy19} all point towards \ri\ being surrounded by a dense---an active---medium, possibly still its wind nebula. \re, on the other hand, shows no evidence of a surrounding compact source either from the continuous emission~\cite{marcote20} or from the spectral properties \cite{pilia20,chawla20,pleunis21a,pastor21}.

Interpretation arising only from two extremely active repeaters may suffer from a sort of discovery bias that one should be careful about. However, despite the many differences which are emerging between the two sources, the frequency behaviour has some common ground and it is worth discussing. 
Observational campaigns have been carried out either using different telescopes at the same time or observing  different bands simultaneously from the same telescope. In the second case, the boundary conditions of the observations were exactly the same in both bands. 
Despite extensive observational hours, only one instance is reported of a burst observed simultaneously at two frequency bands: \ri\ was observed on  MJD 57648 both by Arecibo at 1.5 GHz and by the Very Large Array (VLA) at 3 GHz \cite{law17}.
The flux density of the Arecibo burst was an order of magnitude less than that seen by the VLA. During the same campaign, three other
bursts from \ri\ had similar observing coverage
but were not detected simultaneously. 
The bright radio burst from the galactic magnetar (see Section \ref{sec:intro}) with its detection in two separate bands is currently the only analogue to this phenomenon, even though in that case the fluence of the 1.5 GHz burst was double the one(s) at 600 MHz.
The only other case where emission was detected by two telescopes at the same time was on MJD 58836 during an observation of \re\ at low frequencies \cite{chawla20} involving CHIME and the Robert C. Byrd Green Bank Telescope (GBT). In the latter case, the bandwidths were contiguous and the emission was observed on the lower $\sim$50~MHz of CHIME's band and the full extension of GBT's 100 MHz band. The burst for which coincident emission was detected was the only burst detected at the top of the GBT band (400 MHz), with a
fluence of $\sim$49 Jy ms in the GBT band and a fluence of $\sim$1 Jy ms in the CHIME band. 
In total, ten unique bursts were observed by the two instruments during this campaign. A total of 6 out of 7 GBT bursts were not detected by CHIME, even though their GBT fluences were higher than, or comparable to, the corresponding 95\% confidence fluence threshold. 

\begin{figure}[H]
    \includegraphics[width=0.7\textwidth]{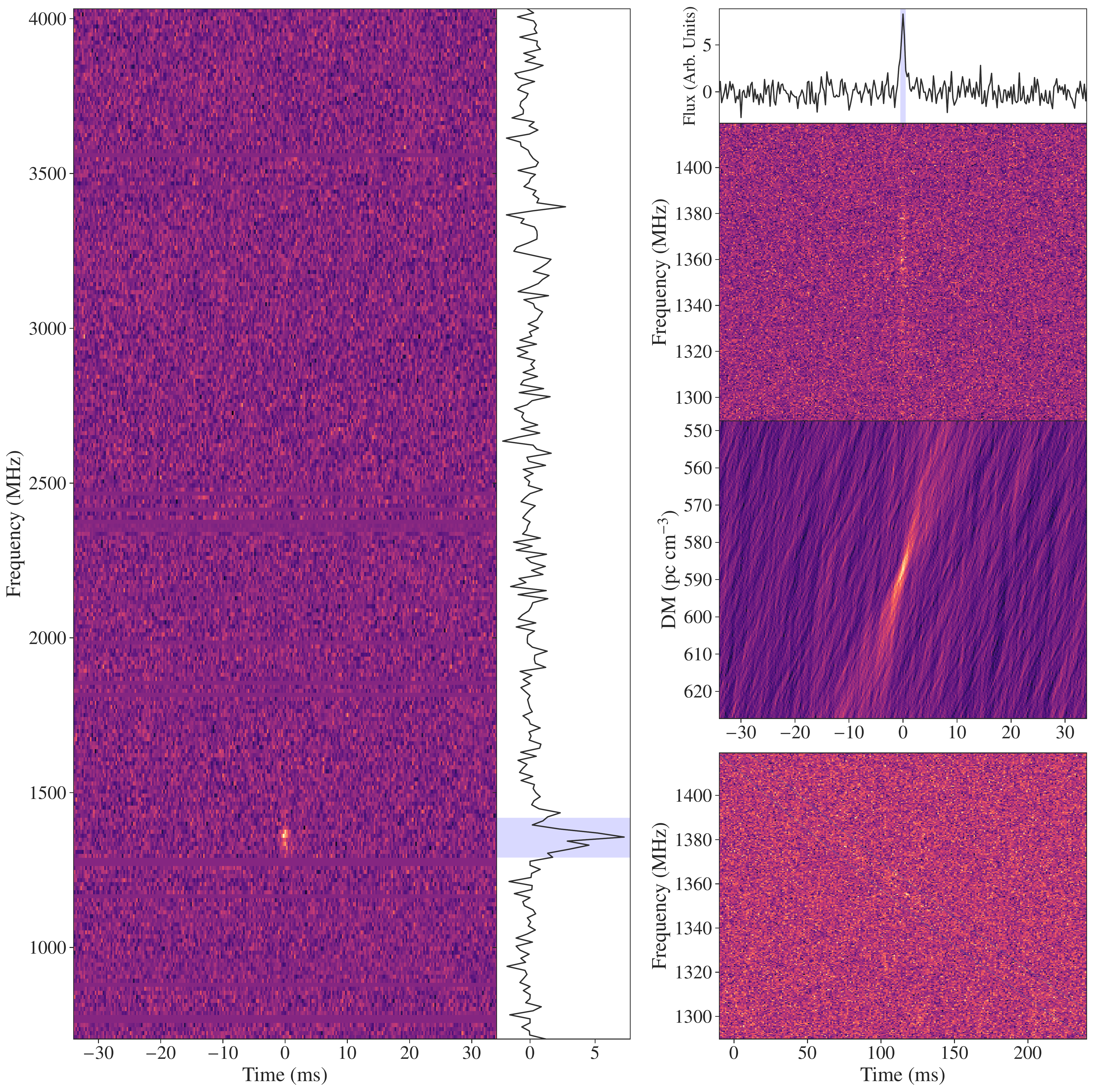}
    \caption{Figure 3 from \cite{kumar21} ``Extremely band-limited repetition from a fast radio burst source'', reproduced with permission from Pravir Kumar; published by MNRAS (RAS Publishing), 2021. The left plot shows the full band of the UWL receiver at Parkes with the FRB barely distinguishable from interference at frequencies below 1500 MHz. The right plot shows the profile, its band-limited dynamic spectrum and the diagnostic plots of DM vs time and dispersed dynamic spectrum hinting to the real FRB nature of the source.}
    \label{fig:kumar}
\end{figure}

Conversely, what is becoming a common feature in most cases, which can be seen if the observation bandwidth is large enough \cite{michilli18,gajjar18, chime8frb}, is that a significant portion of the flux of the
burst lies within a very narrow band. 
A similar behaviour is noticed even with smaller bandwidths when FRBs are detected at the edge of the band, and it is apparent that it is intrinsic and not influenced by sensitivity issues with the bandpass.
An extreme example of this (which will also represent an observational challenge for the new instruments in the coming years) is the detection of a repeat burst from FRB 20190711A with the ultra wide-band low (UWL) receiver at the Parkes telescope \cite{kumar21}: while the UWL bandwidth spans the 0.7--4.0 GHz range, emission from this source was only observed over a band of 65 MHz (see Figure \ref{fig:kumar}).
Interestingly, band-limited flux knots have recently been observed also in the Crab pulsar twin, PSR J0540--6919, in its giant pulse emission \cite{geyer21}.

It seems that a pattern is emerging whereby FRB emission {  {in repeating sources (it is still difficult to make the case for one-off events)}} only comes in specific ``spectral islands''. In the case of the periodic repeaters \ri\ and \re\, adjacent ``islands'' are active at different phases of the active cycle, going downwards in frequency with time (see Section \ref{sec:lowest}). 
It will be interesting to understand whether this is also a common behaviour. One promising source in this respect is repeater FRB20201124A. An extremely active phase was observed for this FRB mainly at CHIME's bands and L-bands between March and May 2021 \cite{piro21,marthi21,lanman21}. These observations, despite the richness of bursts, have not shown evident periodicities, nor did they highlight a clearly ``chromatic'' active window.

\section{Understanding the Nature of FRBs}
\label{sec:models}

The high brightness temperatures ($ T_b \sim 10^{34-37}$~K for standard FRB parameters~\cite{zhang20Nat}) and the duration of the bursts have soon led to the idea that FRB emission must be coherent, meaning that the radiation by relativistic electrons is enhanced with respect to the total emission that would be observed if electrons radiated independently. 
The observed power, for extragalactic sources, suggests a compact object must be responsible for FRB emission.
It is not clear, as yet, whether one, two (or more?) classes of FRBs exist. Some FRBs have never showed repetition despite extensive searches (see e.g.,~\cite{petroff15,shannon18}). At least one case exists, however, where one single burst was observed by ASKAP \cite{shannon18}, and two more were identified after extensive coverage (80h) by GBT \cite{kumar19} which were 500 times weaker than the original burst.
In addition, as already pointed out, contrary to the rates observed at higher frequencies, the very active \ri\ has only been seen once by CHIME. 
These findings show that different telescopes (/bands) can be better suited for the observations of different FRBs, but also that many hours of follow-up from sensitive telescopes would be needed to answer the question, unless a clear answer is provided by alternative measurements, i.e., the~detection of an association between an FRB and a cataclysmic event. 
In the following, we will mostly focus on mechanisms that can reproduce multiple bursts.
With the detection of repeated bursts, models flourished and, until recently, more models than FRBs could be counted \cite{platts19}. The rapid growth of detections changed this proportion and made many models converge to a general consensus whereby magnetars are the origin of repeater FRBs. Strong magnetic fields and acceleration of relativistic particles can be responsible for bright coherent emission.
The detection of a possible FRB from SGR\,1935$+$2154 has provided a direct connection between {  {FRBs and magnetars}}, even though {  {this connection}} still lacks {  {the substantiative evidence of an extragalactic event}} (see Section \ref{sec:intro}). 

Magnetars, as we see them in our galaxy, are seemingly isolated neutron stars whose emission can be explained by invoking magnetic fields of exceptional intensity and complex morphology, at least in some components of their structure. Although magnetars are relatively rare amongst neutron stars in the Milky Way, the fact that they were discovered (and are usually observed) in transitory phases of increased activity intensity (outbursts) makes their real number likely much higher.
We currently know of little more than a handful of galactic magnetars emitting in the radio band (see e.g.,~\cite{kaspi17,esposito21}). The radio emission in magnetars is temporary in nature. It is activated during an X-ray outburst and, in some cases, it outlives the X-ray flux enhancement.
During the active radio window, pulsed emission is observed but, in contrast to one of the distinctive features of ordinary rotation-powered pulsars, the pulse shape of a magnetar is not stable. Magnetars have been observed in radio frequencies up to very high frequencies (290 GHz, \cite{torne17}), showing a much harder spectrum than pulsars: $S \propto \nu^{-0.5}$ or flatter, with $S$ representing the magnetar's flux density. 

\begin{figure}[H]
    \includegraphics[width=0.75\textwidth]{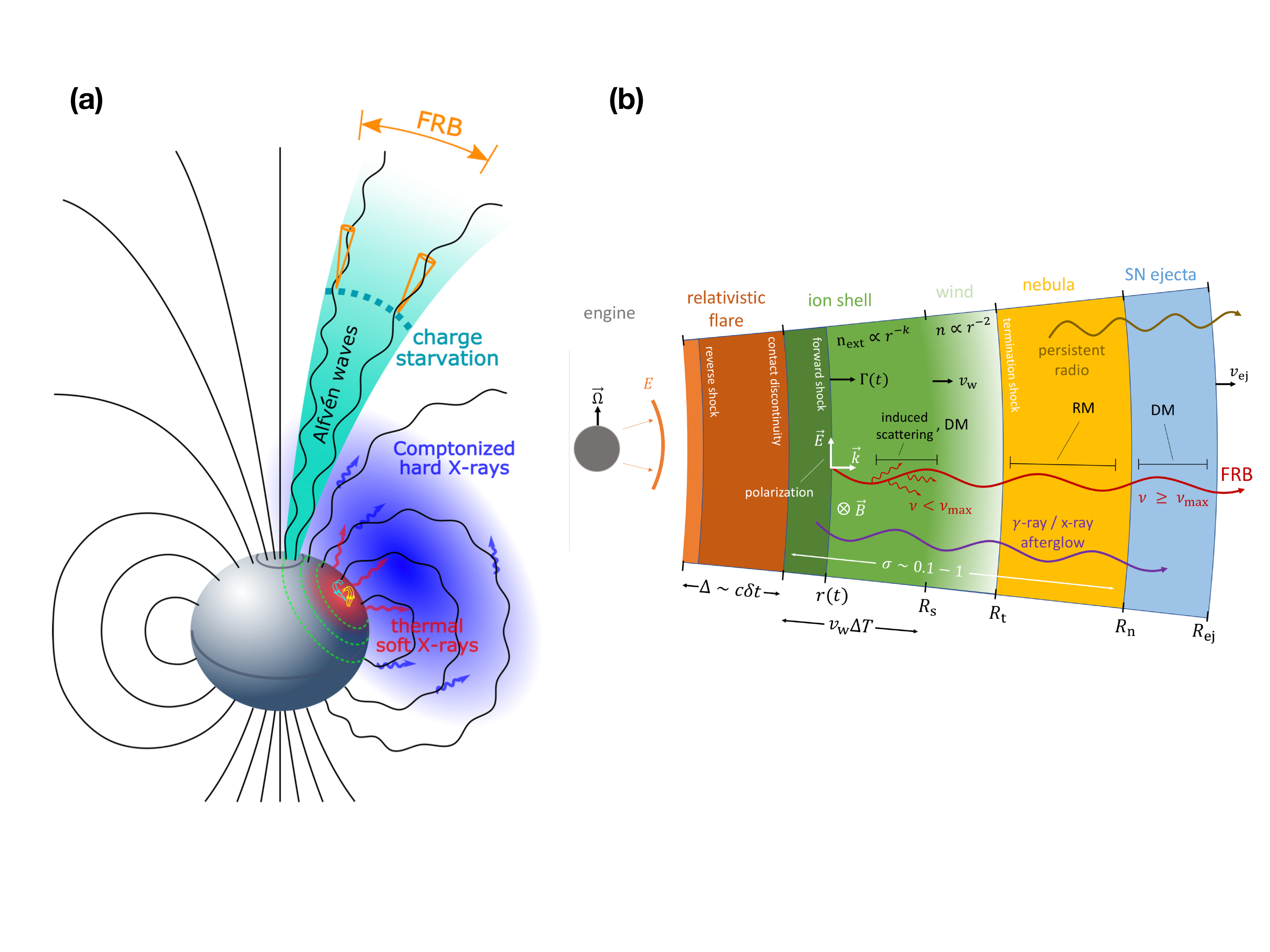}
    \caption{{Left panel: Figure 2 (left panel)} from \cite{lkz20} ``A unified picture of Galactic and cosmological fast radio bursts'', reproduced with permission from Wenbin Lu, published by MNRAs (RAS Publishing), 2020. Right panel: Figure 1 from \cite{metzger19} ``Fast radio bursts as synchrotron maser emission from decelerating relativistic blast waves'', reproduced with permission from Brian Metzger, published by MNRAS (RAS Publishing), 2019. A comparison of the two most general scenarios for magnetar models: (\textbf{a}) pulsar-like or magnetospheric models; (\textbf{b})~GRB-like or external models. See text for details.}
    \label{fig:zhang}
\end{figure}

The mechanisms through which magnetars could be powering FRBs can be divided into two broad classes of models (see Figure \ref{fig:zhang}).
One class advocates magnetospheric emission from magnetars (pulsar-like models) \cite{popov10,lyutikov20,lkz20}. One way to generate bursts within the magnetosphere is to have a magnetic pulse from the inner magnetosphere travelling towards the outer magnetosphere and triggering magnetic reconnection \cite{lyubarsky20}. \textls[-15]{This in turn generates high-frequency fast magnetosonic waves that eventually escape as electromagnetic waves.
An alternative way to produce them is  coherent curvature radiation by bunches of charged particles~\cite{kumar17,ghisellini18}.} The particles would have to be contained in a region of a wavelength size and accelerated
synchronously in such a way that their electric fields add up in
phase.  One critical requirement for this model is that there should be a strong electric field parallel to the magnetic field in the
emitting region. 
As the bunches stream out along the open field line region, the characteristic frequency of curvature radiation normally drops. If our line of sight sweeps the discrete ``spark-like'' bunches in adjacent field lines, softer emission is observed at later times. This provides a possible interpretation to downward-drifting subpulses.
A sort of radius-to-frequency mapping resembling what is seen in drifting subpulses, {  {typical of pulsar emission (see, e.g.,} \cite{rudsuth75}),} can finally be achieved also in the so-called low-twist models~\cite{wadiasingh19} where pair cascades are created by dislocations and oscillations of the magnetic field at the neutron star surface, provided that the background charge density is sufficiently~low.

Another class of models invokes synchrotron maser as the mechanism responsible for FRBs  (GRB-like models )\cite{lyubarsky14,metzger17,marmet18}. 
These models require either internal or external shocks, which are expected to arise from relativistic flares that may be
ejected during magnetar outbursts. Some common requirements of all synchrotron maser models are that the Lorentz factor should be greater than 1 and that the magnetic field must be relatively low to produce the observed $\sim$\,GHz radio emission.
\textls[-15]{In one scenario~\cite{metzger19} the emission comes from the interaction of the new burst with the surrounding medium which has been ionised and partly magnetised by previous bursting activity, and it can therefore be activated, in cycles, and with different fluence levels, depending on the history of the interactions. 
As the shock from one burst propagates in the medium, the maser frequency decreases with time, as the plasma density progressively decreases with increasing radius. 
Sub-bursts, especially very short duration ones~\cite{nimmo21, majid21,michilli18}}, are not predicted by this model and would actually be hard to reconcile with it. However, the frequency down-drifting observed in some FRBs also from bursts that are separated by some time (e.g.,~the behaviour of periodic repeaters) can be explained by this framework.
In an alternative scenario, the burst is produced, with a much higher efficiency, when the ultra-relativistic flare ejecta collides with the pulsar
wind nebula \cite{lyubarsky14}.

A discrimination amongst the two macro-classes of magnetar models and then their sub-classes will probably be made possible by more accurate localisation, which can help pinpoint the region surrounding the FRB or, more broadly, as in the case of FRB 20200120E localised within a globular cluster in the nearby galaxy M81 \cite{bhardwaj21,kirsten21}, can help us probe the formation channels of FRB progenitors.
On the other hand, a complementary way through which we can characterise the environment surrounding the FRB engine is with the help of low-frequency observations in the radio range.  

\section{The Role of the Local Environment}
\label{sec:implications}

\ri\ was found in a star-forming region of a dwarf galaxy \cite{chatterjee17,tendulkar17}, and a persistent source was spotted by the VLBI observations \cite{marcote17}, coincident with its position, extending for $<$0.7\,pc, with flux density $S=180~\upmu$Jy. The likely presence of a nebula surrounding the central engine of \ri\ was corroborated by the evidence in favour of a high magneto-ionic environment in the vicinity of the source \cite{michilli18}.
 Polarisation measurements, and, in particular, the high value and large variations of its rotation measure, seemed to imply that \ri\ is located in a dense environment. 
An interacting nebula well fits the picture of a newly-born magnetar emitting its first cries. The presence of the nebula could also easily explain why the activity of \ri\ seemed to be conspicuous at high frequencies and to be lacking at low ones. Ref. \cite{metzger19} predicted that, indeed, the duty cycle of lower frequency bursts had to be smaller and that emission at those frequencies might not be visible in the first years after the birth of the magnetar because the surrounding medium would be opaque to radiation below 1 GHz.

Ref. \cite{raviloeb19} analysed in more detail the impact of a dense surrounding medium on the radio emission of FRBs at different frequencies. If a cutoff frequency could be determined, the properties of the
immediate progenitor environments of FRBs could be constrained and, in turn, also of the progenitors themselves.
The presence of such a cutoff would also influence fluence and redshift distributions of FRBs at different frequencies.
They considered propagation effects that can suppress the
observed emission from FRBs at frequencies lower than a turnover frequency $\nu_{peak}$ making no assumptions about the intrinsic FRB emission mechanism.
Plasma absorption can suppress emissions at\linebreak $\nu \leq \nu_p \sim 90$\,MHz, where $\nu_p$ is the plasma frequency, and it is dependent on the electron density so that it can change for relativistic electrons. The Razin--Tsytovich effect widens the emission cone of relativistic beaming from individual electrons, and it should affect frequencies lower than $\sim$3\,GHz depending also on electron density and the magnetic field of the emission region. This effect, however, can only provide a lower limit to the characteristic frequency. Stimulated Raman scattering affects beamed emission from sources with high brightness temperatures at frequencies $\nu \leq 130$\,MHz, depending on electron density and temperature.
In the presence of higher photon and electron densities, induced Compton scattering (ICS) would make radio photons lose significant energies to thermal electrons. The spectral changes induced by ICS affects all frequencies for sources with sufficiently high brightness temperatures. The presence of induced Compton scattering would place strong constraints on the electron density in the immediate environment surrounding the source. Finally, free-free absorption of the radiation is expected from a thermal plasma surrounding the source, and it would suppress frequencies $\nu \leq 300$\,MHz depending on electron temperature and emission measurements of the plasma.

The abrupt cutoff at 400 MHz, according to the initial observational evidence, could be explained by absorption. 
It was possible, however, that multiple mechanisms concurred to define critical frequencies and influence the spectrum of FRBs. Different mechanisms at play would lead to different limiting frequencies, and an interplay of them could result in different cutoff frequencies.
Plasma lensing \cite{cordes17}, for example, could be a cause for the enhancement of emissions on one side of the spectrum and suppression on the other, which could strongly influence the spectral measurements.
One prediction of the described scenarios was the detection of events at higher redshift for low-frequency observations, as these would be probing the population below $\nu_{peak}$ and would be sampling a flat fluence distribution.

\section{Low-Frequency Studies}

Until 2013, FRBs had only been discovered in archival surveys, and it was only in 2014 that real time detections started and could be complemented by rapid multi-wavelength follow-up.
All FRBs discovered prior to 2015 were observed at 1.4 GHz using single dish antennas with relatively poor angular resolution. This implied that the observing telescope alone did not have the capability to perform localisation, albeit with little accuracy, and, consequently, even follow-up was a huge effort with little chances of success. 
The first detection below 1.4 GHz came from the archival data of the Green Bank Hydrogen Mapping Survey \cite{masui15} between 700 and 900 MHz. The survey had a time resolution of 1.024 ms and a frequency resolution of 49 kHz, allowing for good sensitivity out to $DM\sim 2000$\,\pc. FRB\,110523 was identified as a real astrophysical source out of $\sim 6500$ candidates at $DM=623$\,\pc. 
Based on experience with pulsars, scattering was expected, and it was indeed observed at these frequencies, with a time constant $\tau_{sc} = 1.66$\,ms at 800 MHz. 
The presence of scintillation, along with scattering, implied strong scattering near the source which, in turn, hinted to the fact that the screen must be close to the emitting source. The observed scattering was too strong to be caused by the disk of the host galaxy and should therefore be attributed to either a strongly-scattering compact nebula or to the dense inner regions of the host galaxy. 
Based on this single detection out of 660 hours of archival data, ref.~\cite{connor16} concluded that steep, non-Euclidean, distributions for the $N(>F) \propto F^{-\gamma}$ relation with $\gamma > 2.2$ could be excluded at the 95\% confidence level. They further  determined a detection rate for GBT observations similar to those of this survey in $\sim$ 0--$\sim5$ events per survey.
Based on this rate, they also attempted estimates for future surveys at similar frequencies, with the caveat that an interplay of different factors influences this estimate, i.e.,~the underlying FRB fluence and scattering distributions to be combined with a survey’s thermal
sensitivity, fluence completeness, and observing frequency. 

\subsection*{How to Look for FRBs at Low Frequencies}
The details of how FRB observations are generally performed can be found in various references (see, e.g.,~\cite{petroff19}). Here we will discuss the specific aspects of and challenges to low-frequency observations.
FRB observations require a good compromise between time and frequency resolution. With widths of some milliseconds, the optimal time resolution has to be of an order or lower than $\sim$ms. The frequency resolution, on the other hand, will constrain the maximum DM that one is sensitive to. 
The time delay caused by DM has a quadratic dependence on the inverse of the observing frequency:
\begin{equation}
     t= \mathcal{D} \times {\frac{DM}{ \nu^2}}.
\end{equation}

Here, $\mathcal{D} = e^2/2\pi m_e c \simeq 4.15\times 10^3$\,MHz$^2$\,pc$^{-1}$\,cm$^3$\,s is the so-called dispersion constant (see \cite{lk04} and discussion in \cite{kulkarni20}), and it depends on physical constants. It follows that the delay across a single channel can be relevant at low frequencies when compared to the corresponding time at an infinite frequency:
\begin{equation}
     t_{DM}\simeq 8.3 \times 10^6 \mathrm{ms} \times DM \times \Delta \nu \times \nu^{-3},
\end{equation}
where $\Delta \nu$ is the width of the channel and $\nu$ is the observing frequency.
For this reason, a fine channelisation is required, taking into account the residual intra-channel smearing that could wrap the signal within the channel and prevent detection.

The time delay caused by DM smearing across the full band of the observation is corrected for by using a well-known technique from pulsar astronomy which was also fundamental for the dawn of FRB studies. This technique, called {\it dedispersion}, ensures that appropriate time delays are applied to each channel as a function of the DM of the source, so that the received signals arrive at the output of each channel at the same time. If the DM is not known a priori, it is necessary to create different time series, each with the correction corresponding to a specific DM, and then search for burst, i.e.,~using matched-filtering techniques \cite{cm03}.
Ref. \cite{cm03} paved the way for such studies, and they provided a
framework to determine the parameters to be used for this blind search in order to obtain an acceptable signal loss as a function of the DM step size, the centre frequency and bandwidth of the observations and the width of a Gaussian pulse $W_{obs}$. From their Equations (12)
and (13):
\begin{equation}
    {S(\delta DM) \over S} = {\sqrt{\pi}\over{2}}\zeta^{-1} \mathrm{erf} \zeta
\end{equation}
where
\begin{equation}
    \zeta=6.91 \times 10^{-3} \delta DM {\Delta\nu_{\rm{MHz}}\over{W_{obs}\nu^{3}_{\rm{GHz}}}}.
\end{equation}

The technique described so far is also currently referred to as {\it incoherent dedispersion}, in contrast with {\it coherent dedispersion} \cite{hanrick75}. Coherent dedispersion acts on the complex raw data incoming from the telescope, and it recovers the intrinsic phase of the signal at the source by applying an appropriate transfer function of the interstellar medium. 
Coherent dedispersion has proven to be fundamental for the recovery of ms-long signals at low frequencies, as it can intrinsically compensate for the intra-channel smearing. However, this task requires that the complex voltages are recorded in a baseband format and temporarily stored. It is now a task that can be done routinely on software using GPUs, but it is computationally demanding and is thus mainly performed when the DM of the source is known a priori.
A hybrid search approach has been proposed by some authors \cite{bassa+16}, whereby a number of coarse steps of coherent dedispersion are interlaced with a finer grid of incoherent dedispersions in between. This method will be particularly useful when following up repeating sources of FRBs with known DMs. 

\section{Low-Frequency Surveys---Single Dishes}
\label{sec:dishes}
\subsection{GBT}
\paragraph{Drift scan survey}
At 350 MHz, a significant fraction of the transient sky (10,000 deg$^2$) was covered in 2007 by the GBT drift-scan survey \cite{boyles13,lynch13}, with a total of 1491 hours of observations, an instantaneous field of view of about 0.3 deg$^2$ and a $10-\sigma$ sensitivity for pulses of width 3 ms of about 35 mJy. 
At this frequency, an FRB was expected
to have a peak flux well in excess of this threshold even out to $z > 1$. 
Ref. \cite{lorimer13} predicted that scattering at these frequencies should be below 10 ms for a DM of a few hundred \pc. For a limiting DM of 500 \pc, assuming that 20\% of this DM can be attributed to our galaxy and the FRB host, they inferred a redshift limit $z<0.33$ for this survey (following the approximate intergalactic medium scaling law $DM \sim 1200z$ \pc\ from \cite{ioka03,inoue04}). One FRB of order was expected for the entire
survey. None were found. 

\paragraph{GBNCC}
In 2009, a more systematic survey of the sky accessible to GBT started observations which shall be completed in 2021. The Green Bank North Celestial Cap survey (GBNCC,~\cite{stovall14}) was devised for the discovery of pulsars also through single pulses, and it targeted the full sky observable from the GBT at 350 MHz. The data spanned 100 MHz of bandwidth split into 4096 frequency channels. Each pointing on the sky observed for 120 s and sampled with a 81.92-$\upmu$s time resolution. At the beginning, the data were searched to a maximum DM of 500\pc\ but, following the discovery of FRBs, the maximum DM for the search was increased to 3000 \pc. Ref. \cite{chawla17} presented an analysis of all pointings up to May 2016. For those dedispersed to 500\,\pc\ the data were analysed in search for FRBs only when the predictions by NE2001 towards that direction in the galaxy did not exceed 100\,\pc. The minimum detectable fluence for the GBNCC survey assuming no scattering and a mean DM of 756\,\pc was 3.15 Jy ms.

This first pass at the GBNCC survey, totalling an on-sky time of 61 days (at\linebreak $DM_{max}=3000$\,\pc) plus 23 days (at $DM_{max}=500$\,\pc), yielded no detections.
Scattering was investigated as a possible cause for the non-detections at these frequencies, as it was difficult to match them with the predicted rate from the Parkes surveys \cite{kp15,champion16,crawford16} if the mean of the scattering time distribution was set to be the observed 6.7 ms at 1 GHz. According to the NE2001 model, the directions sampled by the GBNCC were not affected by significant galactic scattering, with the ISM accounting for, at most, a 10\,ms scattering towards most pointings. The contribution of the IGM could also be excluded as a cause for significant scattering both in relation to the detection of FRB 110523 at 800 MHz from \cite{masui15} and from the predictions of \cite{katz16}, who pointed out that the low densities of the IGM make it unlikely as the location of a scattering screen.
On the other hand, the area surrounding the FRB, if dense enough, could produce all the scattering needed to render 350 MHz observations by the GBNCC unsuccessful. This could either be the result of a dense nebula or a supernova around the source or its location near the galactic center.
Alternatively to external factors, the non-detections could be attributable to the intrinsic spectral properties of FRBs.
The survey yielded a limit on the spectral index for the FRB distribution $\alpha$ \textgreater~0.35. The constraint depends on the cosmological distribution of the sources, and it was obtained with the caveat that the $-N -F$ has a Euclidean standard distribution of 3/2.  
Further constraints obtained from the scattering simulations \cite{chawla17} obtained a limit on the spectral index $\alpha > -0.3$, valid only in the absence of free-free absorption (see Section \ref{sec:implications}). 
The upper limit on the event rate of FRBs obtained from this survey was $R_{FRB} <3620$\,sky$^{-1}$\,day$^{-1}$ for a flux limit of 0.63 Jy.
 
\paragraph{First GBNCC detection} 
The first detection of an FRB from the GBNCC came only much later \cite{parent20} with the detection of FRB 20200125A. This was the first blind discovery of an FRB below 400 MHz. 
The burst had a $DM = 174$\,pc well in excess of the galactic contribution in this sky direction. No scatter broadening was noticeable in the profile even though the low S/N of the pulse could be affecting the measurement. 
The survey completed 90\% of the full sky coverage above GBT in July 2020. The detection of one event in 45.5 days of total on-sky time yielded an update in the FRB rate computed by \cite{chawla17}: $R_{FRB} <5500$\,sky$^{-1}$\,day$^{-1}$ above a flux density of 0.39 Jy for a 5 ms burst. The two limits are compatible at the 95\% level, with the last one being slightly higher, possibly due to the improved survey sensitivity in terms of search algorithms. 
This rate was also consistent with the previous 330 MHz limits obtained by \cite{deneva13,rajwade20a} (see following Sections). The authors cautioned, however, that this rate might be underestimated because the sensitivity to narrow-band bursts, such as those coming from repeaters, would be diluted in band-averaged searches.
 
\subsection{Arecibo}
Similarly to the GBT drift scan survey, the Arecibo telescope started a drift scan survey at 327 MHz for pulsars and transients after the commissioning of their new low-frequency receiver during a maintenance period in 2010, which was later converted into a full proposal with an observing time of 400 hours per year. 
A comprehensive analysis of the data up to 52\% of Phase 1 (declinations from  $-1\deg$ to $28 \deg$) was presented by \cite{deneva13}. Data were searched up to $DM= 1000$\,\pc\ with no detections of FRBs.


\subsection{Lovell}
In 2016, the Lovell telescope underwent a maintenance period, during which observations were performed in the parked position and dedicated to a drift scan survey at 332~MHz for a cumulative time on sky of 58 days \cite{rajwade20a}. The observations were performed using 0.5~MHz frequency channels and a time resolution of 256 $\upmu$s. Dedispersion was performed up to 1000 \pc.
The observation band was clouded by severe radio frequency interference (RFI), which resulted in the impressive number of  $\sim$1000 single pulse candidates in every 5 minutes of observations. The use of clustering and grouping algorithms in the search pipeline based on {\tt heimdall} \cite{barsdell12} and then of machine learning classifiers from the {\tt FETCH} software package \cite{agarwal20fetch} reduced visual inspection to a total of 675 candidates in $\sim$1400\,h of observations. None of these candidates were confirmed as real FRBs.
 The survey covered 0.61 $\deg^2$ of sky at any instance in time, as calculated from the primary beam, and its limiting sensitivity was estimated to be 45.9 Jy ms.
  This made non-detections by this survey constrained in terms of an upper limit to the rate of FRB events at 332~MHz. Similarly to what was estimated by \cite{chawla17} for the GBNCC survey, the expected rate was, in this case, $R_{FRB} <5500$\,sky$^{-1}$\,day$^{-1}$, assuming a uniform distribution of FRBs in a Euclidean~universe.

\section{Low-Frequency Surveys---Arrays}
\label{sec:arrays}
All-sky detectors were the first exploited radio telescopes. In time, however, they have progressively been abandoned in favour of larger and larger single dishes. 
The last decade has witnessed a revitalised input for smaller dishes or even much simpler designs (i.e.,~dipoles) combined into aperture arrays or transit telescopes. This was partly due to the hardware costs being more contained and because we have reached the limit of how much bigger dishes can be, but also to pave the way to the future Square Kilometer Array (SKA) era by understanding how to deal with a vast amount of signals and an unprecedented data flow.  The complexity and the cost of these arrays of small elements rely more on the software side in terms of development, data handling and supercomputing resources.

\begin{figure}[H]
    \includegraphics[width=0.7\textwidth]{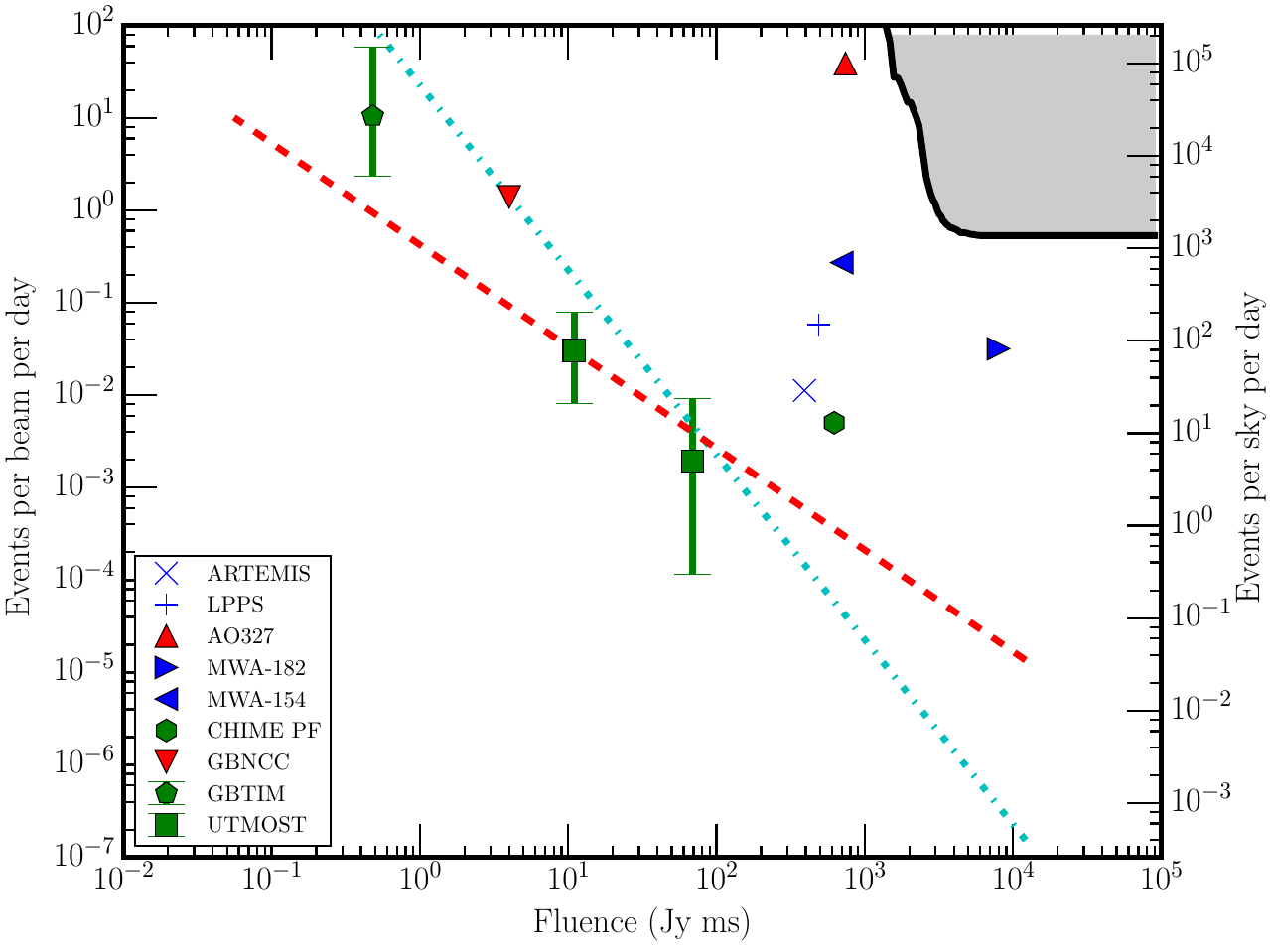}
    \caption{Figure 7 from \cite{terveen19} ``The FRATS project: real-time searches for fast radio bursts and other fast transients with LOFAR at 135 MHz'', Credit: S. ter Veen, A\&A, vol 621, page 57, 2019, reproduced with permission \copyright\ ESO. reproduced with permission from. It summarises the limits on FRB event rates ($R_{FRB}$) versus fluence, normalised to a pulse width of 8\,ms. They were obtained at different low frequencies from either the detections (symbols with error bars) or non-detections (other symbols), described in these sections in the era before CHIME. The color coding represents different frequency bands: green for $\sim$800\,MHz; red for $\sim$350\,MHz; blue for $<$200\,MHz plus the limits obtained by the FRATS survey given by the shaded grey region. The red dashed line represents $\gamma = -1$ for the FRB distribution, while the green dotted line represents a Euclidean distribution with $\gamma = -3/2$.}
    \label{fig:terveen}
\end{figure}

An efficient survey would ideally combine high sensitivity, a large total field-of-view (FoV) and a high angular resolution per FoV element.  By detecting FRBs with a low-frequency radio interferometer and at the same time being able to retrieve short snapshot image data, FRB positions can be constrained to high accuracy ($<$1 arcmin), enabling host galaxy associations and deep constraints on multi-wavelength counterparts.
Based on the specifics of the surveys performed at 350 MHz, \cite{lorimer13} predicted a detection rate above 30 Jy with $DM<500$\,\pc, for 150 MHz surveys with large fields of view of $\sim$1 event per day per 30\,deg$^2$.
Indeed, simulations of how the constraints obtained from the previously listed non-detections fit with those from surveys at low frequencies gave consistent results (see Figure \ref{fig:terveen}).

\subsection{LOFAR}

The era of aperture arrays looking for FRBs started with the first sky survey performed by the LOw Frequency ARray (LOFAR) \cite{vanhaarlem+13}.  
LOFAR
is a radio interferometric array that operates at very low frequencies (10--240 MHz). Each LOFAR station is composed of two sets of antennas: the low-band
antennas (LBA) operating between 10 and 90 MHz, and
the high-band antennas (HBA) operating between 110 and
250 MHz. Currently, LOFAR is composed of 24 core stations and 14 remote stations in the Netherlands, as well as 13~international stations.
LOFAR’s beamformed modes \cite{stappers11} can provide an extremely wide FoV, $>$10\,deg$^2$, as well as the ability to constrain positions to a few arcminutes.

\paragraph{LPPS and LOTAS}
During the 2008--2012 commissioning of LOFAR, two pilot pulsar surveys looked for pulsars and fast transients \cite{coenen14}.  The LOFAR Pilot Pulsar Survey (LPPS) employed incoherent beam-forming, providing lower raw sensitivity, with signal-to-noise ratios (S/N) increasing only according to the square root of the number of stations being added but, on the other hand, encompassing a large FoV which would favour observations of bright, rare events. The second survey, the LOFAR Tied-Array Survey (LOTAS), employed coherent (tied-array) beamforming using only the  innermost six stations of the LOFAR core. Coherent beamforming offered maximum instantaneous sensitivity, scaling linearly with the number of stations summed, while allowing only for limited FoV, which made this survey less constraining for FRB searches.
LPPS single-pulse data were searched at DMs between 2 and 3000 \pc down to a limiting S/N of 10, and no candidate FRBs were found. The sky rate of FRBs derived from LPPS non-detection, based on its limiting sensitivity and time resolution (0.66 ms), was $R_{FRB} <150$\,sky$^{-1}$\,day$^{-1}$ above a fluence $F>107$\,Jy\,ms.
This limit increases by a factor $\sqrt1.125n$, where $n$ is the number of bins over which the burst might be smeared due to dispersion at high redshifts. The corresponding upper limit on the volumetric rate was calculated as $\Phi_{FRB}<2.5\times10^5$\,Gpc$^{-3}$\,yr$^{-1}$ assuming a spectral index $\alpha = -2$.

\paragraph{Rawlings Array} In the following years, many dedicated surveys were put together to detect FRBs in real time.
It was one of LOFAR's single stations, i.e.,~the Rawlings Array in the UK, which was the first to publish a specific low-frequency survey dedicated to FRB detection \cite{karastergiou15}. The project aimed at the real-time detection of fast transients and, in order to do so, it exploited the Advanced Radio Transient Event Monitor and Identification System (ARTEMIS) backend. ARTEMIS put together the hardware and software required to continuously monitor the large LOFAR FoV and detect FRBs in real time by applying HPC techniques.
In total, $\sim$1500\,h of data were acquired by the Rawlings Array and mirrored, for the most part, by the companion LOFAR international station of Nancay in France (FR606) for coincidencing and better handling of radio frequency interference (RFI). Data were searched up to a $DM = 320$\,\pc. Following \cite{lorimer13,ioka03,inoue04}, the maximum redshift corresponding to this limit was 0.17, with an average of 0.13, with large associated uncertainties.
No FRB was detected, and this set another upper limit on the sky rate of FRBs at 145 MHz, with $R_{FRB} =29$\,sky$^{-1}$\,day$^{-1}$ above a fluence $F>310$\,Jy\,ms for a time resolution of 5 ms. 
The detection limits and volume sampled by the survey were based on the assumption that the FRBs are standard candles with broadband emission. 
Spectral index constraints may therefore have a stronger influence than the sampled DM space in this case.  The non-detections implied a lower limit on the spectral index of FRBs of $\alpha \geq 0.1^{+0.1}_{-0.2}$. 
In this framework, the limiting surveyed volume ($3.3 \times 10^7$\,Mpc$^3$) and the time resolution of the survey made an expectancy of detections for this survey of $\sim 1$ FRB. 
The deduced spectral properties, at odds with those of pulsars having a steep (negative) spectral index, pointed towards either an intrinsically more narrow-band emission process, or one characterised by a frequency-dependent emission geometry.

\paragraph{FRATS}
The  Fast RAdio Transient Search (FRATS,~\cite{terveen19}) project was developed for LOFAR with the aim to detect and localise FRBs and possibly other transients. FRATS used real-time triggering algorithms to save beamformed data relative to possible bursts and then perform offline imaging on those data. FRATS was sensitive to bursts 2--128\,ms wide, with a limiting sensitivity to fluences of $F=1$\,kJy\,ms. 
Three surveys were performed starting in 2013, running commensally with other observing projects at LOFAR. The first two surveys observed between 119 and 151 MHz for a total on-sky time of 68 and 99\,h, respectively. DMs up to 120\,\pc\ were searched in the first survey and up to 500\,\pc\ in the second.
The third survey, started in 2014, exploited the imaging domain on timescales from seconds to minutes, and it was mainly used to validate the interferometric mode capabilities of FRATS during commensal observations.
No FRBs were detected by these pilot surveys. 
The first one yielded an upper limit on the FRB sky rate $R_{FRB} < 1500$\,sky$^{-1}$\,day$^{-1}$ above a fluence $F>1.6$\,kJy\,ms for an 8-ms burst (taking into account possible scattering effects). The second one yielded an upper limit on the FRB sky rate $R_{FRB} < 1400$\,sky$^{-1}$\,day$^{-1}$ above a fluence $F>6.0$\,kJy\,ms for an 8-ms burst. The corresponding upper limit on the volumetric rate was calculated as $\Phi_{FRB}<134\times10^5$\,Gpc$^{-3}$\,day$^{-1}$,
 assuming an Euclidean Universe and a spectral index $\alpha = -1.1$.
FRATS' pilot surveys had a limiting sensitivity which did not allow a substantial improvement over previous LOFAR rates, but they demonstrated the capability of the system to perform beamforming plus imaging searches, to be extended to the full array.

\subsection{MWA}

The Murchinson Wide-Field Array (MWA) \cite{tingay13} is a low-frequency radio interferometer located at the CSIRO Murchison Radioastronomy Observatory in Western Australia. 128 tiles form the array, which extends over an area of 3 km diameter. MWA can observe between 80 and 300 MHz, with 30 MHz of processed bandwidth for both linear polarisations.
Additionally, in the case of MWA, like for LOFAR, different types of searches can be performed, with a trade-off in processing between coherent
searches (maintaining the high data rate, sensitivity, and angular resolution at the expense of computational complexity) and incoherent searches (less sensitive, lower data rates, and worse angular resolution, but greatly reduced computing complexity). 

Two companion surveys \cite{tingay15,rowlinson16} were performed soon after the first LOFAR surveys, using MWA. 
The two pilot surveys on FRBs both used the same dataset targeting one specific field, which was originally acquired for an epoch of reonisation observations. 
They used complementary analysis strategies, which not only allowed them to search the dataset more thoroughly but which would also, in case of detection, provide independent confirmation.
In~\cite{tingay15}, 10.5\,h were analysed in standard imaging mode with high time resolution. The observations had 2-s integration times and 40 kHz frequency resolution with dual polarisation correlation. Images of the full MWA FoV were generated at arcminute angular resolution. Dedispersion was then applied to the dynamic spectrum produced from each resolution element in the images. The temporal resolution of this survey was likely not optimal for FRB searches, but it was smaller than the DM delay at these frequencies across the full band for the maximum searched $DM=700$\pc.
Ref. \cite{tingay15} privileged sensitivity by sacrificing the surveyed area: the predictions from \cite{ttw13} applied to the observational parameters relevant to this survey were that MWA may be able to detect between 0 and 10~FRBs per 10~h of observation, depending on the spectral index of the FRB emission: $\sim$9.3 for $\alpha=-2$, $\sim$1.6 for $\alpha=-1$, and $\sim$0.2 for $\alpha=0$. These predictions were also based on the standard candle luminosity model of \cite{lorimer13}. Given the non-detection of any FRB emission and their survey sensitivity to a limiting fluence of 700 Jy ms at 150 MHz, they concluded that $\alpha=-2$ could be rejected with high confidence ($>$99\%) and $\alpha=-1$ could be rejected with moderate confidence ($>$79\%). A limit on the spectral index was then derived as $\alpha>-1.2$. The corresponding upper limit on the FRB rate for $\alpha=0$ was $R_{FRB} <700$\,sky$^{-1}$\,day$^{-1}$ above a fluence $F>700$\,Jy\,ms, which was in agreement with previous results and compatible with the Euclidean distribution of FRBs, $N(>$$F) \propto F^{-3/2}$.

Ref.~\cite{rowlinson16}, on the other hand, sacrificed sensitivity to increase the amount of surveyed area. They analysed 78\,h of imaging data of the same field using 2\,min integration times divided over 4 $\sim 30$\,s snapshots. The resulting FoV was 452\,deg$^2$.
One potential candidate stood out in the analysis, and it was verified both against different pipelines and also using the method from \cite{tingay15}. This last comparison highlighted a faint candidate in the images, but no candidates in the dedispersed time series, implying that the candidate was not likely related to an FRB, as was later confirmed.
The non-detection of FRBs in this survey placed a limit of $R_{FRB} <82$\,sky$^{-1}$\,day$^{-1}$ at 182 MHz, above a fluence of $F>7980$\,Jy\,ms, assuming a flat spectral index and a constant distribution of FRBs. Given that this survey was aimed at a direction well away from the galactic plane, the sum of the galactic plus host distribution was estimated as $\sim 100$\,\pc based on the NE2001 model and the calculations of \cite{karastergiou15}. This implied, for the maximum searched $DM=700$\,\pc, a limiting redshift $z\sim 0.5$.


\subsection{Large Phased Array}
We note that detection of 9 plus 51 bursts at 111 MHz using the  Large Phased Array of the Pushchino Radio Astronomy Observatory has been claimed by \cite{fedrod19,fedrod21}. However, the observational setup of their system (a coarse frequency resolution of 78~kHz over a tiny observation bandwidth of 2.5~MHz, and a sampling time not faster than 12.5~ms) required the use of a template matching approach in order to see the bursts. Although the authors do their best to support the validity of this methodology, its use is very limited so far in the context of FRB searches, and the statistics of the false-positive are not completely assessed. Moreover, the claimed detection is very hard to reconcile with the stringent limits imposed by all other non-detections at similar frequencies, just described, derived by using well-consolidated procedures.

\section{Low-Frequency Surveys-Transit Telescopes}
With large FoVs in mind and the need for a long time on sky, the late 2010s saw in the transit telescopes the answers to the needs of FRB science.
\subsection{UTMOST}
The first low-frequency interferometer to have FRBs as its primary science goal was the Molonglo Observatory Synthesis Telescope (MOST). The 50-year-old instrument in Australia was refurbished in 2015 with a digital backend system and increased bandwidth to transform it into a burst finding machine \cite{caleb16,bailes17}. 
Two fully steerable East-West (E-W)-aligned cylindrical paraboloid reflectors cover a collecting area of 18,000 m$^2$ . The telescope operates at 843 MHz with a 30-MHz bandwidth. A relatively small investment in hardware and a more consistent investment in software upgrades made the UTMOST a sensitive instrument for FRB searches, with 128 frequency channels over the 30-MHz bandwidth and a time resolution of $\sim 655\upmu$s.
The selection of the sources and of the duration of the observation was also deferred to software based on predefined rules and real-time feedback from the data. 
Two surveys were performed during the commissioning of the system while the telescope was still running at a low fraction of the final sensitivity \cite{caleb16}. With the first one operating only over 16 MHz and the second with the upgrade to the full 31.25-MHz bandwidth, an average band of 16 MHz was conservatively considered for the cumulative results. 
The two surveys searched for FRBs up to $DM = 2000$\,pc and excluded all candidates at $DM<100$\,pc.
The first survey was more a validation of the system, with the sensitivity as low as 7\% of the final design one. No FRBs were detected in 467\,h of observations, implying a $2\sigma$ upper limit $R_{FRB} <1000$\,sky$^{-1}$\,day$^{-1}$ at 843 MHz above a fluence $F>23$\,Jy\,ms.
The second survey reached 14\% of the design sensitivity, and further upgrades made it  more sensitive by a factor of 2 than the first survey.
No FRBs were detected down to fluence limits of 11 Jy ms after spending 225\,h on sky, which yielded the same $2\sigma$ upper limit on the FRB rate, but this time at a lower fluence threshold. Both calculations were done on the assumption of a flat spectral index and a Euclidean distribution of the bursts.
A comparison with Parkes' detections set a limit on the spectral index $\alpha > -3.2$.

\paragraph{UTMOST detections}
A third survey, carried out from February to November 2016, covered an effective time span of 159 days and yielded the detection of three FRBs \cite{caleb17}. This was the first detection of FRBs using interferometers, and was achieved while the system was at only about 15\% of its theoretical sensitivity: for the fully upgraded
instrument, $S_{min} = 1.6$ Jy ms for a $10\sigma$ pulse. The event rate at 800 MHz derived from these detections was $R_{FRB} = 78$\,sky$^{-1}$\,day$^{-1}$ above a fluence $F>11$\,Jy\,ms.

Yet another survey was performed between June 2017 and December 2018, reaching an on-sky time of 344 days \cite{farah19}. The sensitivity of the system was improved compared to previous surveys both in hardware and in software: on one side by making the E-W arm not steerable anymore, as that caused failures on a regular basis, and on the other hand by improving the real-time detection and classification capabilities of the pipelines.
Five new bursts were discovered with this setup, enabling a look back to the raw data and the in-depth study of the time and frequency structure.
These detections yielded an improved event rate $R_{FRB} = 98$\,sky$^{-1}$\,day$^{-1}$ above a fluence $F>8$\,Jy\,ms. This rate was somewhat below the expectation from a scaling of $R_{FRB}$ as obtained by Parkes and ASKAP \cite{bhandari18,shannon18}, assuming that the average spectral index for FRBs is flat. These results would also not match the steep negative spectral index ($\alpha = -1.6$) predicted by \cite{macquart19}, advocating for a turnover around 1 GHz.


\subsection{Chime and the ``Industrial Revolution''}
\label{sec:chime}

The Canadian Hydrogen Intensity Mapping Experiment (CHIME, \cite{chime_frb_18}) consists of 4 stationary 20-m
wide and 100-m long cylindrical paraboloidal reflectors, aligned in the North-South (N-S) direction. Originally designed to map baryon acoustic oscillations, its large ($>$200 deg$^2$2) FoV, large collecting area,
wide radio bandwidth (400--800 MHz), and powerful correlator (which provides
1024 independent beams within the telescope’s primary beam) made it a potentially very interesting telescope for FRB searches. CHIME was therefore equipped with an FRB backend, and a series of pipelines were realised
in order to take full advantage of the large instantaneous
FoV of CHIME while maintaining the full sensitivity. The project undertook an unprecedented computational challenge to deal with the coherent beamforming of 1024 beams, a time resolution of 1 ms, dedispersion performed in real time and real time identification of candidates in the raw data in order to allow rapid discarding of empty data and rapid validation of good data for further processing offline.

\begin{figure}[H]
    \includegraphics[width=0.7\textwidth]{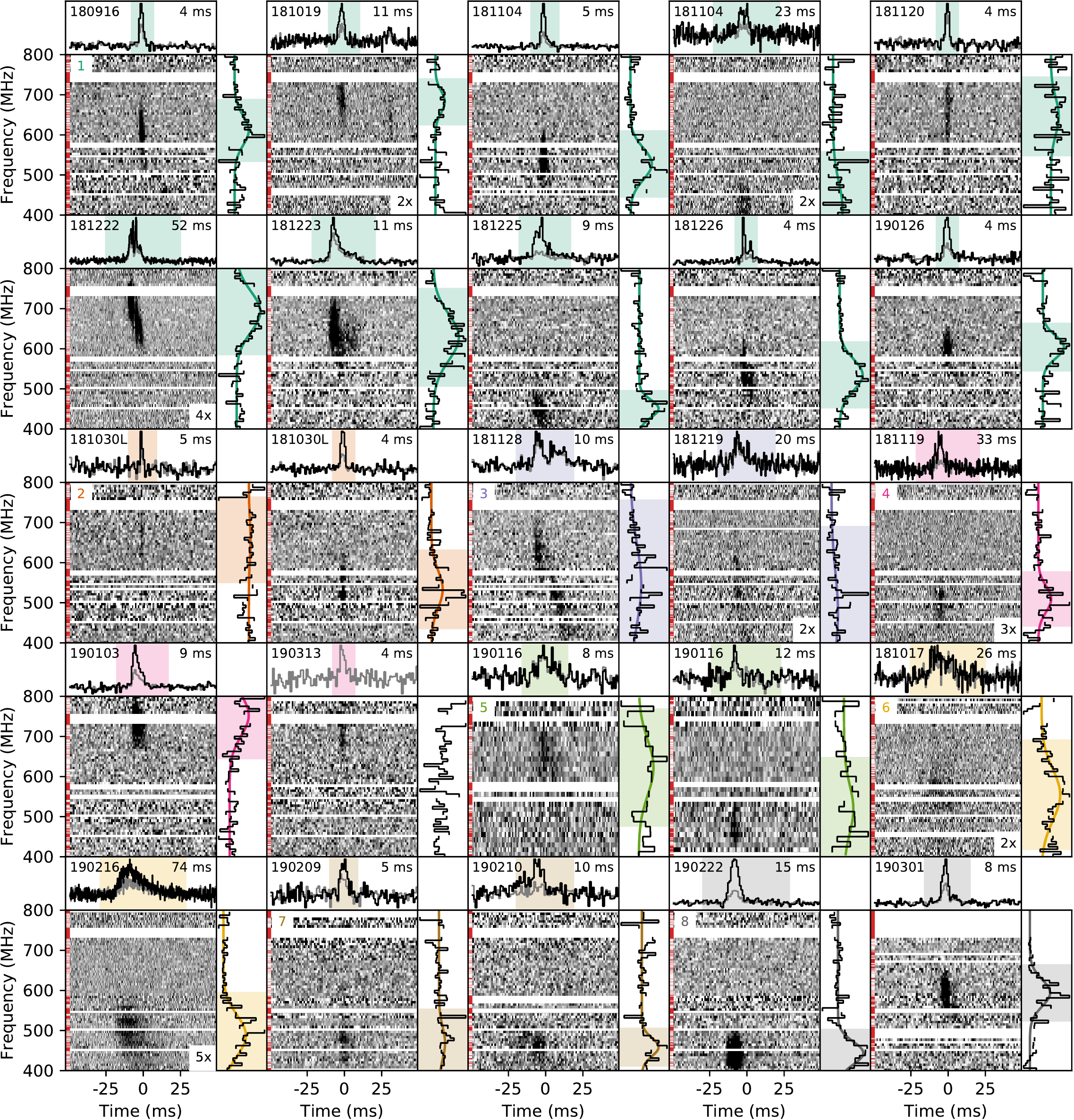}
    \caption{{Figure 3 from} \cite{chime8frb} ``CHIME/FRB Discovery of Eight New Repeating Fast Radio Burst Sources'' reproduced with permission from Emmanuel Fonseca, published by ApJL (IOP Publishing), 2019. Profiles, dynamic spectra and spectral distributions are shown for eight new repeaters discovered by CHIME. The color coding separates bursts from different sources. \re, in green, was the first source in this work, showing its prolific activity compared to the other repeaters.}
    \label{fig:chime}
\end{figure}

The effort paid off and CHIME soon became the top performing FRB machine.
As early as during its pre-commissioning, between July and August 2018, CHIME detected 13~bursts~\cite{chime_13_19}. These were the first detections below 800 MHz. A total of 7 out of 13 bursts were observed down to the lower edge of the band, and only some of them showed visible scatter broadening. A possible DM-scattering correlation was seen in this sample, as the bursts which showed more scattering were the ones with the highest DMs (up to $DM=1007$\,\pc).
Amongst the first 13 FRBs from CHIME, one was a repeating source \cite{chime_1r_19}.
The difference in pace imposed by CHIME was soon clear: in about one year, eight \cite{chime8frb} and then nine \cite{chime9rep} more repeaters were published.  
With CHIME's long dwelling time on the sky and sensitivity (95\% median completeness threshold at $F=5$ Jy ms), detection of repetition from a source was much more straightforward than using single dishes, especially different ones with different sensitivities or even different bands. Indeed, among these repeaters was the first for which periodicity was detected \cite{chime_period_20}, \re\ (see Figure \ref{fig:chime} showing the first eight repeaters detections).
The publication of the first catalog from CHIME \cite{chime_cat1_21}, with data from the first year of operations, added 474 new non-repeaters to this count, bringing CHIME's FRB population to more than 492 FRBs discovered in 1 year, i.e.,~an average of more than 1 per day.
As of September 2021, CHIME/FRB has reported two more repeaters' discoveries: the first, FRB20200120E \cite{bhardwaj21}, which was soon localised to a globular cluster within the nearby galaxy M81 \cite{kirsten21}; the second, FRB20201124A \cite{lanman21}, which became very active in both the P and L bands during the spring of 2021, allowing for extensive coverage in both radio and other frequencies.

\paragraph{First CHIME Catalog}
The known FRB population in literature now stands at $\sim$600 one-off events and two dozen repeaters, thanks in large part to CHIME/FRB.
Population studies have been complicated by comparison among data from different facilities, each with their own systematics and their own limiting survey parameters. The presence of a database with a statistically relevant number of sources all coming from the same experiment and, therefore, with uniform boundary conditions, is an exciting prospect for most of these studies.
The CHIME sample too, however, comes with caveats which should be taken into account for further studies. The event reconstruction can be biased by properties of the burst, i.e., a~strong scattering affecting the measurement of the intrinsic width of the burst; or by instrumental effects, i.e., the~time resolution limiting the same 
burst widths' measurement. Selection effects strongly affect scattering or fluence completeness measurements, while they are less impactful for other properties (e.g., DM).

The catalog itself already provided a number of key answers to long-term questions in the FRB panorama. The sky distribution appears uniform for CHIME FRBs (also in~\cite{josephy21}), contrary to previous claims of a possible depletion of FRBs at mid-galactic latitudes~\cite{petroff14}. 
A comparison between repeaters and one-off events was pursued, with the caveat that only the discovery burst was taken into account for each repeater, as the following ones may have been selected with lower detection thresholds.
Repeating FRBs currently account for $\sim$4\% of the known population, but many repeaters have only been seen twice.
This suggests that many currently one-off sources may be seen to produce a second (or third, etc.) burst, given enough follow-up (as in \cite{kumar19}).
The DM, S/N and (where unbiased) fluence and flux distributions of repeater and one-off events do not appear to belong to different distributions. Width distributions, on the other hand, do appear different between repeaters and one-offs (also in \cite{pleunis21b}), with repeaters more regularly showing a sub-burst structure. Bandwidth distributions also appear different, with repeaters typically occupying a smaller fractional bandwidth.
Scattering times are expected to be different if the local medium plays a major contribution to it and if the hypothesis that the environment surrounding repeaters and one-offs is different. However, no difference was found. Finally, the correlation between the scattering time and the extragalactic DM contribution was assessed, and it demonstrated that the two populations do not differ.
The conclusions reached in the Catalog from the overall comparison is that the local and global (host galaxy) environments around repeaters and non repeaters do not point towards different origins of the two species. Their instrinsic properties, however, do show significant variations, which could point either to different emission mechanisms or to a pulse morphology correlating with the repetition rate \cite{connor20}. It is not uncommon, in astrophysical sources such as, e.g., neutron stars, to find a wide variety of observational behaviours, determined by age or evolution but also intrinsic properties of the single source.

\begin{figure}[H]
    \includegraphics[width=0.5\textwidth]{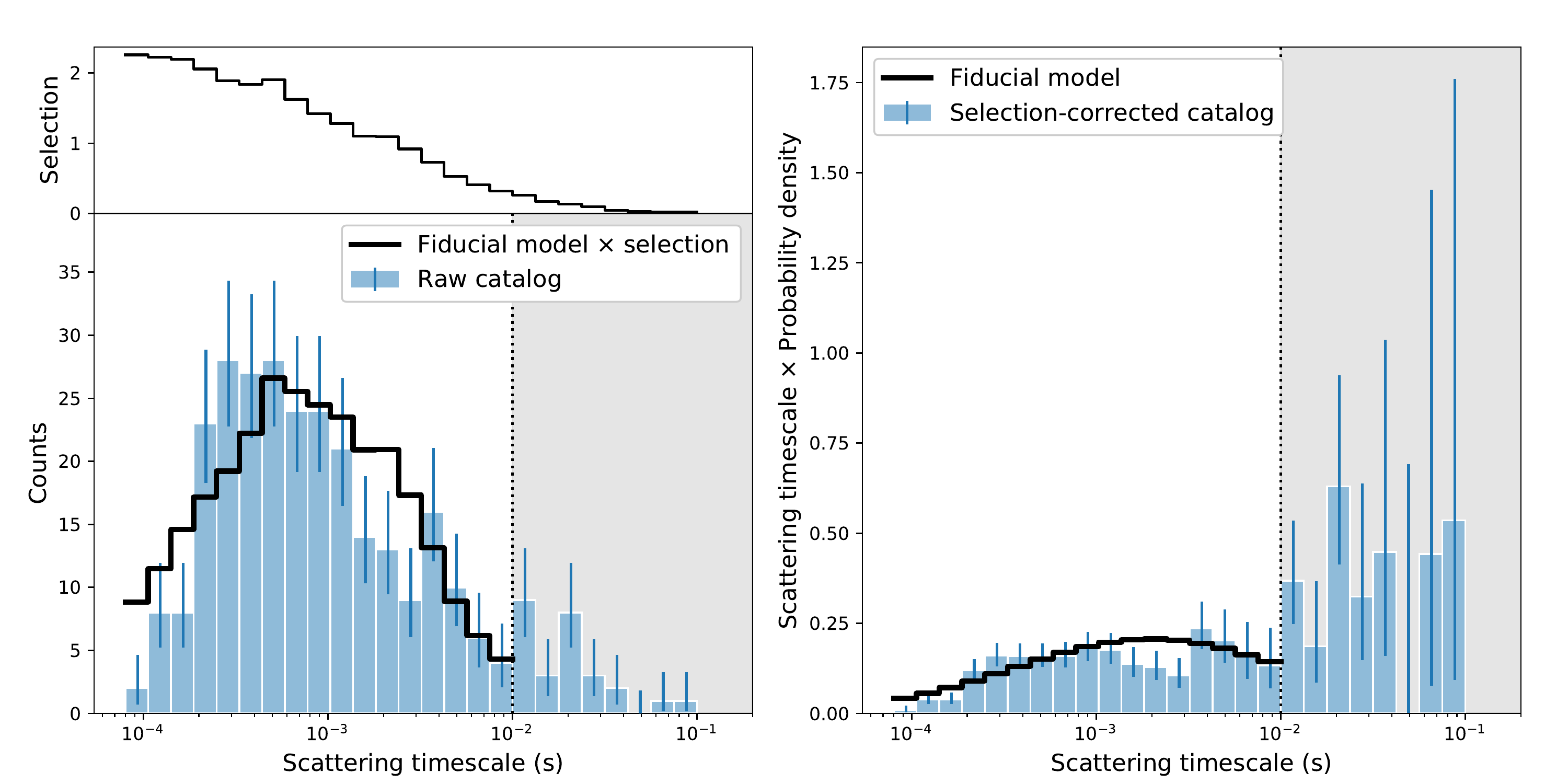}
    \caption{Figure 17 from \cite{chime_cat1_21} ``The First CHIME/FRB Fast Radio Burst Catalog'', reproduced with permission from Kiyoshi Masui, submitted to ApJS. It shows scattering times before and after correction for systematics adopted following the results of the injection process developed to control biases of the survey. It is evident that a large fraction of highly scattered events might be missed by CHIME for scattering \mbox{$>$10 ms.}}
    \label{fig:cat1}
\end{figure}

The first CHIME catalog confirmed the compatibility of the FRB fluence distribution ($\gamma=-1.4\pm 0.11$ with a Euclidean universe with $N(>F) \propto F^{-3/2}$. In this assumption, the derived FRB rate was $R_{FRB} = 820$\,sky$^{-1}$\,day$^{-1}$ above a fluence $F>5$\,Jy\,ms. This rate was validated for $DM>100$\,\pc, because lower values are still believed to be compatible with galactic sources and were excluded from the sample used for this calculation, and for scattering times $\tau < 10$\,ms at 600 MHz, as the sample of detected FRBs from CHIME appears significantly depleted for scattering times above this value. Hence, a significant population of highly scattered FRBs could still be undetected (see Figure \ref{fig:cat1}).
A comparison of this rate with the ones predicted by the other detections at 350 MHz, 800 MHz and 1.4~GHz shows that they are consistent in the framework of a flat spectral index $\alpha=0$ and suggests that there is no need to invoke a spectral turnover below 1 GHz as hinted at by the UTMOST detections \cite{farah19}. 
A conversion of DM to redshift based on the Macquart relation~\cite{macquart20} showed that it is unlikely that CHIME's FRB population can probe the high-redshift end of the universe, implying that low-frequency observations are probably not ideal for cosmology purposes.

\subsection{Northern Cross}
The effort of transit telescopes in the quest for more on-sky time to look for FRBs has recently been joined by the Northern Cross (NC), the oldest Italian radio telescope. The North-South arm of the NC was recently refurbished for space debris activities, and the 64 cylinders are being modified in order to group the
signals of sixteen dipoles together, providing four analogue signals per cylinder. The final effective collecting area of the NC at completion of the upgrade is 8000 deg$^2$. Further hardware and software upgrades made the instrument optimal for FRB studies \cite{locatelli20}. The NC observes at 408 MHz, with a bandwidth of 16 MHz. Observations are performed with native time resolution of 1\,$\upmu$s and 21 coarse channels, but the storage of the baseband data and a flexible post-processing approach guarantees optimisation of the time and frequency resolution depending on the searched parameter space.
The NC is currently observing with 8 out of its 64 cylinders and performing a targeted monitoring of known repeating FRBs and active magnetars in transit. The targeted strategy allows for a longer dwelling time on sky per source than is possible for survey instruments (i.e.,~from $\sim$0.5 to $\sim$3 h per source per day). This has led to the first detections (from \re), obtained with a configuration still using only 6~cylinders \cite{bernardi21}.
The final goal is to use the NC as a survey instrument. Ref. \cite{locatelli20} predicts the detection of one FRB every three days in the final configuration.

\section{First Detections below 400 MHz}
The detection of periodicity from \re\ \cite{chime_period_20} was a milestone not only per se, but also because it meant that more telescope time could be invested, more productively, in FRB-pointed observations by sensitive instruments. 
Indeed, the first detection of bursts at 350 MHz were obtained with the Green Bank Telescope (GBT) on 19 December  2019 and 20 January 2020 \cite{chawla20}. These dates had been chosen within the active window of \re\ and emission was detected, as expected, during some of the days of the cycle.
The Sardinia Radio Telescope (SRT) also pointed at \re\ as soon as periodicity was announced~\cite{chime_period_20}. Observations were performed on 20--24 February 2020 and four {bursts}~{(the fourth one was found in a later analysis and will be published elsewhere)} were detected, all happening during the very first hour of the 30-h long multiwavelength campaign on the periodic repeater \cite{pilia20}. 
Soon afterwards, the upgraded Giant Meterwave Radio Telescope (uGMRT) also detected bursts in the frequency interval from 500 to 300 MHz. The first four bursts were detected on 23 and 24 March 2020 \cite{sand20}. They were all more luminous in the upper part of the band, but emission was also always visible down to 300 MHz. 

\begin{figure}[H]
    \includegraphics[width=0.7\textwidth]{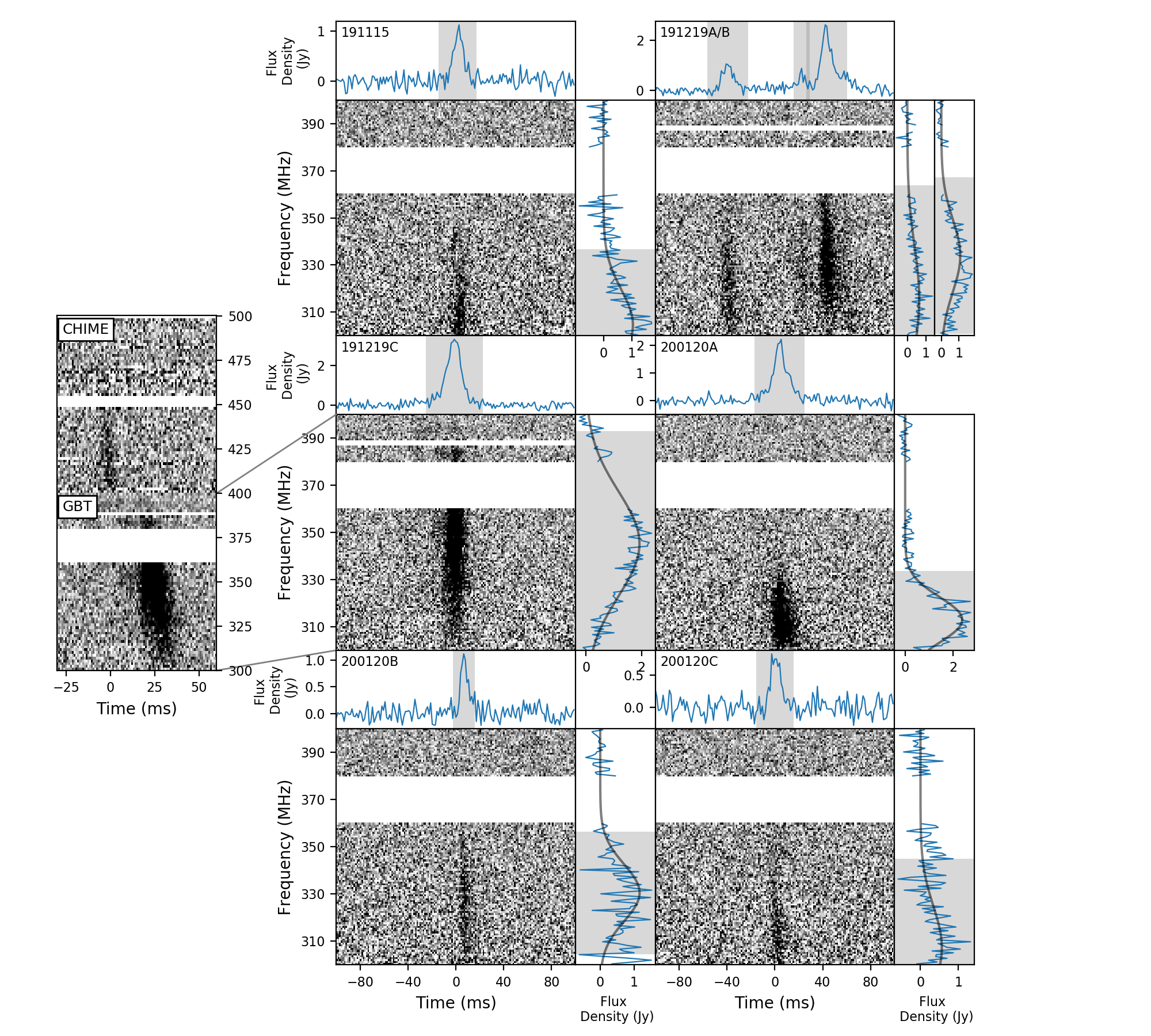}
    \caption{Figure 2 from \cite{chawla20} ``Detection of Repeating FRB\,180916.J0158$+65$ Down to Frequencies of 300 MHz'' reproduced with permission from Pragya Chawla, published by ApJL (IOP Publishing), 2020. It shows profiles, dynamic spectra and on-pulse spectrum for \re's bursts detected by GBT at 350 MHz. The inset shows the burst that was observed both in CHIME and GBT bands.}
    \label{fig:chawla}
\end{figure}

The characteristics of the bursts observed down to 300 MHz were similar in all cases. 
GBT observed \re\ in the period from 15 November 2019 to 20 January 2020. Data spanned 100 MHz of bandwidth centered at 350 MHz, but the upper 20 MHz of the band were unusable due to persistent RFI. Full--Stokes data were recorded for 512 frequency channels
at a cadence of 20.48 $\upmu$s.
The seven bursts that were detected from GBT had DMs consistent with the DM measured by CHIME (348.82 \pc) in all but two cases. 
One of the bursts detected was a double burst within the time span of 100 ms. The other burst where drifting was observed was the only burst that was seen simultaneously by GBT and CHIME in adjacent bandwidths (see Figure \ref{fig:chawla}). A delay between the two signals was present (after correcting for telescopes' different paths), and this made it possible to calculate an instantaneous drift rate $\delta\nu/\delta t = -4.2$\,MHz\,ms$^{-1}$, compatible with what was found in previous cases for \ri\ \cite{hessels19,josephy19}.
The sample of GBT bursts, checked against the activity phases of \re, showed no monotonic variation in DM, fluence, burst width, scattering timescale or emission frequency with phase or time.
No scattering was observed from any of the bursts, leading to a 95\% confidence upper limit of 1.7 ms on the scattering timescale of the source at 350 MHz.
Despite there being no scattering detected at 350 MHz, no burst was detected in a simultaneous LOFAR campaign, which was active when the bursts were detected by GBT. 
In the hypothesis that the properties of the detected bursts could be directly applied to the emission at 150 MHz, a fluence detection limit of 21 Jy ms was derived for LOFAR in case of no pulse broadening. Conversely, if a scatter-broadening upper limit of 50 ms is assumed (with a typical Kolmogorov spectrum with dependence $\nu^{-4}$), the 90\% completeness of LOFAR observations becomes 106 Jy ms, but this limit alone would not be enough to explain the non-detections.

SRT observed at a central frequency of 328 MHz, with a bandwidth of 80 MHz.  Due to RFI, only the lower 64 MHz of the band were used. Acquisitions were performed in baseband mode, and the data were then coherently dedispersed to 349 MHz, channelised to 250 kHz resolution and decimated to a time resolution of 128\,$\upmu$s. Observations were performed at different times of the day during five days spanning the active interval in order to cover simultaneous observations with the NC (radio), the Telescopio Nazionale Galileo (TNG) and Asiago (optical), XMM-Newton, NICER and Integral (X-rays) telescopes. SRT detected the three bursts all within $\sim$1\,h from the start of the campaign. The three bursts spanned the full band, had widths around 10 ms and fluences ranging from 37 to 13 Jy ms, and had no detectable structure, which was also checked against DM optimisation.
Additionally, in the case of SRT, the bursts showed no detectable scattering: the 2$\sigma$ upper limits $\tau_{sc} < 10$\,ms at $328$\,MHz are compatible with no (or a very low level of) scattering. This result fits in nicely
with the DM$-\tau_{sc}$ correlation {reported by} \cite{chime_13_19} {in their Figure 2.} 
On the other hand, for $\tau_{sc}$ values approaching the upper limits reported by SRT, the burst energy would be diluted over 0.8--1.0\,s at frequencies around 150 MHz, easily causing the non-detection of similar bursts with LOFAR.
SRT observations were simultaneously carried out at 328 MHz and at 1.5 GHz. No burst was detected at 1.5 GHz during the whole campaign. A deeper search was performed around the time of the 328-MHz bursts, taking into account the delay due to dispersion, but no emissions were present down to a sensitivity limit of 600~mJy ms. This non-detection at L-band set an upper limit  on the burst spectral index $\alpha < -1$ for the brightest burst, assuming the nominal SRT sensitivity at 1.5 GHz. 
A modulation due to plasma lenses \cite{cordes17} remained an open  possibility, although the lack of simultaneous detection at 1.5~GHz made it not obvious.

The 4 bursts detected by GMRT in its Band-3 (250--500 MHz) \cite{sand20} were all simple bursts with no structure and no scattering apparent. Their fluences ranged from 1 to 15 Jy ms, and their widths were also consistently around 10 ms. 
It is worth noting that 12 more bursts were observed later on the same day by another program running at the uGMRT \cite{marthi20}. In the latter case, the uGMRT Band-4 receiver at 550--750 MHz was used to sample a subset of the CHIME band. Significantly, most of the bursts are seen only in the lower part of the band, but still at higher frequencies than previous observations on the same date. 
Similar observations were also carried out on 9 March 2020 and 30 June 2020 with 0 and 3 detections, respectively, all near the peak of the active phase within the 16.35-d period.  The fluences of uGMRT Band-4 bursts ranged from 48 to 0.1 Jy ms, placing the faintest burst among the faintest ever observed. Some show double peaks or downward frequency structures, which were investigated through DM optimisation.

The detection of emission from \re\
down to 300 MHz is consistent with the non-detection of FRBs with several surveys in this frequency range (see Section \ref{sec:dishes}) if a less dense
circumburst environment, low scattering timescale and the proximity of the
source all conspire to make its emission particularly detectable.
As discussed in Section \ref{sec:implications}, the presence of an overdensity in the form of a nebula or the expanding supernova from the aftermath of the birth of the neutron star could be responsible for the absorption of radiation below 400 MHz.
With this detection below 400 MHz, \cite{chawla20} calculated the optical depth due to free-free absorption in the case of an ionized nebula with a $DM < 70$\pc\ implied by the observations of \cite{marcote20}. In this context the extent of the nebula should be $L>>0.02$\,pc, which excluded a young supernova remnant or a hyper-compact HII region.
All the datasets at 350 MHz confirmed that the emission from \re\ was localised within the activity interval defined by CHIME observations. In addition, the bursts appear clustered at a certain time of the active phase for emission at a specific frequency interval.

\section{Lowest Frequency Detections of FRBs}
\label{sec:lowest}

\subsection{Targeted Non-Detections}
Following the non-detections by the lowest frequencies surveys (see Section \ref{sec:arrays}), some more attempts were made on targeted sources.
MWA  was shadowing Parkes observations when FRB 150814 was detected \cite{keane16}, but did not detect a counterpart. The resulting $3\sigma$ fluence upper limit of 1050\,Jy\,ms at 185\,MHz implied a spectral index limit $\alpha > -3$.
Ref. \cite{sokolowski18} ran an MWA campaign shadowing ASKAP observations at 1.4 GHz which encompassed the detection of 7 FRBs with ASKAP. MWA's high band (170--200 MHz) was used in order to reduce the effects of scatter broadening and RFI contamination. Data were recorded with a frequency resolution of 10\,kHz and time resolution 0.5\,s. Despite the relatively high fluences of the bursts detected by ASKAP, not all shadowings were equally successful, due to either observing limits on the MWA side (i.e., the~sun in the sidelobes of the primary beam) or to the bursts having high intrinsic DMs and, in some cases, a noticeable scattering tail already present at ASKAP frequencies. No burst was detected at MWA frequencies implying a stricter broadband spectral index upper limit of $\alpha > -1$.
Neither scatter broadening nor plasma lensing were considered as viable explanations for the non-detections, while a spectral turnover (intrinsic to the progenitor and not related to absorption from the environment) was favoured.

A multi-frequency campaign was also attempted by LOFAR by targeting \ri\ for 20\,h in shadowing mode with the Effelsberg telescope observing at 1.4\,GHz \cite{houben19}.
LOFAR HBA antennas were used with remote stations, plus one core station were used in imaging mode while the remaining core stations were combined to obtain a tied-array coherent beam. A time resolution of 1.31 ms was adopted and 25600 frequency
channels were produced to cover 78 MHz of bandwidth centered at 150 MHz.
Nine bursts were detected by Effelsberg during the simultaneous observations, but none were seen at LOFAR frequencies.
A limit on the spectral index was derived for broadband
instantaneous emission of $\alpha> -1.2$, which was higher than the one obtained by the non-detections of \cite{karastergiou15}, but less model-dependent.
\ri\ was in a highly active state during these observations, and the non-detections in the LOFAR band could be explained by the band-limitedness of the bursts. However, considerations regarding the varying central frequency of its emission, as observed by \cite{law17}, led to the possibility of a shift of the peak of the emission towards the LOFAR band.
In this case, detections would not happen simultaneously, and a ``statistical'' spectral index could be derived from fluence limits of different instruments, taking into account their diverse observational setups. Many assumptions go into the calculation of this index, but a first comparison of the Effelsberg data with GBT and VLA data from~\cite{law17} gave an indication of a negative spectral index for the overall distribution of burst energies versus frequency. By comparing the Effelsberg data to the LOFAR upper limits, ref.~\cite{houben19} obtained $\alpha_{stat} > -0.5$, indicating a possible flattening of this spectrum. 

\subsection{Targeted Detections}
The first detection of bursts below 300 MHz was finally also achieved by targeting \re\ with LOFAR \cite{pleunis21a,pastor21}. The knowledge of an active window 
helped in reaching this goal, together with the favourable physical conditions of \re, likely different from those of other sources like, e.g., \ri. 

Ref. \cite{pleunis21a} presents the detection of 18 bursts which were observed out of 112 hours of observations on 128 sessions between June 2019 and August 2020. 15 activity  cycles were sampled by the observations. For half of the observations, total intensity data were acquired over the bandwidth
110--188 MHz and were sampled with a frequency resolution of
3 kHz and and time resolution of 983 $\upmu$s. The other half of the observations were recorded as complex voltages, with
195 kHz frequency resolution and 5 $\upmu$s time resolution. To maximize sensitivity towards possibly narrow-band
radio bursts, the time series were created for the full band as well as sub-banded with different samplings of the band. 
A total of 14 of the bursts were found in the full-bandwidth data, while 4 fainter bursts were identified in the subbanded data segments. The datasets where bursts were discovered were cross-checked for possible weaker candidates using {\tt FETCH} \cite{agarwal20fetch}.
 All  bursts were band-limited, with spectral widths ranging from 20 to 50 MHz and fluences ranging from 300 to 30 Jy ms.  The temporal width of the bursts varied between 40 ms
for bursts peaking in the top of the LOFAR band to 160~ms near the bottom of the LOFAR
band. 
Some bursts were observed all the way down to the bottom of the band at 110 MHz. They confirmed that the scattering relative to this source is quite low: $\tau_{sc} \sim 40~\upmu$s, a value that was consistent with extrapolation from the frequency scintillation scaling found at 1.4\,GHz by \cite{marcote20}. 
Broadening due to dispersion smearing  was limited by the use of coherent dedispersion. 
Even after dedispersion to the best-fit DM, the brightest LOFAR bursts showed
residual time delays towards lower observing frequencies and, in one case, broadening towards decreasing frequencies. The absence of a visible burst substructure, or possibly a limit in the resolution, made it difficult to pinpoint this behaviour to DM underestimation, multi-path scattering, or sad trombone effects.
The drift toward later times increases toward lower frequencies \cite{hessels19,josephy19}, and the drift could be about 10 ms per  50 MHz at these frequencies. Indeed, many bursts from
\ri\ show asymmetric burst profiles regardless of scattering (see Figure \ref{fig:hessels}) \cite{hessels19}.



Ref. \cite{pastor21} carried out simultaneous observations with LOFAR and the APERTIF system at the Westerbork Synthesis Radio Telescope (WSRT) for a total superposition of 57.6\,h. APERTIF observations were performed during different phases of the periodicity cycle of \re, aiming at a validation of CHIME's inferred activity window at different frequencies. LOFAR, on the other hand, was only operational in proximity of the expected maxima of the activity.
As this data represented a subset of those used in~\cite{pleunis21a}, they confirmed the detection of nine bursts in the LOFAR band. None of them were simultaneously observed by APERTIF down to a fluence limit of 0.5\,Jy\,ms. 
Conversely, APERTIF found 54~bursts over $\sim$388\,h of observations, none of which were with coincident detections.
This analysis of the LOFAR data confirmed the detection of bursts down to the lowest frequencies, with, in those cases, a scattering time $\tau_{sc} \sim$~45\,ms.
Ref. \cite{pastor21} derived a first lower limit on the FRB rate by combining previous non-detections with the hypothesis that only \re\ emits at this band: $R_{FRB} > 3-450$\,sky$^{-1}$\,day$^{-1}$ at 150 MHz above a fluence $F>5$\,Jy\,ms. By extending this limit with Euclidean fluence scaling, the rate becomes $R_{FRB} =90-1400$\,sky$^{-1}$\,day$^{-1}$.
These rates are indicative of the fact that LOFAR detections showed an apparent much higher activity rate per corresponding fluence at 150 MHz with respect to 1.4 GHz.

Both works were an additional important confirmation that the environment around \re\ did not show evidence for overdensities and, on the contrary, excess DM or scattering contributions at these low frequencies seem to be relative to the ISM plasma or to unresolved structures rather than to the local environment or \re's host galaxy. 
Free-free absorption and induced Compton scattering were shown to be a negligible contribution down to 110 MHz, even though nothing can be said about the presence of lower frequencies' turnovers. 
One further aspect of the periodicity of \re\ that became clear with the wealth of multi-frequency observations from high to low frequencies was the ``chromaticity'' of the active window (see, e.g., Extended Figure 3 of \cite{pastor21}). If compared to the $\sim$5\,d activity cycle spanned by CHIME detections, the L-band active window seemed to consistently lead the peak phase (see also \cite{marcote20,aggarwal20}), while the lower frequencies trailed this phase. Ref.~\cite{pleunis21a} refined the period of \re\ to $16.33$\,d and deduced a shift of $\sim$0.2 in phase for LOFAR's peak activity with respect to the peak at CHIME's frequencies.

\section{Future Prospects}

The development of a highly specialised and highly active framework for low-frequency observations of FRBs has been detailed in this review, with a particular focus on the advancements in our understanding of the phenomenon that have come from this perspective.
One of the, perhaps unexpected, achievements of the low-frequency observations has been the in-depth characterisation of the behaviour of close-by (even galactic, in one instance) FRBs. 
The analyses of the wide-band spectral properties of FRBs have allowed us to characterise, on different scales, the bursting behaviour (with chromatic evolution in time), the local environment (constraining the presence of an absorbing medium) and the cosmic distribution of events.
{  {The current findings seem to show that, for low-frequency studies, it is better to trace the local universe (see, e.g.,} \cite{chime_cat1_21}). {This is consistent with the agreement of all the observed rates and their limits with a Euclidean distribution of the FRB fluences.  On the other hand, inconsistencies with this distribution have been obtained for the sample of FRBs detected by ASKAP at 1.5 GHz:} $\gamma = 2.2$ \cite{james19}. {One possible bias in this view is the limiting sensitivity of most of the low-frequency surveys performed so far, or a DM completeness limit that is lower for current lower frequency studies.  
 This would, in turn, imply that FRBs observed preferentially at low frequencies might not be interesting for cosmological studies.
 Theoretical predictions, such as those of} \cite{raviloeb19}, {argue in disfavour of this statement. They assume a double power-law spectrum for FRB fluences, which implies a peak frequency dependent on both the intrinsic emission mechanism and possible low-frequency emission suppression. The rate of FRBs detected at high redshifts is expected to be higher at low frequencies, as the correction of the observed fluence for the redshifted frequency is negative. Therefore, more distant FRBs observed close to their spectral peak are brighter than would be expected, assuming their luminosity distance relation. This would be especially true if there was a cutoff on FRB emission as, in that case, the observations below the frequency maximising the event rate would detect more faint distant events and fewer bright nearby events.
 The large availability of arrays at these frequencies enabling simultaneous imaging observations will either confirm or deny the observational findings and theoretical predictions. This will also occur thanks to the much wider future availability of precise localisations.}}

While CHIME has demonstrated its leading role in the field, it has also confirmed that the key to a deeper knowledge of the origin and properties of FRBs lies in the dedicated on-sky time and large FoV that an instrument can devote to the quest. New and future instruments coming online (e.g.,~SKA-low) can and ought to complement CHIME's capabilities while occupying still-missed frequency bands, sky regions, and exploiting higher temporal or frequency resolutions.
The technological advancements of the last years have recently allowed, for most of the low-frequency studies, the possibility to record and temporarily store baseband data. Coherent dedispersion is a fundamental detection step at low frequencies, but the availability of the raw voltages also enables the deepest frequency- or time-resolved studies and polarisation studies.

Detections of multiple bursts enveloped in a larger burst structure, of wider bursts at the lowest frequencies and, on the other hand, of substructures at the ns level, imply a still-fertile ground for vastly unexplored areas of the transients' parameter space.
One important niche for low-frequency studies will remain the detection of nearby sources, repeaters in particular. Multiwavelength efforts have increased in parallel with the discovery of predictable active windows from repeaters, but have so far been unsuccessful (see e.g.,~\cite{pilia20, scholz20} in connection to the first 350 MHz studies of \re). These studies~\cite{nicastro21} will be more fruitful by targeting the nearest sources, even though constraints on the activity rates and energies will factor in with an important role. 
More instruments and more real-time capabilities, both of detection, classification and of community alerts, will probably lead to a detection in the next few years.

\vspace{6pt} 

\funding{This research received no external funding.}

\institutionalreview{{ Not applicable}}

\informedconsent{{ Not applicable}}



\conflictsofinterest{The authors declare no conflict of interest.} 





\end{paracol}
\reftitle{References}


\def\apj{ApJ}
\def\apjl{ApJL}
\def\nat{Nature}
\def\mnras{MNRAS}
\def\aap{A\&A}
\def\physrep{Phys. Rep.}


\begin{thebibliography}{999}

\bibitem[{Lorimer} \em{et~al.}(2007){Lorimer}, {Bailes}, {McLaughlin},
  {Narkevic}, and {Crawford}]{lorimer07}
{Lorimer}, D.R.; {Bailes}, M.; {McLaughlin}, M.A.; {Narkevic}, D.J.;
  {Crawford}, F.
\newblock {A Bright Millisecond Radio Burst of Extragalactic Origin}.
\newblock {\em Science} {\bf 2007}, {\em 318},~777,
  doi:10.1126/science.1147532.

\bibitem[{Cordes} and {Lazio}(2002)]{cordeslazio02}
{Cordes}, J.M.; {Lazio}, T.J.W.
\newblock {NE2001.I. A New Model for the Galactic Distribution of Free
  Electrons and its Fluctuations}.
\newblock {\em arXiv} {\bf 2002}, arXiv:0207156.


\bibitem[{Yao} \em{et~al.}(2017){Yao}, {Manchester}, and {Wang}]{ymw17}
{Yao}, J.M.; {Manchester}, R.N.; {Wang}, N.
\newblock {A New Electron-density Model for Estimation of Pulsar and FRB
  Distances}.
\newblock {\em \apj} {\bf 2017}, {\em 835},~29,
doi:10.3847/1538-4357/835/1/29.

\bibitem[{Spitler} \em{et~al.}(2014){Spitler}, {Cordes}, {Hessels}, {Lorimer},
  {McLaughlin}, {Chatterjee}, {Crawford}, {Deneva}, {Kaspi}, {Wharton},
  {Allen}, {Bogdanov}, {Brazier}, {Camilo}, {Freire}, {Jenet},
  {Karako-Argaman}, {Knispel}, {Lazarus}, {Lee}, {van Leeuwen}, {Lynch},
  {Ransom}, {Scholz}, {Siemens}, {Stairs}, {Stovall}, {Swiggum},
  {Venkataraman}, {Zhu}, {Aulbert}, and {Fehrmann}]{spitler14}
{Spitler}, L.G.; {Cordes}, J.M.; {Hessels}, J.W.T.; {Lorimer}, D.R.;
  {McLaughlin}, M.A.; {Chatterjee}, S.; {Crawford}, F.; {Deneva}, J.S.;
  {Kaspi}, V.M.; et~al.
\newblock {Fast Radio Burst Discovered in the Arecibo Pulsar ALFA Survey}.
\newblock {\em \apj} {\bf 2014}, {\em 790},~101,
 doi:10.1088/0004-637X/790/2/101.

\bibitem[{Spitler} \em{et~al.}(2016){Spitler}, {Scholz}, {Hessels}, {Bogdanov},
  {Brazier}, {Camilo}, {Chatterjee}, {Cordes}, {Crawford}, {Deneva}, {Ferdman},
  {Freire}, {Kaspi}, {Lazarus}, {Lynch}, {Madsen}, {McLaughlin}, {Patel},
  {Ransom}, {Seymour}, {Stairs}, {Stappers}, {van Leeuwen}, and
  {Zhu}]{spitler16}
{Spitler}, L.G.; {Scholz}, P.; {Hessels}, J.W.T.; {Bogdanov}, S.; {Brazier},
  A.; {Camilo}, F.; {Chatterjee}, S.; {Cordes}, J.M.; {Crawford}, F.; {Deneva},
  J.; et~al.
\newblock {A repeating fast radio burst}.
\newblock {\em \nat} {\bf 2016}, {\em 531},~202--205,
  doi:10.1038/nature17168.

\bibitem[{Keane} \em{et~al.}(2012){Keane}, {Stappers}, {Kramer}, and
  {Lyne}]{keane12}
{Keane}, E.F.; {Stappers}, B.W.; {Kramer}, M.; {Lyne}, A.G.
\newblock {On the origin of a highly dispersed coherent radio burst}.
\newblock {\em \mnras} {\bf 2012}, {\em 425},~L71--L75,
doi:10.1111/j.1745-3933.2012.01306.x.

\bibitem[{Thornton} \em{et~al.}(2013){Thornton}, {Stappers}, {Bailes},
  {Barsdell}, {Bates}, {Bhat}, {Burgay}, {Burke-Spolaor}, {Champion}, {Coster},
  {D'Amico}, {Jameson}, {Johnston}, {Keith}, {Kramer}, {Levin}, {Milia}, {Ng},
  {Possenti}, and {van Straten}]{thornton13}
{Thornton}, D.; {Stappers}, B.; {Bailes}, M.; {Barsdell}, B.; {Bates}, S.;
  {Bhat}, N.D.R.; {Burgay}, M.; {Burke-Spolaor}, S.; {Champion}, D.J.;
  {Coster}, P.; et~al.
\newblock {A Population of Fast Radio Bursts at Cosmological Distances}.
\newblock {\em Science} {\bf 2013}, {\em 341},~53--56,
doi:10.1126/science.1236789.

\bibitem[{Ravi} \em{et~al.}(2015){Ravi}, {Shannon}, and {Jameson}]{ravi15}
{Ravi}, V.; {Shannon}, R.M.; {Jameson}, A.
\newblock {A Fast Radio Burst in the Direction of the Carina Dwarf Spheroidal
  Galaxy}.
\newblock {\em \apjl} {\bf 2015}, {\em 799},~L5,
 doi:10.1088/2041-8205/799/1/L5.

\bibitem[{Petroff} \em{et~al.}(2015){Petroff}, {Johnston}, {Keane}, {van
  Straten}, {Bailes}, {Barr}, {Barsdell}, {Burke-Spolaor}, {Caleb}, {Champion},
  {Flynn}, {Jameson}, {Kramer}, {Ng}, {Possenti}, and {Stappers}]{petroff15}
{Petroff}, E.; {Johnston}, S.; {Keane}, E.F.; {van Straten}, W.; {Bailes}, M.;
  {Barr}, E.D.; {Barsdell}, B.R.; {Burke-Spolaor}, S.; {Caleb}, M.; {Champion},
  D.J.; et~al.
\newblock {A survey of FRB fields: Limits on repeatability}.
\newblock {\em \mnras} {\bf 2015}, {\em 454},~457--462,
doi:10.1093/mnras/stv1953.

\bibitem[{Totani}(2013)]{totani13}
{Totani}, T.
\newblock {Cosmological Fast Radio Bursts from Binary Neutron Star Mergers}.
\newblock \emph{Publ. Astron. Soc. Jpn.} {\bf 2013}, {\em 65},~L12,
doi:10.1093/pasj/65.5.L12.

\bibitem[{Lingam} and {Loeb}(2017)]{lingam17}
{Lingam}, M.; {Loeb}, A.
\newblock {Fast Radio Bursts from Extragalactic Light Sails}.
\newblock {\em \apjl} {\bf 2017}, {\em 837},~L23,
doi:10.3847/2041-8213/aa633e.

\bibitem[{Katz}(2017)]{katz17}
{Katz}, J.I.
\newblock {FRB as products of accretion disc funnels}.
\newblock {\em \mnras} {\bf 2017}, {\em 471},~L92--L95,
 doi:10.1093/mnrasl/slx113.

\bibitem[{Platts} \em{et~al.}(2019){Platts}, {Weltman}, {Walters}, {Tendulkar},
  {Gordin}, and {Kandhai}]{platts19}
{Platts}, E.; {Weltman}, A.; {Walters}, A.; {Tendulkar}, S.P.; {Gordin},
  J.E.B.; {Kandhai}, S.
\newblock {A living theory catalogue for fast radio bursts}.
\newblock {\em \physrep} {\bf 2019}, {\em 821},~1--27,
doi:10.1016/j.physrep.2019.06.003.

\bibitem[{The CHIME/FRB Collaboration} \em{et~al.}(2021){The CHIME/FRB
  Collaboration}, {:}, {Amiri}, {Andersen}, {Bandura}, {Berger}, {Bhardwaj},
  {Boyce}, {Boyle}, {Brar}, {Breitman}, {Cassanelli}, {Chawla}, {Chen},
  {Cliche}, {Cook}, {Cubranic}, {Curtin}, {Deng}, {Dobbs}, {Fengqiu}, {Dong},
  {Eadie}, {Fandino}, {Fonseca}, {Gaensler}, {Giri}, {Good}, {Halpern}, {Hill},
  {Hinshaw}, {Josephy}, {Kaczmarek}, {Kader}, {Kania}, {Kaspi}, {Landecker},
  {Lang}, {Leung}, {Li}, {Lin}, {Masui}, {Mckinven}, {Mena-Parra},
  {Merryfield}, {Meyers}, {Michilli}, {Milutinovic}, {Mirhosseini},
  {M{\"u}nchmeyer}, {Naidu}, {Newburgh}, {Ng}, {Patel}, {Pen}, {Petroff},
  {Pinsonneault-Marotte}, {Pleunis}, {Rafiei-Ravandi}, {Rahman}, {Ransom},
  {Renard}, {Sanghavi}, {Scholz}, {Shaw}, {Shin}, {Siegel}, {Sikora}, {Singh},
  {Smith}, {Stairs}, {Tan}, {Tendulkar}, {Vanderlinde}, {Wang}, {Wulf}, and
  {Zwaniga}]{chime_cat1_21}
{The CHIME/FRB Collaboration}; {Amiri}, M.; {Andersen}, B.C.; {Bandura},
  K.; {Berger}, S.; {Bhardwaj}, M.; {Boyce}, M.M.; {Boyle}, P.J.; {Brar}, C.;
  {Breitman}, D.; et~al.
\newblock {The First CHIME/FRB Fast Radio Burst Catalog}.
\newblock {\em arXiv} {\bf 2021}, arXiv:2106.04352.


\bibitem[{CHIME/FRB Collaboration} \em{et~al.}(2018){CHIME/FRB Collaboration},
  {Amiri}, {Bandura}, {Berger}, {Bhardwaj}, {Boyce}, {Boyle}, {Brar},
  {Burhanpurkar}, {Chawla}, {Chowdhury}, {Cliche}, {Cranmer}, {Cubranic},
  {Deng}, {Denman}, {Dobbs}, {Fandino}, {Fonseca}, {Gaensler}, {Giri},
  {Gilbert}, {Good}, {Guliani}, {Halpern}, {Hinshaw}, {H{\"o}fer}, {Josephy},
  {Kaspi}, {Landecker}, {Lang}, {Liao}, {Masui}, {Mena-Parra}, {Naidu},
  {Newburgh}, {Ng}, {Patel}, {Pen}, {Pinsonneault-Marotte}, {Pleunis}, {Rafiei
  Ravandi}, {Ransom}, {Renard}, {Scholz}, {Sigurdson}, {Siegel}, {Smith},
  {Stairs}, {Tendulkar}, {Vanderlinde}, and {Wiebe}]{chime_frb_18}
{CHIME/FRB Collaboration}; {Amiri}, M.; {Bandura}, K.; {Berger}, P.;
  {Bhardwaj}, M.; {Boyce}, M.M.; {Boyle}, P.J.; {Brar}, C.; {Burhanpurkar}, M.;
  {Chawla}, P.; et~al.
\newblock {The CHIME Fast Radio Burst Project: System Overview}.
\newblock {\em \apj} {\bf 2018}, {\em 863},~48,
  doi:10.3847/1538-4357/aad188.

\bibitem[{CHIME/FRB Collaboration} \em{et~al.}(2019){CHIME/FRB Collaboration},
  {:}, {Andersen}, {Band ura}, {Bhardwaj}, {Boubel}, {Boyce}, {Boyle}, {Brar},
  {Cassanelli}, {Chawla}, {Cubranic}, {Deng}, {Dobbs}, {Fandino}, {Fonseca},
  {Gaensler}, {Gilbert}, {Giri}, {Good}, {Halpern}, {H{\"o}fer}, {Hill},
  {Hinshaw}, {Josephy}, {Kaspi}, {Kothes}, {Landecker}, {Lang}, {Li}, {Lin},
  {Masui}, {Mena-Parra}, {Merryfield}, {Mckinven}, {Michilli}, {Milutinovic},
  {Naidu}, {Newburgh}, {Ng}, {Patel}, {Pen}, {Pinsonneault-Marotte}, {Pleunis},
  {Rafiei-Ravandi}, {Rahman}, {Ransom}, {Renard}, {Scholz}, {Siegel}, {Singh},
  {Smith}, {Stairs}, {Tendulkar}, {Tretyakov}, {Vanderlinde}, {Yadav}, and
  {Zwaniga}]{chime8_19}
{CHIME/FRB Collaboration}; {Andersen}, B.C.; {Band ura}, K.; {Bhardwaj},
  M.; {Boubel}, P.; {Boyce}, M.M.; {Boyle}, P.J.; {Brar}, C.; {Cassanelli}, T.;
  {Chawla}, P.; et~al.
\newblock {CHIME/FRB Detection of Eight New Repeating Fast Radio Burst
  Sources}.
\newblock {\em arXiv} {\bf 2019}, arXiv:1908.03507.


\bibitem[{Inoue}(2004)]{inoue04}
{Inoue}, S.
\newblock {Probing the cosmic reionization history and local environment of
  gamma-ray bursts through radio dispersion}.
\newblock {\em \mnras} {\bf 2004}, {\em 348},~999--1008,
doi:10.1111/j.1365-2966.2004.07359.x.

\bibitem[{Ioka}(2003)]{ioka03}
{Ioka}, K.
\newblock {The Cosmic Dispersion Measure from Gamma-Ray Burst Afterglows:
  Probing the Reionization History and the Burst Environment}.
\newblock {\em \apjl} {\bf 2003}, {\em 598},~L79--L82,
doi:10.1086/380598.

\bibitem[{Macquart} \em{et~al.}(2020){Macquart}, {Prochaska}, {McQuinn},
  {Bannister}, {Bhandari}, {Day}, {Deller}, {Ekers}, {James}, {Marnoch},
  {Os{\l}owski}, {Phillips}, {Ryder}, {Scott}, {Shannon}, and
  {Tejos}]{macquart20}
{Macquart}, J.P.; {Prochaska}, J.X.; {McQuinn}, M.; {Bannister}, K.W.;
  {Bhandari}, S.; {Day}, C.K.; {Deller}, A.T.; {Ekers}, R.D.; {James}, C.W.;
  {Marnoch}, L.; et~al.
\newblock {A census of baryons in the Universe from localized fast radio
  bursts}.
\newblock {\em \nat} {\bf 2020}, {\em 581},~391--395,
 doi:10.1038/s41586-020-2300-2.

\bibitem[{Marcote} \em{et~al.}(2020){Marcote}, {Nimmo}, {Hessels}, {Tendulkar},
  {Bassa}, {Paragi}, {Keimpema}, {Bhardwaj}, {Karuppusamy}, {Kaspi}, {Law},
  {Michilli}, {Aggarwal}, {Andersen}, {Archibald}, {Bandura}, {Bower}, {Boyle},
  {Brar}, {Burke-Spolaor}, {Butler}, {Cassanelli}, {Chawla}, {Demorest},
  {Dobbs}, {Fonseca}, {Giri}, {Good}, {Gourdji}, {Josephy}, {Kirichenko},
  {Kirsten}, {Landecker}, {Lang}, {Lazio}, {Li}, {Lin}, {Linford}, {Masui},
  {Mena-Parra}, {Naidu}, {Ng}, {Patel}, {Pen}, {Pleunis}, {Rafiei-Ravandi},
  {Rahman}, {Renard}, {Scholz}, {Siegel}, {Smith}, {Stairs}, {Vanderlinde}, and
  {Zwaniga}]{marcote20}
{Marcote}, B.; {Nimmo}, K.; {Hessels}, J.W.T.; {Tendulkar}, S.P.; {Bassa},
  C.G.; {Paragi}, Z.; {Keimpema}, A.; {Bhardwaj}, M.; {Karuppusamy}, R.;
  {Kaspi}, V.M.; et~al.
\newblock {A repeating fast radio burst source localized to a nearby spiral
  galaxy}.
\newblock {\em \nat} {\bf 2020}, {\em 577},~190--194,
 doi:10.1038/s41586-019-1866-z.

\bibitem[{CHIME/FRB Collaboration} \em{et~al.}(2020){CHIME/FRB Collaboration},
  {Amiri}, {Andersen}, {Bandura}, {Bhardwaj}, {Boyle}, {Brar}, {Chawla},
  {Chen}, {Cliche}, {Cubranic}, {Deng}, {Denman}, {Dobbs}, {Dong}, {Fand ino},
  {Fonseca}, {Gaensler}, {Giri}, {Good}, {Halpern}, {Hessels}, {Hill},
  {H{\"o}fer}, {Josephy}, {Kania}, {Karuppusamy}, {Kaspi}, {Keimpema},
  {Kirsten}, {Landecker}, {Lang}, {Leung}, {Li}, {Lin}, {Marcote}, {Masui},
  {Mckinven}, {Mena-Parra}, {Merryfield}, {Michilli}, {Milutinovic},
  {Mirhosseini}, {Naidu}, {Newburgh}, {Ng}, {Nimmo}, {Paragi}, {Patel}, {Pen},
  {Pinsonneault-Marotte}, {Pleunis}, {Rafiei-Ravandi}, {Rahman}, {Ransom},
  {Renard}, {Sanghavi}, {Scholz}, {Shaw}, {Shin}, {Siegel}, {Singh}, {Smegal},
  {Smith}, {Stairs}, {Tendulkar}, {Tretyakov}, {Vanderlinde}, {Wang}, {Wang},
  {Wulf}, {Yadav}, and {Zwaniga}]{chime_period_20}
{CHIME/FRB Collaboration}; {Amiri}, M.; {Andersen}, B.C.; {Bandura}, K.M.;
  {Bhardwaj}, M.; {Boyle}, P.J.; {Brar}, C.; {Chawla}, P.; {Chen}, T.;
  {Cliche}, J.F.; et~al.
\newblock {Periodic activity from a fast radio burst source}.
\newblock {\em arXiv} {\bf 2020}, arXiv:2001.10275,


\bibitem[{Scholz} \em{et~al.}(2016){Scholz}, {Spitler}, {Hessels},
  {Chatterjee}, {Cordes}, {Kaspi}, {Wharton}, {Bassa}, {Bogdanov}, {Camilo},
  {Crawford}, {Deneva}, {van Leeuwen}, {Lynch}, {Madsen}, {McLaughlin},
  {Mickaliger}, {Parent}, {Patel}, {Ransom}, {Seymour}, {Stairs}, {Stappers},
  and {Tendulkar}]{scholz16}
{Scholz}, P.; {Spitler}, L.G.; {Hessels}, J.W.T.; {Chatterjee}, S.; {Cordes},
  J.M.; {Kaspi}, V.M.; {Wharton}, R.S.; {Bassa}, C.G.; {Bogdanov}, S.;
  {Camilo}, F.; et~al.
\newblock {The Repeating Fast Radio Burst FRB 121102: Multi-wavelength
  Observations and Additional Bursts}.
\newblock {\em \apj} {\bf 2016}, {\em 833},~177,
doi:10.3847/1538-4357/833/2/177.

\bibitem[{CHIME/FRB Collaboration} \em{et~al.}(2019){CHIME/FRB Collaboration},
  {Amiri}, {Bandura}, {Bhardwaj}, {Boubel}, {Boyce}, {Boyle}, {. Brar},
  {Burhanpurkar}, {Cassanelli}, {Chawla}, {Cliche}, {Cubranic}, {Deng},
  {Denman}, {Dobbs}, {Fandino}, {Fonseca}, {Gaensler}, {Gilbert}, {Gill},
  {Giri}, {Good}, {Halpern}, {Hanna}, {Hill}, {Hinshaw}, {H{\"o}fer},
  {Josephy}, {Kaspi}, {Landecker}, {Lang}, {Lin}, {Masui}, {Mckinven},
  {Mena-Parra}, {Merryfield}, {Michilli}, {Milutinovic}, {Moatti}, {Naidu},
  {Newburgh}, {Ng}, {Patel}, {Pen}, {Pinsonneault-Marotte}, {Pleunis},
  {Rafiei-Ravandi}, {Rahman}, {Ransom}, {Renard}, {Scholz}, {Shaw}, {Siegel},
  {Smith}, {Stairs}, {Tendulkar}, {Tretyakov}, {Vanderlinde}, and
  {Yadav}]{zhang18}
{CHIME/FRB Collaboration}; {Amiri}, M.; {Bandura}, K.; {Bhardwaj}, M.;
  {Boubel}, P.; {Boyce}, M.M.; {Boyle}, P.J.; {Brar}, C.; {Burhanpurkar}, M.;
  {Cassanelli}, T.; et~al.
\newblock {{A second source of repeating fast radio bursts}}.
\newblock {\em \nat} {\bf 2019}, {\em 566},~235--238,
doi:10.1038/s41586-018-0864-x.

\bibitem[{Rajwade} \em{et~al.}(2020){Rajwade}, {Mickaliger}, {Stappers},
  {Bassa}, {Breton}, {Karastergiou}, and {Keane}]{rajwade20a}
{Rajwade}, K.M.; {Mickaliger}, M.B.; {Stappers}, B.W.; {Bassa}, C.G.; {Breton},
  R.P.; {Karastergiou}, A.; {Keane}, E.F.
\newblock {Limits on absorption from a 332-MHz survey for fast radio bursts}.
\newblock {\em \mnras} {\bf 2020}, {\em 493},~4418--4427,
doi:10.1093/mnras/staa616.

\bibitem[{Cruces} \em{et~al.}(2021){Cruces}, {Spitler}, {Scholz}, {Lynch},
  {Seymour}, {Hessels}, {Gouiff{\'e}s}, {Hilmarsson}, {Kramer}, and
  {Munjal}]{cruces21}
{Cruces}, M.; {Spitler}, L.G.; {Scholz}, P.; {Lynch}, R.; {Seymour}, A.;
  {Hessels}, J.W.T.; {Gouiff{\'e}s}, C.; {Hilmarsson}, G.H.; {Kramer}, M.;
  {Munjal}, S.
\newblock {Repeating behaviour of FRB 121102: Periodicity, waiting times, and
  energy distribution}.
\newblock {\em \mnras} {\bf 2021}, {\em 500},~448--463,
doi:10.1093/mnras/staa3223.

\bibitem[{Israel} \em{et~al.}(2016){Israel}, {Esposito}, {Rea}, {Coti Zelati},
  {Tiengo}, {Campana}, {Mereghetti}, {Rodriguez Castillo}, {G{\"o}tz},
  {Burgay}, {Possenti}, {Zane}, {Turolla}, {Perna}, {Cannizzaro}, and
  {Pons}]{israel16}
{Israel}, G.L.; {Esposito}, P.; {Rea}, N.; {Coti Zelati}, F.; {Tiengo}, A.;
  {Campana}, S.; {Mereghetti}, S.; {Rodriguez Castillo}, G.A.; {G{\"o}tz}, D.;
  {Burgay}, M.; et~al.
\newblock {The discovery, monitoring and environment of SGR J1935+2154}.
\newblock {\em \mnras} {\bf 2016}, {\em 457},~3448--3456,
doi:10.1093/mnras/stw008.

\bibitem[{Burgay} \em{et~al.}(2014){Burgay}, {Israel}, {Rea}, {Possenti}, {Coti
  Zelati}, {Esposito}, {Mereghetti}, and {Tiengo}]{burgay14}
{Burgay}, M.; {Israel}, G.L.; {Rea}, N.; {Possenti}, A.; {Coti Zelati}, F.;
  {Esposito}, P.; {Mereghetti}, S.; {Tiengo}, A.
\newblock {Parkes upper limits on the pulsed radio emission of SGR 1935+2154}.
\newblock {\em  Astron. Telegr.} {\bf 2014}, {\em 6371},~1.

\bibitem[{Palmer}(2020)]{palmer20}
{Palmer}, D.M.
\newblock {A Forest of Bursts from SGR 1935+2154}.
\newblock {\em  Astron. Telegr.} {\bf 2020}, {\em 13675},~1.

\bibitem[{The CHIME/FRB Collaboration} \em{et~al.}(2020){The CHIME/FRB
  Collaboration}, {:}, {Andersen}, {Band ura}, {Bhardwaj}, {Bij}, {Boyce},
  {Boyle}, {Brar}, {Cassanelli}, {Chawla}, {Chen}, {Cliche}, {Cook},
  {Cubranic}, {Curtin}, {Denman}, {Dobbs}, {Dong}, {Fandino}, {Fonseca},
  {Gaensler}, {Giri}, {Good}, {Halpern}, {Hill}, {Hinshaw}, {H{\"o}fer},
  {Josephy}, {Kania}, {Kaspi}, {Landecker}, {Leung}, {Li}, {Lin}, {Masui},
  {Mckinven}, {Mena-Parra}, {Merryfield}, {Meyers}, {Michilli}, {Milutinovic},
  {Mirhosseini}, {M{\"u}nchmeyer}, {Naidu}, {Newburgh}, {Ng}, {Patel}, {Pen},
  {Pinsonneault-Marotte}, {Pleunis}, {Quine}, {Rafiei-Ravandi}, {Rahman},
  {Ransom}, {Renard}, {Sanghavi}, {Scholz}, {Shaw}, {Shin}, {Siegel}, {Singh},
  {Smegal}, {Smith}, {Stairs}, {Tan}, {Tendulkar}, {Tretyakov}, {Vanderlinde},
  {Wang}, {Wulf}, and {Zwaniga}]{chime20sgr}
{The CHIME/FRB Collaboration}; {Andersen}, B.C.; {Band ura}, K.M.;
  {Bhardwaj}, M.; {Bij}, A.; {Boyce}, M.M.; {Boyle}, P.J.; {Brar}, C.;
  {Cassanelli}, T.; {Chawla}, P.; et~al.
\newblock {A bright millisecond-duration radio burst from a Galactic magnetar}.
\newblock {\em arXiv} {\bf 2020}, arXiv:2005.10324,


\bibitem[{Bochenek} \em{et~al.}(2020){Bochenek}, {Ravi}, {Belov}, {Hallinan},
  {Kocz}, {Kulkarni}, and {McKenna}]{bochenek20}
{Bochenek}, C.D.; {Ravi}, V.; {Belov}, K.V.; {Hallinan}, G.; {Kocz}, J.;
  {Kulkarni}, S.R.; {McKenna}, D.L.
\newblock {A fast radio burst associated with a Galactic magnetar}.
\newblock {\em arXiv} {\bf 2020}, arXiv:2005.10828.


\bibitem[{Mereghetti} \em{et~al.}(2020){Mereghetti}, {Savchenko}, {Gotz},
  {Ducci}, {Ferrigno}, {Bozzo}, {Borkowski}, and {Bazzano}]{mereghetti20}
{Mereghetti}, S.; {Savchenko}, V.; {Gotz}, D.; {Ducci}, L.; {Ferrigno}, C.;
  {Bozzo}, E.; {Borkowski}, J.; {Bazzano}, A.
\newblock {SGR 1935+2154: INTEGRAL hard X-ray counterpart of radio burst}.
\newblock {\em GRB Coord. Netw.} {\bf 2020}, {\em 27668},~1.

\bibitem[{Tavani} \em{et~al.}(2020){Tavani}, {Casentini}, {Ursi}, {Verrecchia},
  {Addis}, {Antonelli}, {Argan}, {Barbiellini}, {Baroncelli}, {Bernardi},
  {Bianchi}, {Bulgarelli}, {Caraveo}, {Cardillo}, {Cattaneo}, {Chen}, {Costa},
  {Del Monte}, {Di Cocco}, {Di Persio}, {Donnarumma}, {Evangelista}, {Feroci},
  {Ferrari}, {Fioretti}, {Fuschino}, {Galli}, {Gianotti}, {Giuliani},
  {Labanti}, {Lazzarotto}, {Lipari}, {Longo}, {Lucarelli}, {Magro},
  {Marisaldi}, {Mereghetti}, {Morelli}, {Morselli}, {Naldi}, {Pacciani},
  {Parmiggiani}, {Paoletti}, {Pellizzoni}, {Perri}, {Perotti}, {Piano},
  {Picozza}, {Pilia}, {Pittori}, {Puccetti}, {Pupillo}, {Rapisarda},
  {Rappoldi}, {Rubini}, {Setti}, {Soffitta}, {Trifoglio}, {Trois},
  {Vercellone}, {Vittorini}, {Giommi}, and {D' Amico}]{tavani20b}
{Tavani}, M.; {Casentini}, C.; {Ursi}, A.; {Verrecchia}, F.; {Addis}, A.;
  {Antonelli}, L.A.; {Argan}, A.; {Barbiellini}, G.; {Baroncelli}, L.;
  {Bernardi}, G.; et~al.
\newblock {An X-Ray Burst from a Magnetar Enlightening the Mechanism of Fast
  Radio Bursts}.
\newblock {\em arXiv} {\bf 2020}, arXiv:2005.12164.


\bibitem[{Li} \em{et~al.}(2020){Li}, {Lin}, {Xiong}, {Ge}, {Li}, {Li}, {Lu},
  {Zhang}, {Tuo}, {Nang}, {Zhang}, {Xiao}, {Chen}, {Song}, {Xu}, {Liu}, {Jia},
  {Cao}, {Zhang}, {Qu}, {Liao}, {Zhao}, {Tan}, {Nie}, {Zhao}, {Zheng}, {Zheng},
  {Luo}, {Cai}, {Li}, {Xue}, {Bu}, {Chang}, {Chen}, {Chen}, {Chen}, {Chen},
  {Chen}, {Cui}, {Cui}, {Deng}, {Dong}, {Du}, {Fu}, {Gao}, {Gao}, {Gao}, {Gu},
  {Guan}, {Guo}, {Han}, {Huang}, {Huo}, {Jiang}, {Jiang}, {Jin}, {Jin}, {Kong},
  {Li}, {Li}, {Li}, {Li}, {Li}, {Li}, {Li}, {Liang}, {Liu}, {Liu}, {Liu},
  {Liu}, {Liu}, {Lu}, {Lu}, {Luo}, {Ma}, {Meng}, {Ou}, {Sai}, {Shang}, {Song},
  {Sun}, {Tao}, {Wang}, {Wang}, {Wang}, {Wang}, {Wang}, {Wen}, {Wu}, {Wu},
  {Wu}, {Xiao}, {Yang}, {Yang}, {Yang}, {Yang}, {Yi}, {Yin}, {You}, {Zhang},
  {Zhang}, {Zhang}, {Zhang}, {Zhang}, {Zhang}, {Zhang}, {Zhang}, {Zhang},
  {Zhang}, {Zhang}, {Zhang}, {Zhang}, {Zhang}, {Zhang}, {Zhang}, {Zhou},
  {Zhou}, {Zhu}, {Zhu}, and {Zhuang}]{hxmt}
{Li}, C.K.; {Lin}, L.; {Xiong}, S.L.; {Ge}, M.Y.; {Li}, X.B.; {Li}, T.P.; {Lu},
  F.J.; {Zhang}, S.N.; {Tuo}, Y.L.; {Nang}, Y.; et~al.
\newblock {Identification of a non-thermal X-ray burst with the Galactic
  magnetar SGR 1935+2154 and a fast radio burst with Insight-HXMT}.
\newblock {\em arXiv} {\bf 2020}, arXiv:2005.11071.


\bibitem[{Ridnaia} \em{et~al.}(2020){Ridnaia}, {Svinkin}, {Frederiks}, {Bykov},
  {Popov}, {Aptekar}, {Golenetskii}, {Lysenko}, {Tsvetkova}, {Ulanov}, and
  {Cline}]{ridnaia20}
{Ridnaia}, A.; {Svinkin}, D.; {Frederiks}, D.; {Bykov}, A.; {Popov}, S.;
  {Aptekar}, R.; {Golenetskii}, S.; {Lysenko}, A.; {Tsvetkova}, A.; {Ulanov},
  M.; {Cline}, T.
\newblock {A peculiar hard X-ray counterpart of a Galactic fast radio burst}.
\newblock {\em arXiv} {\bf 2020}, arXiv:2005.11178.
 

\bibitem[{Bhardwaj} \em{et~al.}(2021){Bhardwaj}, {Gaensler}, {Kaspi},
  {Landecker}, {Mckinven}, {Michilli}, {Pleunis}, {Tendulkar}, {Andersen},
  {Boyle}, {Cassanelli}, {Chawla}, {Cook}, {Dobbs}, {Fonseca}, {Kaczmarek},
  {Leung}, {Masui}, {Mnchmeyer}, {Ng}, {Rafiei-Ravandi}, {Scholz}, {Shin},
  {Smith}, {Stairs}, and {Zwaniga}]{bhardwaj21}
{Bhardwaj}, M.; {Gaensler}, B.M.; {Kaspi}, V.M.; {Landecker}, T.L.; {Mckinven},
  R.; {Michilli}, D.; {Pleunis}, Z.; {Tendulkar}, S.P.; {Andersen}, B.C.;
  {Boyle}, P.J.; et~al.
\newblock {A Nearby Repeating Fast Radio Burst in the Direction of M81}.
\newblock {\em \apjl} {\bf 2021}, {\em 910},~L18,
 doi:10.3847/2041-8213/abeaa6.

\bibitem[{Macquart} and {Ekers}(2018{\natexlab{a}})]{meI18}
{Macquart}, J.P.; {Ekers}, R.D.
\newblock {Fast radio burst event rate counts-I. Interpreting the
  observations}.
\newblock {\em \mnras} {\bf 2018}, {\em 474},~1900--1908,
 doi:10.1093/mnras/stx2825.

\bibitem[{Macquart} and {Ekers}(2018{\natexlab{b}})]{meII18}
{Macquart}, J.P.; {Ekers}, R.
\newblock {FRB event rate counts-II. Fluence, redshift, and dispersion
  measure distributions}.
\newblock {\em \mnras} {\bf 2018}, {\em 480},~4211--4230,
 doi:10.1093/mnras/sty2083.

 

\bibitem[{Bhat} \em{et~al.}(2004){Bhat}, {Cordes}, {Camilo}, {Nice}, and
  {Lorimer}]{bhat04}
{Bhat}, N.D.R.; {Cordes}, J.M.; {Camilo}, F.; {Nice}, D.J.; {Lorimer}, D.R.
\newblock {Multifrequency Observations of Radio Pulse Broadening and
  Constraints on Interstellar Electron Density Microstructure}.
\newblock {\em \apj} {\bf 2004}, {\em 605},~759--783,
  doi:10.1086/382680.

\bibitem[{Lorimer} \em{et~al.}(2013){Lorimer}, {Karastergiou}, {McLaughlin},
  and {Johnston}]{lorimer13}
{Lorimer}, D.R.; {Karastergiou}, A.; {McLaughlin}, M.A.; {Johnston}, S.
\newblock {On the detectability of extragalactic fast radio transients.}
\newblock {\em \mnras} {\bf 2013}, {\em 436},~L5--L9,
 doi:10.1093/mnrasl/slt098.

\bibitem[{Bates} \em{et~al.}(2013){Bates}, {Lorimer}, and {Verbiest}]{bates13}
{Bates}, S.D.; {Lorimer}, D.R.; {Verbiest}, J.P.W.
\newblock {The pulsar spectral index distribution}.
\newblock {\em \mnras} {\bf 2013}, {\em 431},~1352--1358,
 doi:10.1093/mnras/stt257.

\bibitem[{Macquart} and {Koay}(2013)]{mk13}
{Macquart}, J.P.; {Koay}, J.Y.
\newblock {Temporal Smearing of Transient Radio Sources by the Intergalactic
  Medium}.
\newblock {\em \apj} {\bf 2013}, {\em 776},~125,
 doi:10.1088/0004-637X/776/2/125.

\bibitem[{Cordes} \em{et~al.}(2021){Cordes}, {Ocker}, and
  {Chatterjee}]{cordes21}
{Cordes}, J.M.; {Ocker}, S.K.; {Chatterjee}, S.
\newblock {Redshift Estimation and Constraints on Intergalactic and
  Interstellar Media from Dispersion and Scattering of Fast Radio Bursts}.
\newblock {\em arXiv} {\bf 2021}, arXiv:2108.01172.

\bibitem[{The CHIME/FRB Collaboration} \em{et~al.}(2021){The CHIME/FRB
  Collaboration}, {Andersen}, {Bandura}, {Bhardwaj}, {Boyle}, {Brar},
  {Breitman}, {Cassanelli}, {Chatterjee}, {Chawla}, {Cliche}, {Cubranic},
  {Curtin}, {Deng}, {Dobbs}, {Dong}, {Fonseca}, {Gaensler}, {Giri}, {Good},
  {Hill}, {Josephy}, {Kaczmarek}, {Kader}, {Kania}, {Kaspi}, {Leung}, {Li},
  {Lin}, {Masui}, {Mckinven}, {Mena-Parra}, {Merryfield}, {Meyers}, {Michilli},
  {Naidu}, {Newburgh}, {Ng}, {Ordog}, {Patel}, {Pearlman}, {Pen}, {Petroff},
  {Pleunis}, {Rafiei-Ravandi}, {Rahman}, {Ransom}, {Renard}, {Sanghavi},
  {Scholz}, {Shaw}, {Shin}, {Siegel}, {Singh}, {Smith}, {Stairs}, {Tan},
  {Tendulkar}, {Vanderlinde}, {Wiebe}, {Wulf}, and {Zwaniga}]{chime_ms_period}
{The CHIME/FRB Collaboration}; {Andersen}, B.C.; {Bandura}, K.; {Bhardwaj},
  M.; {Boyle}, P.J.; {Brar}, C.; {Breitman}, D.; {Cassanelli}, T.;
  {Chatterjee}, S.; {Chawla}, P.; et~al.
\newblock {Sub-second periodicity in a fast radio burst}.
\newblock {\em arXiv} {\bf 2021}, arXiv:2107.08463.

\bibitem[{Day} \em{et~al.}(2020){Day}, {Deller}, {Shannon}, {Qiu(邱昊)},
  {Bannister}, {Bhandari}, {Ekers}, {Flynn}, {James}, {Macquart}, {Mahony},
  {Phillips}, and {Xavier Prochaska}]{day20}
{Day}, C.K.; {Deller}, A.T.; {Shannon}, R.M.; {Qiu}, H.; {Bannister},
  K.W.; {Bhandari}, S.; {Ekers}, R.; {Flynn}, C.; {James}, C.W.; {Macquart},
  J.P.; et~al.
\newblock {High time resolution and polarization properties of ASKAP-localized
  fast radio bursts}.
\newblock {\em \mnras} {\bf 2020}, {\em 497},~3335--3350,
doi:10.1093/mnras/staa2138.

\bibitem[{Pleunis} \em{et~al.}(2021){Pleunis}, {Good}, {Kaspi}, {Mckinven},
  {Ransom}, {Scholz}, {Bandura}, {Bhardwaj}, {Boyle}, {Brar}, {Cassanelli},
  {Chawla}, {Fengqiu}, {Dong}, {Fonseca}, {Gaensler}, {Josephy}, {Kaczmarek},
  {Leung}, {Lin}, {Masui}, {Mena-Parra}, {Michilli}, {Ng}, {Patel},
  {Rafiei-Ravandi}, {Rahman}, {Sanghavi}, {Shin}, {Smith}, {Stairs}, and
  {Tendulkar}]{pleunis21b}
{Pleunis}, Z.; {Good}, D.C.; {Kaspi}, V.M.; {Mckinven}, R.; {Ransom}, S.M.;
  {Scholz}, P.; {Bandura}, K.; {Bhardwaj}, M.; {Boyle}, P.J.; {Brar}, C.;
et~al.
\newblock {Fast Radio Burst Morphology in the First CHIME/FRB Catalog}.
\newblock {\em arXiv} {\bf 2021}, arXiv:2106.04356.

\bibitem[{Hessels} \em{et~al.}(2019){Hessels}, {Spitler}, {Seymour}, {Cordes},
  {Michilli}, {Lynch}, {Gourdji}, {Archibald}, {Bassa}, {Bower}, {Chatterjee},
  {Connor}, {Crawford}, {Deneva}, {Gajjar}, {Kaspi}, {Keimpema}, {Law},
  {Marcote}, {McLaughlin}, {Paragi}, {Petroff}, {Ransom}, {Scholz}, {Stappers},
  and {Tendulkar}]{hessels19}
{Hessels}, J.W.T.; {Spitler}, L.G.; {Seymour}, A.D.; {Cordes}, J.M.;
  {Michilli}, D.; {Lynch}, R.S.; {Gourdji}, K.; {Archibald}, A.M.; {Bassa},
  C.G.; {Bower}, G.C.; et~al.
\newblock {FRB 121102 Bursts Show Complex Time-Frequency Structure}.
\newblock {\em \apjl} {\bf 2019}, {\em 876},~L23,
 doi:10.3847/2041-8213/ab13ae.
 
\bibitem[{Cordes} and {McLaughlin}(2003)]{cm03}
{Cordes}, J.M.; {McLaughlin}, M.A.
\newblock {Searches for Fast Radio Transients}.
\newblock {\em \apj} {\bf 2003}, {\em 596},~1142--1154,
doi:10.1086/378231.

\bibitem[{Gajjar} \em{et~al.}(2020){Gajjar}, {Siemion}, {Price}, ., {Li}, {Li},
  {Li}, {Li}, {Li}, {Li}, {Li}, {Li}, {Li}, {Li}, {Liang}, {Liu}, {Liu}, {Liu},
  {Liu}, {Liu}, {Liu}, {Lu}, {Lu}, {Lu}, {Luo}, {Luo}, {Ma}, {Ma}, {Meng},
  {Nang}, {Nie}, {Oui}, {Qu}, {Sai}, {Shang}, {Song}, {Song}, {Sun}, {Tani},
  {Tao}, {Tuo}, {Wang}, {Wang}, {Wang}, {Wang}, {Wang}, {Wen}, {Wu}, {Wu},
  {Wu}, {Xiao}, {Xiao}, {Xu}, {Yang}, {Yang}, {Yang}, {Yi}, {Yin}, {You},
  {Zhang}, {Zhang}, {Zhang}, {Zhang}, {Zhang}, {Zhang}, {Zhang}, {Zhang},
  {Zhang}, {Zhang}, {Zhang}, {Zhang}, {Zhang}, {Zhang}, {Zhang}, {Zhang},
  {Zhang}, {Zheng}, {Zhou}, {Zhou}, {Zhu}, {Zhu}, and {Zhuang}]{gajjar18}
{Gajjar}, V.; {Siemion}, A.P.V.; {Price}, D.C.; {Li}, B.; {Li}, C.K.; {Li},
  M.S.; {Li}, T.P.; {Li}, W.; {Li}, X.; {Li}, X.B.; et~al.
\newblock {A search for prompt gamma-ray counterparts to fast radio bursts in
  the Insight-HXMT data}.
\newblock {\em arXiv} {\bf 2020}, arXiv:2003.10889.

\bibitem[{Seymour} \em{et~al.}(2019){Seymour}, {Michilli}, and
  {Pleunis}]{seymour19}
{Seymour}, A.; {Michilli}, D.; {Pleunis}, Z.
\newblock {DM\_phase: Algorithm for correcting dispersion of radio signals}.
 \emph{{Astrophys. Source Code Libr.}}  \textbf{{2019}}, {ascl-1910.004}.

\bibitem[{Bilous} \em{et~al.}(2021){Bilous}, {Griessmeier}, {Pennucci}, {Wu},
  {Bondonneau}, {Kondratiev}, {van Leeuwen}, {Maan}, {Connor}, {Oostrum},
  {Petroff}, {Verbiest}, {Vohl}, {McKee}, {Shaifullah}, {Theureau}, {Ulyanov},
  {Cecconi}, {Coolen}, {Corbel}, {Damstra}, {Denes}, {Girard}, {Hut},
  {Ivashina}, {Konovalenko}, {Kutkin}, {Loose}, {Mulder}, {Ruiter}, {Smits},
  {Tokarsky}, {Vermaas}, {Zakharenko}, {Zarka}, and {Ziemke}]{bilous21}
{Bilous}, A.V.; {Griessmeier}, J.M.; {Pennucci}, T.; {Wu}, Z.; {Bondonneau},
  L.; {Kondratiev}, V.; {van Leeuwen}, J.; {Maan}, Y.; {Connor}, L.; {Oostrum},
  L.C.; et~al.
\newblock {Dual-frequency single-pulse study of PSR B0950+08}.
\newblock {\em arXiv} {\bf 2021}, arXiv:2109.08500.
 
\bibitem[{Majid} \em{et~al.}(2021){Majid}, {Pearlman}, {Prince}, {Wharton},
  {Naudet}, {Bansal}, {Connor}, {Bhardwaj}, and {Tendulkar}]{majid21}
{Majid}, W.A.; {Pearlman}, A.B.; {Prince}, T.A.; {Wharton}, R.S.; {Naudet},
  C.J.; {Bansal}, K.; {Connor}, L.; {Bhardwaj}, M.; {Tendulkar}, S.P.
\newblock {A Bright Fast Radio Burst from FRB 20200120E with Sub-100 Nanosecond
  Structure}.
\newblock {\em \apjl} {\bf 2021}, {\em 919},~L6,
  doi:10.3847/2041-8213/ac1921.

\bibitem[{Nimmo} \em{et~al.}(2021){Nimmo}, {Hessels}, {Kirsten}, {Keimpema},
  {Cordes}, {Snelders}, {Hewitt}, {Karuppusamy}, {Archibald}, {Bezukovs},
  {Bhardwaj}, {Blaauw}, {Buttaccio}, {Cassanelli}, {Conway}, {Corongiu},
  {Feiler}, {Fonseca}, {Forssen}, {Gawronski}, {Giroletti}, {Kharinov},
  {Leung}, {Lindqvist}, {Maccaferri}, {Marcote}, {Masui}, {Mckinven},
  {Melnikov}, {Michilli}, {Mikhailov}, {Ng}, {Orbidans}, {Ould-Boukattine},
  {Paragi}, {Pearlman}, {Petroff}, {Rahman}, {Scholz}, {Shin}, {Smith},
  {Stairs}, {Surcis}, {Tendulkar}, {Vlemmings}, {Wang}, {Yang}, and
  {Yuan}]{nimmo21}
{Nimmo}, K.; {Hessels}, J.W.T.; {Kirsten}, F.; {Keimpema}, A.; {Cordes}, J.M.;
  {Snelders}, M.P.; {Hewitt}, D.M.; {Karuppusamy}, R.; {Archibald}, A.M.;
  {Bezukovs}, V.; et~al.
\newblock {Burst timescales and luminosities link young pulsars and fast radio
  bursts}.
\newblock {\em arXiv} {\bf 2021}, arXiv:2105.11446.

\bibitem[{Zhang}(2020)]{zhang20ApJL}
{Zhang}, B.
\newblock {Fast Radio Bursts from Interacting Binary Neutron Star Systems}.
\newblock {\em \apjl} {\bf 2020}, {\em 890},~L24,
 doi:10.3847/2041-8213/ab7244.

\bibitem[{Michilli} \em{et~al.}(2018){Michilli}, {Seymour}, {Hessels},
  {Spitler}, {Gajjar}, {Archibald}, {Bower}, {Chatterjee}, {Cordes}, {Gourdji},
  {Heald}, {Kaspi}, {Law}, {Sobey}, {Adams}, {Bassa}, {Bogdanov}, {Brinkman},
  {Demorest}, {Fernand ez}, {Hellbourg}, {Lazio}, {Lynch}, {Maddox}, {Marcote},
  {McLaughlin}, {Paragi}, {Ransom}, {Scholz}, {Siemion}, {Tendulkar}, {van
  Rooy}, {Wharton}, and {Whitlow}]{michilli18}
{Michilli}, D.; {Seymour}, A.; {Hessels}, J.W.T.; {Spitler}, L.G.; {Gajjar},
  V.; {Archibald}, A.M.; {Bower}, G.C.; {Chatterjee}, S.; {Cordes}, J.M.;
  {Gourdji}, K.; et~al.
\newblock {An extreme magneto-ionic environment associated with the fast radio
  burst source FRB 121102}.
\newblock {\em \nat} {\bf 2018}, {\em 553},~182--185,
doi:10.1038/nature25149.

\bibitem[{CHIME/FRB Collaboration} \em{et~al.}(2019{\natexlab{a}}){CHIME/FRB
  Collaboration}, {Amiri}, {Bandura}, {Bhardwaj}, {Boubel}, {Boyce}, {Boyle},
  {. Brar}, {Burhanpurkar}, {Cassanelli}, {Chawla}, {Cliche}, {Cubranic},
  {Deng}, {Denman}, {Dobbs}, {Fandino}, {Fonseca}, {Gaensler}, {Gilbert},
  {Gill}, {Giri}, {Good}, {Halpern}, {Hanna}, {Hill}, {Hinshaw}, {H{\"o}fer},
  {Josephy}, {Kaspi}, {Landecker}, {Lang}, {Lin}, {Masui}, {Mckinven},
  {Mena-Parra}, {Merryfield}, {Michilli}, {Milutinovic}, {Moatti}, {Naidu},
  {Newburgh}, {Ng}, {Patel}, {Pen}, {Pinsonneault-Marotte}, {Pleunis},
  {Rafiei-Ravandi}, {Rahman}, {Ransom}, {Renard}, {Scholz}, {Shaw}, {Siegel},
  {Smith}, {Stairs}, {Tendulkar}, {Tretyakov}, {Vanderlinde}, and
  {Yadav}]{chime_1r_19}
{CHIME/FRB Collaboration}; {Amiri}, M.; {Bandura}, K.; {Bhardwaj}, M.;
  {Boubel}, P.; {Boyce}, M.M.; {Boyle}, P.J.; {. Brar}, C.; {Burhanpurkar}, M.;
  {Cassanelli}, T.; et~al.
\newblock {{A second source of repeating fast radio bursts}}.
\newblock {\em \nat} {\bf 2019}, {\em 566},~235--238,
doi:10.1038/s41586-018-0864-x.

\bibitem[{CHIME/FRB Collaboration} \em{et~al.}(2019{\natexlab{b}}){CHIME/FRB
  Collaboration}, {Andersen}, {Bandura}, {Bhardwaj}, {Boubel}, {Boyce},
  {Boyle}, {Brar}, {Cassanelli}, {Chawla}, {Cubranic}, {Deng}, {Dobbs},
  {Fandino}, {Fonseca}, {Gaensler}, {Gilbert}, {Giri}, {Good}, {Halpern},
  {Hill}, {Hinshaw}, {H{\"o}fer}, {Josephy}, {Kaspi}, {Kothes}, {Landecker},
  {Lang}, {Li}, {Lin}, {Masui}, {Mena-Parra}, {Merryfield}, {Mckinven},
  {Michilli}, {Milutinovic}, {Naidu}, {Newburgh}, {Ng}, {Patel}, {Pen},
  {Pinsonneault-Marotte}, {Pleunis}, {Rafiei-Ravandi}, {Rahman}, {Ransom},
  {Renard}, {Scholz}, {Siegel}, {Singh}, {Smith}, {Stairs}, {Tendulkar},
  {Tretyakov}, {Vanderlinde}, {Yadav}, and {Zwaniga}]{chime8frb}
{CHIME/FRB Collaboration}; {Andersen}, B.C.; {Bandura}, K.; {Bhardwaj}, M.;
  {Boubel}, P.; {Boyce}, M.M.; {Boyle}, P.J.; {Brar}, C.; {Cassanelli}, T.;
  {Chawla}, P.; et~al.
\newblock {CHIME/FRB Discovery of Eight New Repeating Fast Radio Burst
  Sources}.
\newblock {\em \apjl} {\bf 2019}, {\em 885},~L24,
 doi:10.3847/2041-8213/ab4a80.

\bibitem[{Li} \em{et~al.}(2021){Li}, {Wang}, {Zhu}, {Zhang}, {Zhang}, {Duan},
  {Zhang}, {Feng}, {Tang}, {Chatterjee}, {Cordes}, {Cruces}, {Dai}, {Gajjar},
  {Hobbs}, {Jin}, {Kramer}, {Lorimer}, {Miao}, {Niu}, {Niu}, {Pan}, {Qian},
  {Spitler}, {Werthimer}, {Zhang}, {Wang}, {Xie}, {Yue}, {Zhang}, {Zhi}, and
  {Zhu}]{li21}
{Li}, D.; {Wang}, P.; {Zhu}, W.W.; {Zhang}, B.; {Zhang}, X.X.; {Duan}, R.;
  {Zhang}, Y.K.; {Feng}, Y.; {Tang}, N.Y.; {Chatterjee}, S.; et~al.
\newblock {A bimodal burst energy distribution of a repeating fast radio burst
  source}.
\newblock {\em arXiv} {\bf 2021}, arXiv:2107.08205.

\bibitem[{Josephy} \em{et~al.}(2019){Josephy}, {Chawla}, {Fonseca}, {Ng},
  {Patel}, {Pleunis}, {Scholz}, {Andersen}, {Bandura}, {Bhardwaj}, {Boyce},
  {Boyle}, {Brar}, {Cubranic}, {Dobbs}, {Gaensler}, {Gill}, {Giri}, {Good},
  {Halpern}, {Hinshaw}, {Kaspi}, {Landecker}, {Lang}, {Lin}, {Masui},
  {Mckinven}, {Mena-Parra}, {Merryfield}, {Michilli}, {Milutinovic}, {Naidu},
  {Pen}, {Rafiei-Ravandi}, {Rahman}, {Ransom}, {Renard}, {Siegel}, {Smith},
  {Stairs}, {Tendulkar}, {Vanderlinde}, {Yadav}, and {Zwaniga}]{josephy19}
{Josephy}, A.; {Chawla}, P.; {Fonseca}, E.; {Ng}, C.; {Patel}, C.; {Pleunis},
  Z.; {Scholz}, P.; {Andersen}, B.C.; {Bandura}, K.; {Bhardwaj}, M.; et~al.
\newblock {CHIME/FRB Detection of the Original Repeating Fast Radio Burst
  Source FRB 121102}.
\newblock {\em \apjl} {\bf 2019}, {\em 882},~L18,
doi:10.3847/2041-8213/ab2c00.

\bibitem[{Pastor-Marazuela} \em{et~al.}(2020){Pastor-Marazuela}, {Connor}, {van
  Leeuwen}, {Maan}, {ter Veen}, {Bilous}, {Oostrum}, {Petroff}, {Straal},
  {Vohl}, {Attema}, {Boersma}, {Kooistra}, {van der Schuur}, {Sclocco},
  {Smits}, {Adams}, {Adebahr}, {de Blok}, {Coolen}, {Damstra}, {D{\'e}nes},
  {Hess}, {van der Hulst}, {Hut}, {Ivashina}, {Kutkin}, {Marcel Loose},
  {Lucero}, {Mika}, {Moss}, {Mulder}, {Norden}, {Oosterloo}, {Orr{\'u}},
  {Ruiter}, and {Wijnholds}]{pastor21}
{Pastor-Marazuela}, I.; {Connor}, L.; {van Leeuwen}, J.; {Maan}, Y.; {ter
  Veen}, S.; {Bilous}, A.; {Oostrum}, L.; {Petroff}, E.; {Straal}, S.; {Vohl},
  D.; et~al.
\newblock {Chromatic periodic activity down to 120 MHz in a Fast Radio Burst}.
\newblock {\em arXiv} {\bf 2020}, arXiv:2012.08348.

\bibitem[{Pearlman} \em{et~al.}(2020){Pearlman}, {Majid}, {Prince}, {Nimmo},
  {Hessels}, {Naudet}, and {Kocz}]{pearlman20}
{Pearlman}, A.B.; {Majid}, W.A.; {Prince}, T.A.; {Nimmo}, K.; {Hessels},
  J.W.T.; {Naudet}, C.J.; {Kocz}, J.
\newblock {Multiwavelength Radio Observations of Two Repeating Fast Radio Burst
  Sources: FRB 121102 and FRB 180916.J0158+65}.
\newblock {\em \apjl} {\bf 2020}, {\em 905},~L27,
doi:10.3847/2041-8213/abca31.

\bibitem[{Katz}(2021)]{katz21}
{Katz}, J.I.
\newblock {The environment of FRB 121102 and possible relation to SGR/PSR
  J1745-2900}.
\newblock {\em \mnras} {\bf 2021}, {\em 501},~L76--L79,
 doi:10.1093/mnrasl/slaa202.

\bibitem[{Marcote} \em{et~al.}(2017){Marcote}, {Paragi}, {Hessels}, {Keimpema},
  {van Langevelde}, {Huang}, {Bassa}, {Bogdanov}, {Bower}, {Burke-Spolaor},
  {Butler}, {Campbell}, {Chatterjee}, {Cordes}, {Demorest}, {Garrett}, {Ghosh},
  {Kaspi}, {Law}, {Lazio}, {McLaughlin}, {Ransom}, {Salter}, {Scholz},
  {Seymour}, {Siemion}, {Spitler}, {Tendulkar}, and {Wharton}]{marcote17}
{Marcote}, B.; {Paragi}, Z.; {Hessels}, J.W.T.; {Keimpema}, A.; {van
  Langevelde}, H.J.; {Huang}, Y.; {Bassa}, C.G.; {Bogdanov}, S.; {Bower}, G.C.;
  {Burke-Spolaor}, S.; et~al.
\newblock {The Repeating Fast Radio Burst FRB 121102 as Seen on Milliarcsecond
  Angular Scales}.
\newblock {\em \apjl} {\bf 2017}, {\em 834},~L8,
 doi:10.3847/2041-8213/834/2/L8.

\bibitem[{Pilia} \em{et~al.}(2020){Pilia}, {Burgay}, {Possenti}, {Ridolfi},
  {Gajjar}, {Corongiu}, {Perrodin}, {Bernardi}, {Naldi}, {Pupillo},
  {Ambrosino}, {Bianchi}, {Burtovoi}, {Casella}, {Casentini}, {Cecconi},
  {Ferrigno}, {Fiori}, {Gendreau}, {Ghedina}, {Naletto}, {Nicastro}, {Ochner},
  {Palazzi}, {Panessa}, {Papitto}, {Pittori}, {Rea}, {Rodriguez Castillo},
  {Savchenko}, {Setti}, {Tavani}, {Trois}, {Trudu}, {Turatto}, {Ursi},
  {Verrecchia}, and {Zampieri}]{pilia20}
{Pilia}, M.; {Burgay}, M.; {Possenti}, A.; {Ridolfi}, A.; {Gajjar}, V.;
  {Corongiu}, A.; {Perrodin}, D.; {Bernardi}, G.; {Naldi}, G.; {Pupillo}, G.;
  et~al.
\newblock {The lowest frequency Fast Radio Bursts: Sardinia Radio Telescope
  detection of the periodic FRB 180916 at 328 MHz}.
\newblock {\em arXiv} {\bf 2020}, arXiv:2003.12748.

\bibitem[{Chawla} \em{et~al.}(2020){Chawla}, {Andersen}, {Bhardwaj}, {Fonseca},
  {Josephy}, {Kaspi}, {Michilli}, {Pleunis}, {Bandura}, {Bassa}, {Boyle},
  {Brar}, {Cassanelli}, {Cubranic}, {Dobbs}, {Dong}, {Gaensler}, {Good},
  {Hessels}, {Land ecker}, {Leung}, {Li}, {Lin}, {Masui}, {Mckinven},
  {Mena-Parra}, {Merryfield}, {Meyers}, {Naidu}, {Ng}, {Patel},
  {Rafiei-Ravandi}, {Rahman}, {Sanghavi}, {Scholz}, {Shin}, {Smith}, {Stairs},
  {Tendulkar}, and {Vanderlinde}]{chawla20}
{Chawla}, P.; {Andersen}, B.C.; {Bhardwaj}, M.; {Fonseca}, E.; {Josephy}, A.;
  {Kaspi}, V.M.; {Michilli}, D.; {Pleunis}, Z.; {Bandura}, K.M.; {Bassa}, C.G.;
  et~al.
\newblock {Detection of Repeating FRB 180916.J0158+65 Down to Frequencies of
  300 MHz}.
\newblock {\em \apjl} {\bf 2020}, {\em 896},~L41,
 doi:10.3847/2041-8213/ab96bf.

\bibitem[{Pleunis} \em{et~al.}(2021){Pleunis}, {Michilli}, {Bassa}, {Hessels},
  {Naidu}, {Andersen}, {Chawla}, {Fonseca}, {Gopinath}, {Kaspi}, {Kondratiev},
  {Li}, {Bhardwaj}, {Boyle}, {Brar}, {Cassanelli}, {Gupta}, {Josephy},
  {Karuppusamy}, {Keimpema}, {Kirsten}, {Leung}, {Marcote}, {Masui},
  {Mckinven}, {Meyers}, {Ng}, {Nimmo}, {Paragi}, {Rahman}, {Scholz}, {Shin},
  {Smith}, {Stairs}, and {Tendulkar}]{pleunis21a}
{Pleunis}, Z.; {Michilli}, D.; {Bassa}, C.G.; {Hessels}, J.W.T.; {Naidu}, A.;
  {Andersen}, B.C.; {Chawla}, P.; {Fonseca}, E.; {Gopinath}, A.; {Kaspi}, V.M.;
  et~al.
\newblock {LOFAR Detection of 110-188 MHz Emission and Frequency-dependent
  Activity from FRB 20180916B}.
\newblock {\em \apjl} {\bf 2021}, {\em 911},~L3,
doi:10.3847/2041-8213/abec72.

\bibitem[{Law} \em{et~al.}(2017){Law}, {Abruzzo}, {Bassa}, {Bower},
  {Burke-Spolaor}, {Butler}, {Cantwell}, {Carey}, {Chatterjee}, {Cordes},
  {Demorest}, {Dowell}, {Fender}, {Gourdji}, {Grainge}, {Hessels}, {Hickish},
  {Kaspi}, {Lazio}, {McLaughlin}, {Michilli}, {Mooley}, {Perrott}, {Ransom},
  {Razavi-Ghods}, {Rupen}, {Scaife}, {Scott}, {Scholz}, {Seymour}, {Spitler},
  {Stovall}, {Tendulkar}, {Titterington}, {Wharton}, and {Williams}]{law17}
{Law}, C.J.; {Abruzzo}, M.W.; {Bassa}, C.G.; {Bower}, G.C.; {Burke-Spolaor},
  S.; {Butler}, B.J.; {Cantwell}, T.; {Carey}, S.H.; {Chatterjee}, S.;
  {Cordes}, J.M.; et~al.
\newblock {A Multi-telescope Campaign on FRB 121102: Implications for the FRB
  Population}.
\newblock {\em \apj} {\bf 2017}, {\em 850},~76,
  doi:10.3847/1538-4357/aa9700.

\bibitem[{Kumar} \em{et~al.}(2021){Kumar}, {Shannon}, {Flynn}, {Os{\l}owski},
  {Bhandari}, {Day}, {Deller}, {Farah}, {Kaczmarek}, {Kerr}, {Phillips},
  {Price}, {Qiu}, and {Thyagarajan}]{kumar21}
{Kumar}, P.; {Shannon}, R.M.; {Flynn}, C.; {Os{\l}owski}, S.; {Bhandari}, S.;
  {Day}, C.K.; {Deller}, A.T.; {Farah}, W.; {Kaczmarek}, J.F.; {Kerr}, M.;
 et~al.
\newblock {Extremely band-limited repetition from a fast radio burst source}.
\newblock {\em \mnras} {\bf 2021}, {\em 500},~2525--2531,
doi:10.1093/mnras/staa3436.

\bibitem[{Geyer} \em{et~al.}(2021){Geyer}, {Serylak}, {Abbate}, {Bailes},
  {Buchner}, {Chilufya}, {Johnston}, {Karastergiou}, {Main}, {van Straten}, and
  {Shamohammadi}]{geyer21}
{Geyer}, M.; {Serylak}, M.; {Abbate}, F.; {Bailes}, M.; {Buchner}, S.;
  {Chilufya}, J.; {Johnston}, S.; {Karastergiou}, A.; {Main}, R.; {van
  Straten}, W.; {Shamohammadi}, M.
\newblock {The Thousand-Pulsar-Array programme on MeerKAT - III. Giant pulse
  characteristics of PSR J0540-6919}.
\newblock {\em \mnras} {\bf 2021}, {\em 505},~4468--4482,
  doi:10.1093/mnras/stab1501.

\bibitem[{Piro} \em{et~al.}(2021){Piro}, {Bruni}, {Troja}, {O'Connor},
  {Panessa}, {Ricci}, {Zhang}, {Burgay}, {Dichiara}, {Lee}, {Lotti}, {Niu},
  {Pilia}, {Possenti}, {Trudu}, {Xu}, {Zhu}, {Kutyrev}, and {Veilleux}]{piro21}
{Piro}, L.; {Bruni}, G.; {Troja}, E.; {O'Connor}, B.; {Panessa}, F.; {Ricci},
  R.; {Zhang}, B.; {Burgay}, M.; {Dichiara}, S.; {Lee}, K.J.; et~al.
\newblock {The Fast Radio Burst FRB 20201124A in a star forming region:
  constraints to the progenitor and multiwavelength counterparts}.
\newblock {\em arXiv} {\bf 2021}, arXiv:2107.14339.

\bibitem[{Marthi} \em{et~al.}(2021){Marthi}, {Bethapudi}, {Main}, {Lin},
  {Spitler}, {Wharton}, {Li}, {Gautam}, {Pen}, and {Hilmarsson}]{marthi21}
{Marthi}, V.R.; {Bethapudi}, S.; {Main}, R.A.; {Lin}, H.H.; {Spitler}, L.G.;
  {Wharton}, R.S.; {Li}, D.Z.; {Gautam}, T.G.; {Pen}, U.L.; {Hilmarsson}, G.H.
\newblock {Burst properties of the highly active FRB 20201124A using uGMRT}.
\newblock {\em arXiv} {\bf 2021}, arXiv:2108.00697.

\bibitem[{Lanman} \em{et~al.}(2021){Lanman}, {Andersen}, {Chawla}, {Josephy},
  {Kaspi}, {Bandura}, {Bhardwaj}, {Boyle}, {Brar}, {Breitman}, {Cassanelli},
  {Dong}, {Fonseca}, {Gaensler}, {Good}, {Kaczmarek}, {Leung}, {Masui},
  {Meyers}, {Ng}, {Patel}, {Pearlman}, {Petroff}, {Pleunis}, {Rafiei-Ravandi},
  {Rahman}, {Sanghavi}, {Scholz}, {Shin}, {Stairs}, {Tendulkar}, and
  {Zwaniga}]{lanman21}
{Lanman}, A.E.; {Andersen}, B.C.; {Chawla}, P.; {Josephy}, A.; {Kaspi}, V.M.;
  {Bandura}, K.; {Bhardwaj}, M.; {Boyle}, P.J.; {Brar}, C.; {Breitman}, D.;
  et~al.
\newblock {A sudden period of high activity from repeating Fast Radio Burst
  20201124A}.
\newblock {\em arXiv} {\bf 2021}, arXiv:2109.09254.

\bibitem[{Zhang}(2020)]{zhang20Nat}
{Zhang}, B.
\newblock {The physical mechanisms of fast radio bursts}.
\newblock {\em \nat} {\bf 2020}, {\em 587},~45--53,
 doi:10.1038/s41586-020-2828-1.

\bibitem[{Shannon} \em{et~al.}(2018){Shannon}, {Macquart}, {Bannister},
  {Ekers}, {James}, {Os{\l}owski}, {Qiu}, {Sammons}, {Hotan}, {Voronkov},
  {Beresford}, {Brothers}, {Brown}, {Bunton}, {Chippendale}, {Haskins},
  {Leach}, {Marquarding}, {McConnell}, {Pilawa}, {Sadler}, {Troup}, {Tuthill},
  {Whiting}, {Allison}, {Anderson}, {Bell}, {Collier}, {G{\"u}rkan}, {Heald},
  and {Riseley}]{shannon18}
{Shannon}, R.M.; {Macquart}, J.P.; {Bannister}, K.W.; {Ekers}, R.D.; {James},
  C.W.; {Os{\l}owski}, S.; {Qiu}, H.; {Sammons}, M.; {Hotan}, A.W.; {Voronkov},
  M.A.; et~al.
\newblock {The dispersion-brightness relation for fast radio bursts from a
  wide-field survey}.
\newblock {\em \nat} {\bf 2018}, {\em 562},~386--390, doi:10.1038/s41586-018-0588-y.

\bibitem[{Kumar} \em{et~al.}(2019){Kumar}, {Shannon}, {Os{\l}owski}, {Qiu},
  {Bhandari}, {Farah}, {Flynn}, {Kerr}, {Lorimer}, {Macquart}, {Ng},
  {Phillips}, {Price}, and {Spiewak}]{kumar19}
{Kumar}, P.; {Shannon}, R.M.; {Os{\l}owski}, S.; {Qiu}, H.; {Bhandari}, S.;
  {Farah}, W.; {Flynn}, C.; {Kerr}, M.; {Lorimer}, D.R.; {Macquart}, J.P.;
  et~al.
\newblock {Faint Repetitions from a Bright Fast Radio Burst Source}.
\newblock {\em \apjl} {\bf 2019}, {\em 887},~L30,
 doi:10.3847/2041-8213/ab5b08.

\bibitem[{Kaspi} and {Beloborodov}(2017)]{kaspi17}
{Kaspi}, V.M.; {Beloborodov}, A.M.
\newblock {Magnetars}.
\newblock {\em Annu. Rev. Astron. Astrophys.} {\bf 2017}, {\em 55},~261--301,
 doi:10.1146/annurev-astro-081915-023329.

\bibitem[{Esposito} \em{et~al.}(2021){Esposito}, {Rea}, and
  {Israel}]{esposito21}
{Esposito}, P.; {Rea}, N.; {Israel}, G.L.
\newblock {Magnetars: A Short Review and Some Sparse Considerations}.
\newblock In \emph{Astrophysics and Space Science Library}; {Belloni}, T.M.;
  {M{\'e}ndez}, M.; {Zhang}, C., Eds.; Springer: {Berlin/Heidelberg, Germany,} 
 2021, Volume 461, pp. 97--142,
doi:10.1007/978-3-662-62110-3\_3.

\bibitem[{Torne} \em{et~al.}(2017){Torne}, {Desvignes}, {Eatough},
  {Karuppusamy}, {Paubert}, {Kramer}, {Cognard}, {Champion}, and
  {Spitler}]{torne17}
{Torne}, P.; {Desvignes}, G.; {Eatough}, R.P.; {Karuppusamy}, R.; {Paubert},
  G.; {Kramer}, M.; {Cognard}, I.; {Champion}, D.J.; {Spitler}, L.G.
\newblock {Detection of the magnetar SGR J1745-2900 up to 291 GHz with evidence
  of polarized millimetre emission}.
\newblock {\em \mnras} {\bf 2017}, {\em 465},~242--247, doi:10.1093/mnras/stw2757.

\bibitem[{Lu} \em{et~al.}(2020){Lu}, {Kumar}, and {Zhang}]{lkz20}
{Lu}, W.; {Kumar}, P.; {Zhang}, B.
\newblock {A unified picture of Galactic and cosmological fast radio bursts}.
\newblock {\em arXiv} {\bf 2020}, arXiv:2005.06736.

\bibitem[{Popov} and {Postnov}(2010)]{popov10}
{Popov}, S.B.; {Postnov}, K.A.
\newblock {Hyperflares of SGRs as an engine for millisecond extragalactic radio
  bursts}.
\newblock In \emph{Evolution of Cosmic Objects through Their Physical Activity};
  {Harutyunian}, H.A., {Mickaelian}, A.M., {Terzian}, Y., Eds.;  {2010} Yerevan, "Gitutyun" Publishing House of NAS RA 
, pp.~129--132.

\bibitem[{Lyutikov} and {Popov}(2020)]{lyutikov20}
{Lyutikov}, M.; {Popov}, S.
\newblock {Fast Radio Bursts from reconnection events in magnetar
  magnetospheres}.
\newblock {\em arXiv} {\bf 2020}, arXiv:2005.05093.

\bibitem[{Lyubarsky}(2020)]{lyubarsky20}
{Lyubarsky}, Y.
\newblock {Fast Radio Bursts from Reconnection in a Magnetar Magnetosphere}.
\newblock {\em \apj} {\bf 2020}, {\em 897},~1,
 doi:10.3847/1538-4357/ab97b5.

\bibitem[{Kumar} \em{et~al.}(2017){Kumar}, {Lu}, and {Bhattacharya}]{kumar17}
{Kumar}, P.; {Lu}, W.; {Bhattacharya}, M.
\newblock {Fast radio burst source properties and curvature radiation model}.
\newblock {\em \mnras} {\bf 2017}, {\em 468},~2726--2739,
 doi:10.1093/mnras/stx665.

\bibitem[{Ghisellini} and {Locatelli}(2018)]{ghisellini18}
{Ghisellini}, G.; {Locatelli}, N.
\newblock {Coherent curvature radiation and fast radio bursts}.
\newblock {\em \aap} {\bf 2018}, {\em 613},~A61,
 doi:10.1051/0004-6361/201731820.

\bibitem[{Ruderman} and {Sutherland}(1975)]{rudsuth75}
{Ruderman}, M.A.; {Sutherland}, P.G.
\newblock {Theory of pulsars: Polar gaps, sparks, and coherent microwave
  radiation.}
\newblock {\em \apj} {\bf 1975}, {\em 196},~51--72, doi:10.1086/153393.

\bibitem[{Wadiasingh} and {Timokhin}(2019)]{wadiasingh19}
{Wadiasingh}, Z.; {Timokhin}, A.
\newblock {Repeating Fast Radio Bursts from Magnetars with Low Magnetospheric
  Twist}.
\newblock {\em \apj} {\bf 2019}, {\em 879},~4,
 doi:10.3847/1538-4357/ab2240.

\bibitem[{Metzger} \em{et~al.}(2019){Metzger}, {Margalit}, and
  {Sironi}]{metzger19}
{Metzger}, B.D.; {Margalit}, B.; {Sironi}, L.
\newblock {Fast radio bursts as synchrotron maser emission from decelerating
  relativistic blast waves}.
\newblock {\em \mnras} {\bf 2019}, {\em 485},~4091--4106,
 doi:10.1093/mnras/stz700.

\bibitem[{Lyubarsky}(2014)]{lyubarsky14}
{Lyubarsky}, Y.
\newblock {A model for fast extragalactic radio bursts.}
\newblock {\em \mnras} {\bf 2014}, {\em 442},~L9--L13,
doi:10.1093/mnrasl/slu046.

\bibitem[{Metzger} \em{et~al.}(2017){Metzger}, {Berger}, and
  {Margalit}]{metzger17}
{Metzger}, B.D.; {Berger}, E.; {Margalit}, B.
\newblock {Millisecond Magnetar Birth Connects FRB 121102 to Superluminous
  Supernovae and Long-duration Gamma-Ray Bursts}.
\newblock {\em \apj} {\bf 2017}, {\em 841},~14,
 doi:10.3847/1538-4357/aa633d.

\bibitem[{Margalit} and {Metzger}(2018)]{marmet18}
{Margalit}, B.; {Metzger}, B.D.
\newblock {A Concordance Picture of FRB 121102 as a Flaring Magnetar Embedded
  in a Magnetized Ion-Electron Wind Nebula}.
\newblock {\em \apjl} {\bf 2018}, {\em 868},~L4,
 doi:10.3847/2041-8213/aaedad.

\bibitem[{Kirsten} \em{et~al.}(2021){Kirsten}, {Marcote}, {Nimmo}, {Hessels},
  {Bhardwaj}, {Tendulkar}, {Keimpema}, {Yang}, {Snelders}, {Scholz},
  {Pearlman}, {Law}, {Peters}, {Giroletti}, {Paragi}, {Bassa}, {Hewitt},
  {Bach}, {Bezrukovs}, {Burgay}, {Buttaccio}, {Conway}, {Corongiu}, {Feiler},
  {Forss{\'e}n}, {Gawro{\'n}ski}, {Karuppusamy}, {Kharinov}, {Lindqvist},
  {Maccaferri}, {Melnikov}, {Ould-Boukattine}, {Possenti}, {Surcis}, {Wang},
  {Yuan}, {Aggarwal}, {Anna-Thomas}, {Bower}, {Blaauw}, {Burke-Spolaor},
  {Cassanelli}, {Clarke}, {Fonseca}, {Gaensler}, {Gopinath}, {Kaspi}, {Kassim},
  {Lazio}, {Leung}, {Li}, {Lin}, {Masui}, {Mckinven}, {Michilli}, {Mikhailov},
  {Ng}, {Orbidans}, {Pen}, {Petroff}, {Rahman}, {Ransom}, {Shin}, {Smith},
  {Stairs}, and {Vlemmings}]{kirsten21}
{Kirsten}, F.; {Marcote}, B.; {Nimmo}, K.; {Hessels}, J.W.T.; {Bhardwaj}, M.;
  {Tendulkar}, S.P.; {Keimpema}, A.; {Yang}, J.; {Snelders}, M.P.; {Scholz},
  P.; et~al.
\newblock {A repeating fast radio burst source in a globular cluster}.
\newblock {\em arXiv} {\bf 2021}, arXiv:2105.11445.

\bibitem[{Chatterjee} \em{et~al.}(2017){Chatterjee}, {Law}, {Wharton},
  {Burke-Spolaor}, {Hessels}, {Bower}, {Cordes}, {Tendulkar}, {Bassa},
  {Demorest}, {Butler}, {Seymour}, {Scholz}, {Abruzzo}, {Bogdanov}, {Kaspi},
  {Keimpema}, {Lazio}, {Marcote}, {McLaughlin}, {Paragi}, {Ransom}, {Rupen},
  {Spitler}, and {van Langevelde}]{chatterjee17}
{Chatterjee}, S.; {Law}, C.J.; {Wharton}, R.S.; {Burke-Spolaor}, S.; {Hessels},
  J.W.T.; {Bower}, G.C.; {Cordes}, J.M.; {Tendulkar}, S.P.; {Bassa}, C.G.;
  {Demorest}, P.; et~al.
\newblock {A direct localization of a fast radio burst and its host}.
\newblock {\em \nat} {\bf 2017}, {\em 541},~58--61,
 doi:10.1038/nature20797.

\bibitem[{Tendulkar} \em{et~al.}(2017){Tendulkar}, {Bassa}, {Cordes}, {Bower},
  {Law}, {Chatterjee}, {Adams}, {Bogdanov}, {Burke-Spolaor}, {Butler},
  {Demorest}, {Hessels}, {Kaspi}, {Lazio}, {Maddox}, {Marcote}, {McLaughlin},
  {Paragi}, {Ransom}, {Scholz}, {Seymour}, {Spitler}, {van Langevelde}, and
  {Wharton}]{tendulkar17}
{Tendulkar}, S.P.; {Bassa}, C.G.; {Cordes}, J.M.; {Bower}, G.C.; {Law}, C.J.;
  {Chatterjee}, S.; {Adams}, E.A.K.; {Bogdanov}, S.; {Burke-Spolaor}, S.;
  {Butler}, B.J.; et~al.
\newblock {The Host Galaxy and Redshift of the Repeating Fast Radio Burst FRB
  121102}.
\newblock {\em \apjl} {\bf 2017}, {\em 834},~L7.

\bibitem[{Ravi} and {Loeb}(2019)]{raviloeb19}
{Ravi}, V.; {Loeb}, A.
\newblock {Explaining the Statistical Properties of Fast Radio Bursts with
  Suppressed Low-frequency Emission}.
\newblock {\em \apj} {\bf 2019}, {\em 874},~72,
doi:10.3847/1538-4357/ab0748.

\bibitem[{Cordes} \em{et~al.}(2017){Cordes}, {Wasserman}, {Hessels}, {Lazio},
  {Chatterjee}, and {Wharton}]{cordes17}
{Cordes}, J.M.; {Wasserman}, I.; {Hessels}, J.W.T.; {Lazio}, T.J.W.;
  {Chatterjee}, S.; {Wharton}, R.S.
\newblock {Lensing of Fast Radio Bursts by Plasma Structures in Host Galaxies}.
\newblock {\em \apj} {\bf 2017}, {\em 842},~35,
 doi:10.3847/1538-4357/aa74da.

\bibitem[{Masui} \em{et~al.}(2015){Masui}, {Lin}, {Sievers}, {Anderson},
  {Chang}, {Chen}, {Ganguly}, {Jarvis}, {Kuo}, {Li}, {Liao}, {McLaughlin},
  {Pen}, {Peterson}, {Roman}, {Timbie}, {Voytek}, and {Yadav}]{masui15}
{Masui}, K.; {Lin}, H.H.; {Sievers}, J.; {Anderson}, C.J.; {Chang}, T.C.;
  {Chen}, X.; {Ganguly}, A.; {Jarvis}, M.; {Kuo}, C.Y.; {Li}, Y.C.; et~al.
\newblock {Dense magnetized plasma associated with a fast radio burst}.
\newblock {\em \nat} {\bf 2015}, {\em 528},~523--525,
 doi:10.1038/nature15769.

\bibitem[{Connor} \em{et~al.}(2016){Connor}, {Lin}, {Masui}, {Oppermann},
  {Pen}, {Peterson}, {Roman}, and {Sievers}]{connor16}
{Connor}, L.; {Lin}, H.H.; {Masui}, K.; {Oppermann}, N.; {Pen}, U.L.;
  {Peterson}, J.B.; {Roman}, A.; {Sievers}, J.
\newblock {Constraints on the FRB rate at 700--900 MHz}.
\newblock {\em \mnras} {\bf 2016}, {\em 460},~1054--1058,
 doi:10.1093/mnras/stw907.

\bibitem[{Petroff} \em{et~al.}(2019){Petroff}, {Hessels}, and
  {Lorimer}]{petroff19}
{Petroff}, E.; {Hessels}, J.W.T.; {Lorimer}, D.R.
\newblock {Fast radio bursts}.
\newblock {\em  Astron. Astrophys. Rev.} {\bf 2019}, {\em 27},~4,
 doi:10.1007/s00159-019-0116-6.

\bibitem[{Lorimer} and {Kramer}(2004)]{lk04}
{Lorimer}, D.R.; {Kramer}, M.
\newblock {\em {Handbook of Pulsar Astronomy}}; 
 Cambridge University Press: Cambridge, UK,  2004.

\bibitem[{Kulkarni}(2020)]{kulkarni20}
{Kulkarni}, S.R.
\newblock {Dispersion measure: Confusion, Constants \& Clarity}.
\newblock {\em arXiv} {\bf 2020}, arXiv:2007.02886.

\bibitem[{Hankins} and {Rickett}(1975)]{hanrick75}
{Hankins}, T.H.; {Rickett}, B.J.
\newblock {Pulsar signal processing.}
\newblock {\em Methods Comput. Phys.} {\bf 1975}, {\em 14},~55--129, doi:10.1016/B978-0-12-460814-6.50007-3.

\bibitem[{Bassa} \em{et~al.}(2016){Bassa}, {Janssen}, {Karuppusamy}, {Kramer},
  {Lee}, {Liu}, {McKee}, {Perrodin}, {Purver}, {Sanidas}, {Smits}, and
  {Stappers}]{bassa+16}
{Bassa}, C.G.; {Janssen}, G.H.; {Karuppusamy}, R.; {Kramer}, M.; {Lee}, K.J.;
  {Liu}, K.; {McKee}, J.; {Perrodin}, D.; {Purver}, M.; {Sanidas}, S.; {Smits},
  R.; {Stappers}, B.W.
\newblock {LEAP: The Large European Array for Pulsars}.
\newblock {\em \mnras} {\bf 2016}, {\em 456},~2196--2209,
 doi:10.1093/mnras/stv2755.

\bibitem[{Boyles} \em{et~al.}(2013){Boyles}, {Lynch}, {Ransom}, {Stairs},
  {Lorimer}, {McLaughlin}, {Hessels}, {Kaspi}, {Kondratiev}, {Archibald},
  {Berndsen}, {Cardoso}, {Cherry}, {Epstein}, {Karako-Argaman}, {McPhee},
  {Pennucci}, {Roberts}, {Stovall}, and {van Leeuwen}]{boyles13}
{Boyles}, J.; {Lynch}, R.S.; {Ransom}, S.M.; {Stairs}, I.H.; {Lorimer}, D.R.;
  {McLaughlin}, M.A.; {Hessels}, J.W.T.; {Kaspi}, V.M.; {Kondratiev}, V.I.;
  {Archibald}, A.; et~al.
\newblock {The Green Bank Telescope 350 MHz Drift-scan survey. I. Survey
  Observations and the Discovery of 13 Pulsars}.
\newblock {\em \apj} {\bf 2013}, {\em 763},~80,
 doi:10.1088/0004-637X/763/2/80.

\bibitem[{Lynch} \em{et~al.}(2013){Lynch}, {Boyles}, {Ransom}, {Stairs},
  {Lorimer}, {McLaughlin}, {Hessels}, {Kaspi}, {Kondratiev}, {Archibald},
  {Berndsen}, {Cardoso}, {Cherry}, {Epstein}, {Karako-Argaman}, {McPhee},
  {Pennucci}, {Roberts}, {Stovall}, and {van Leeuwen}]{lynch13}
{Lynch}, R.S.; {Boyles}, J.; {Ransom}, S.M.; {Stairs}, I.H.; {Lorimer}, D.R.;
  {McLaughlin}, M.A.; {Hessels}, J.W.T.; {Kaspi}, V.M.; {Kondratiev}, V.I.;
  {Archibald}, A.M.; et~al.
\newblock {The Green Bank Telescope 350 MHz Drift-scan Survey II: Data Analysis
  and the Timing of 10 New Pulsars, Including a Relativistic Binary}.
\newblock {\em \apj} {\bf 2013}, {\em 763},~81,
 doi:10.1088/0004-637X/763/2/81.

\bibitem[{Stovall} \em{et~al.}(2014){Stovall}, {Lynch}, {Ransom}, {Archibald},
  {Banaszak}, {Biwer}, {Boyles}, {Dartez}, {Day}, {Ford}, {Flanigan}, {Garcia},
  {Hessels}, {Hinojosa}, {Jenet}, {Kaplan}, {Karako-Argaman}, {Kaspi},
  {Kondratiev}, {Leake}, {Lorimer}, {Lunsford}, {Martinez}, {Mata},
  {McLaughlin}, {Roberts}, {Rohr}, {Siemens}, {Stairs}, {van Leeuwen},
  {Walker}, and {Wells}]{stovall14}
{Stovall}, K.; {Lynch}, R.S.; {Ransom}, S.M.; {Archibald}, A.M.; {Banaszak},
  S.; {Biwer}, C.M.; {Boyles}, J.; {Dartez}, L.P.; {Day}, D.; {Ford}, A.J.;
et~al.
\newblock {The Green Bank Northern Celestial Cap Pulsar Survey. I. Survey
  Description, Data Analysis, and Initial Results}.
\newblock {\em \apj} {\bf 2014}, {\em 791},~67,
 doi:10.1088/0004-637X/791/1/67.

\bibitem[{Chawla} \em{et~al.}(2017){Chawla}, {Kaspi}, {Josephy}, {Rajwade},
  {Lorimer}, {Archibald}, {DeCesar}, {Hessels}, {Kaplan}, {Karako-Argaman},
  {Kondratiev}, {Levin}, {Lynch}, {McLaughlin}, {Ransom}, {Roberts}, {Stairs},
  {Stovall}, {Swiggum}, and {van Leeuwen}]{chawla17}
{Chawla}, P.; {Kaspi}, V.M.; {Josephy}, A.; {Rajwade}, K.M.; {Lorimer}, D.R.;
  {Archibald}, A.M.; {DeCesar}, M.E.; {Hessels}, J.W.T.; {Kaplan}, D.L.;
  {Karako-Argaman}, C.; et~al.
\newblock {A Search for Fast Radio Bursts with the GBNCC Pulsar Survey}.
\newblock {\em \apj} {\bf 2017}, {\em 844},~140,
 doi:10.3847/1538-4357/aa7d57.

\bibitem[{Keane} and {Petroff}(2015)]{kp15}
{Keane}, E.F.; {Petroff}, E.
\newblock {Fast radio bursts: Search sensitivities and completeness}.
\newblock {\em \mnras} {\bf 2015}, {\em 447},~2852--2856,
 doi:10.1093/mnras/stu2650.

\bibitem[{Champion} \em{et~al.}(2016){Champion}, {Petroff}, {Kramer}, {Keith},
  {Bailes}, {Barr}, {Bates}, {Bhat}, {Burgay}, {Burke-Spolaor}, {Flynn},
  {Jameson}, {Johnston}, {Ng}, {Levin}, {Possenti}, {Stappers}, {van Straten},
  {Thornton}, {Tiburzi}, and {Lyne}]{champion16}
{Champion}, D.J.; {Petroff}, E.; {Kramer}, M.; {Keith}, M.J.; {Bailes}, M.;
  {Barr}, E.D.; {Bates}, S.D.; {Bhat}, N.D.R.; {Burgay}, M.; {Burke-Spolaor},
  S.; et~al.
\newblock {Five new fast radio bursts from the HTRU high-latitude survey at
  Parkes: First evidence for two-component bursts}.
\newblock {\em \mnras} {\bf 2016}, {\em 460},~L30--L34,
doi:10.1093/mnrasl/slw069.

\bibitem[{Crawford} \em{et~al.}(2016){Crawford}, {Rane}, {Tran}, {Rolph},
  {Lorimer}, and {Ridley}]{crawford16}
{Crawford}, F.; {Rane}, A.; {Tran}, L.; {Rolph}, K.; {Lorimer}, D.R.; {Ridley},
  J.P.
\newblock {A search for highly dispersed fast radio bursts in three Parkes
  multibeam surveys}.
\newblock {\em \mnras} {\bf 2016}, {\em 460},~3370--3375,
 doi:10.1093/mnras/stw1233.

\bibitem[{Katz}(2016)]{katz16}
{Katz}, J.I.
\newblock {Inferences from the Distributions of Fast Radio Burst Pulse Widths,
  Dispersion Measures, and Fluences}.
\newblock {\em \apj} {\bf 2016}, {\em 818},~19,
  doi:10.3847/0004-637X/818/1/19.

\bibitem[{Parent} \em{et~al.}(2020){Parent}, {Chawla}, {Kaspi}, {Agazie},
  {Blumer}, {DeCesar}, {Fiore}, {Fonseca}, {Hessels}, {Kaplan}, {Kondratiev},
  {LaRose}, {Levin}, {Lewis}, {Lynch}, {McEwen}, {McLaughlin}, {Mingyar}, {Al
  Noori}, {Ransom}, {Roberts}, {Schmiedekamp}, {Schmiedekamp}, {Siemens},
  {Spiewak}, {Stairs}, {Surnis}, {Swiggum}, and {van Leeuwen}]{parent20}
{Parent}, E.; {Chawla}, P.; {Kaspi}, V.M.; {Agazie}, G.Y.; {Blumer}, H.;
  {DeCesar}, M.; {Fiore}, W.; {Fonseca}, E.; {Hessels}, J.W.T.; {Kaplan}, D.L.;
et~al.
\newblock {First Discovery of a Fast Radio Burst at 350 MHz by the GBNCC
  Survey}.
\newblock {\em \apj} {\bf 2020}, {\em 904},~92,
 doi:10.3847/1538-4357/abbdf6.

\bibitem[{Deneva} \em{et~al.}(2013){Deneva}, {Stovall}, {McLaughlin}, {Bates},
  {Freire}, {Martinez}, {Jenet}, and {Bagchi}]{deneva13}
{Deneva}, J.S.; {Stovall}, K.; {McLaughlin}, M.A.; {Bates}, S.D.; {Freire},
  P.C.C.; {Martinez}, J.G.; {Jenet}, F.; {Bagchi}, M.
\newblock {Goals, Strategies and First Discoveries of AO327, the Arecibo
  All-sky 327 MHz Drift Pulsar Survey}.
\newblock {\em \apj} {\bf 2013}, {\em 775},~51,
 doi:10.1088/0004-637X/775/1/51.

\bibitem[{Barsdell} \em{et~al.}(2012){Barsdell}, {Bailes}, {Barnes}, and
  {Fluke}]{barsdell12}
{Barsdell}, B.R.; {Bailes}, M.; {Barnes}, D.G.; {Fluke}, C.J.
\newblock {Accelerating incoherent dedispersion}.
\newblock {\em \mnras} {\bf 2012}, {\em 422},~379--392,
 doi:10.1111/j.1365-2966.2012.20622.x.

\bibitem[{Agarwal} \em{et~al.}(2020){Agarwal}, {Aggarwal}, {Burke-Spolaor},
  {Lorimer}, and {Garver-Daniels}]{agarwal20fetch}
{Agarwal}, D.; {Aggarwal}, K.; {Burke-Spolaor}, S.; {Lorimer}, D.R.;
  {Garver-Daniels}, N.
\newblock {FETCH: A deep-learning based classifier for fast transient
  classification}.
\newblock {\em \mnras} {\bf 2020}, {\em 497},~1661--1674,
 doi:10.1093/mnras/staa1856.

\bibitem[{ter Veen} \em{et~al.}(2019){ter Veen}, {Enriquez}, {Falcke},
  {Rachen}, {van den Akker}, {Schellart}, {Bonardi}, {Breton}, {Broderick},
  {Corbel}, {Corstanje}, {Eisl{\"o}ffel}, {Grie{\ss}meier}, {H{\"o}randel},
  {van der Horst}, {Law}, {van Leeuwen}, {Nelles}, {Rossetto}, {Rowlinson},
  {Winchen}, and {Zarka}]{terveen19}
{ter Veen}, S.; {Enriquez}, J.E.; {Falcke}, H.; {Rachen}, J.P.; {van den
  Akker}, M.; {Schellart}, P.; {Bonardi}, A.; {Breton}, R.P.; {Broderick},
  J.W.; {Corbel}, S.; et~al.
\newblock {The FRATS project: Real-time searches for fast radio bursts and
  other fast transients with LOFAR at 135 MHz}.
\newblock {\em \aap} {\bf 2019}, {\em 621},~A57, doi:10.1051/0004-6361/201732515.

\bibitem[{van Haarlem} \em{et~al.}(2013){van Haarlem}, {Wise}, {Gunst},
  {Heald}, {McKean}, {Hessels}, {de Bruyn}, {Nijboer}, {Swinbank}, {Fallows},
  {Brentjens}, {Nelles}, {Beck}, {Falcke}, {Fender}, {H{\"o}randel},
  {Koopmans}, {Mann}, {Miley}, {R{\"o}ttgering}, {Stappers}, {Wijers},
  {Zaroubi}, {van den Akker}, {Alexov}, {Anderson}, {Anderson}, {van Ardenne},
  {Arts}, {Asgekar}, {Avruch}, {Batejat}, {B{\"a}hren}, {Bell}, {Bell}, {van
  Bemmel}, {Bennema}, {Bentum}, {Bernardi}, {Best}, {B{\^\i}rzan}, {Bonafede},
  {Boonstra}, {Braun}, {Bregman}, {Breitling}, {van de Brink}, {Broderick},
  {Broekema}, {Brouw}, {Br{\"u}ggen}, {Butcher}, {van Cappellen}, {Ciardi},
  {Coenen}, {Conway}, {Coolen}, {Corstanje}, {Damstra}, {Davies}, {Deller},
  {Dettmar}, {van Diepen}, {Dijkstra}, {Donker}, {Doorduin}, {Dromer}, {Drost},
  {van Duin}, {Eisl{\"o}ffel}, {van Enst}, {Ferrari}, {Frieswijk}, {Gankema},
  {Garrett}, {de Gasperin}, {Gerbers}, {de Geus}, {Grie{\ss}meier}, {Grit},
  {Gruppen}, {Hamaker}, {Hassall}, {Hoeft}, {Holties}, {Horneffer}, {van der
  Horst}, {van Houwelingen}, {Huijgen}, {Iacobelli}, {Intema}, {Jackson},
  {Jelic}, {de Jong}, {Juette}, {Kant}, {Karastergiou}, {Koers}, {Kollen},
  {Kondratiev}, {Kooistra}, {Koopman}, {Koster}, {Kuniyoshi}, {Kramer},
  {Kuper}, {Lambropoulos}, {Law}, {van Leeuwen}, {Lemaitre}, {Loose}, {Maat},
  {Macario}, {Markoff}, {Masters}, {McFadden}, {McKay-Bukowski}, {Meijering},
  {Meulman}, {Mevius}, {Middelberg}, {Millenaar}, {Miller-Jones}, {Mohan},
  {Mol}, {Morawietz}, {Morganti}, {Mulcahy}, {Mulder}, {Munk}, {Nieuwenhuis},
  {van Nieuwpoort}, {Noordam}, {Norden}, {Noutsos}, {Offringa}, {Olofsson},
  {Omar}, {Orr{\'u}}, {Overeem}, {Paas}, {Pand ey-Pommier}, {Pandey}, {Pizzo},
  {Polatidis}, {Rafferty}, {Rawlings}, {Reich}, {de Reijer}, {Reitsma},
  {Renting}, {Riemers}, {Rol}, {Romein}, {Roosjen}, {Ruiter}, {Scaife}, {van
  der Schaaf}, {Scheers}, {Schellart}, {Schoenmakers}, {Schoonderbeek},
  {Serylak}, {Shulevski}, {Sluman}, {Smirnov}, {Sobey}, {Spreeuw}, {Steinmetz},
  {Sterks}, {Stiepel}, {Stuurwold}, {Tagger}, {Tang}, {Tasse}, {Thomas},
  {Thoudam}, {Toribio}, {van der Tol}, {Usov}, {van Veelen}, {van der Veen},
  {ter Veen}, {Verbiest}, {Vermeulen}, {Vermaas}, {Vocks}, {Vogt}, {de Vos},
  {van der Wal}, {van Weeren}, {Weggemans}, {Weltevrede}, {White}, {Wijnholds},
  {Wilhelmsson}, {Wucknitz}, {Yatawatta}, {Zarka}, {Zensus}, and {van
  Zwieten}]{vanhaarlem+13}
{van Haarlem}, M.P.; {Wise}, M.W.; {Gunst}, A.W.; {Heald}, G.; {McKean}, J.P.;
  {Hessels}, J.W.T.; {de Bruyn}, A.G.; {Nijboer}, R.; {Swinbank}, J.;
  {Fallows}, R.; et~al.
\newblock {LOFAR: The LOw-Frequency ARray}.
\newblock {\em \aap} {\bf 2013}, {\em 556},~A2,
 doi:10.1051/0004-6361/201220873.

\bibitem[{Stappers} \em{et~al.}(2011){Stappers}, {Hessels}, {Alexov},
  {Anderson}, {Coenen}, {Hassall}, {Karastergiou}, {Kondratiev}, {Kramer}, {van
  Leeuwen}, {Mol}, {Noutsos}, {Romein}, {Weltevrede}, {Fender}, {Wijers},
  {B{\"a}hren}, {Bell}, {Broderick}, {Daw}, {Dhillon}, {Eisl{\"o}ffel},
  {Falcke}, {Griessmeier}, {Law}, {Markoff}, {Miller-Jones}, {Scheers},
  {Spreeuw}, {Swinbank}, {Ter Veen}, {Wise}, {Wucknitz}, {Zarka}, {Anderson},
  {Asgekar}, {Avruch}, {Beck}, {Bennema}, {Bentum}, {Best}, {Bregman},
  {Brentjens}, {van de Brink}, {Broekema}, {Brouw}, {Br{\"u}ggen}, {de Bruyn},
  {Butcher}, {Ciardi}, {Conway}, {Dettmar}, {van Duin}, {van Enst}, {Garrett},
  {Gerbers}, {Grit}, {Gunst}, {van Haarlem}, {Hamaker}, {Heald}, {Hoeft},
  {Holties}, {Horneffer}, {Koopmans}, {Kuper}, {Loose}, {Maat},
  {McKay-Bukowski}, {McKean}, {Miley}, {Morganti}, {Nijboer}, {Noordam},
  {Norden}, {Olofsson}, {Pandey-Pommier}, {Polatidis}, {Reich},
  {R{\"o}ttgering}, {Schoenmakers}, {Sluman}, {Smirnov}, {Steinmetz}, {Sterks},
  {Tagger}, {Tang}, {Vermeulen}, {Vermaas}, {Vogt}, {de Vos}, {Wijnholds},
  {Yatawatta}, and {Zensus}]{stappers11}
{Stappers}, B.W.; {Hessels}, J.W.T.; {Alexov}, A.; {Anderson}, K.; {Coenen},
  T.; {Hassall}, T.; {Karastergiou}, A.; {Kondratiev}, V.I.; {Kramer}, M.; {van
  Leeuwen}, J.; et~al.
\newblock {Observing pulsars and fast transients with LOFAR}.
\newblock {\em \aap} {\bf 2011}, {\em 530},~A80,
doi:10.1051/0004-6361/201116681.

\bibitem[{Coenen} \em{et~al.}(2014){Coenen}, {van Leeuwen}, {Hessels},
  {Stappers}, {Kondratiev}, {Alexov}, {Breton}, {Bilous}, {Cooper}, {Falcke},
  {Fallows}, {Gajjar}, {Grie{\ss}meier}, {Hassall}, {Karastergiou}, {Keane},
  {Kramer}, {Kuniyoshi}, {Noutsos}, {Os{\l}owski}, {Pilia}, {Serylak},
  {Schrijvers}, {Sobey}, {ter Veen}, {Verbiest}, {Weltevrede}, {Wijnholds},
  {Zagkouris}, {van Amesfoort}, {Anderson}, {Asgekar}, {Avruch}, {Bell},
  {Bentum}, {Bernardi}, {Best}, {Bonafede}, {Breitling}, {Broderick},
  {Br{\"u}ggen}, {Butcher}, {Ciardi}, {Corstanje}, {Deller}, {Duscha},
  {Eisl{\"o}ffel}, {Fender}, {Ferrari}, {Frieswijk}, {Garrett}, {de Gasperin},
  {de Geus}, {Gunst}, {Hamaker}, {Heald}, {Hoeft}, {van der Horst}, {Juette},
  {Kuper}, {Law}, {Mann}, {McFadden}, {McKay-Bukowski}, {McKean}, {Munk},
  {Orru}, {Paas}, {Pandey-Pommier}, {Polatidis}, {Reich}, {Renting},
  {R{\"o}ttgering}, {Rowlinson}, {Scaife}, {Schwarz}, {Sluman}, {Smirnov},
  {Swinbank}, {Tagger}, {Tang}, {Tasse}, {Thoudam}, {Toribio}, {Vermeulen},
  {Vocks}, {van Weeren}, {Wucknitz}, {Zarka}, and {Zensus}]{coenen14}
{Coenen}, T.; {van Leeuwen}, J.; {Hessels}, J.W.T.; {Stappers}, B.W.;
  {Kondratiev}, V.I.; {Alexov}, A.; {Breton}, R.P.; {Bilous}, A.; {Cooper}, S.;
  {Falcke}, H.; et~al.
\newblock {The LOFAR pilot surveys for pulsars and fast radio transients}.
\newblock {\em \aap} {\bf 2014}, {\em 570},~A60,
 doi:10.1051/0004-6361/201424495.

\bibitem[{Karastergiou} \em{et~al.}(2015){Karastergiou}, {Chennamangalam},
  {Armour}, {Williams}, {Mort}, {Dulwich}, {Salvini}, {Magro}, {Roberts},
  {Serylak}, {Doo}, {Bilous}, {Breton}, {Falcke}, {Grie{\ss}meier}, {Hessels},
  {Keane}, {Kondratiev}, {Kramer}, {van Leeuwen}, {Noutsos}, {Os{\l}owski},
  {Sobey}, {Stappers}, and {Weltevrede}]{karastergiou15}
{Karastergiou}, A.; {Chennamangalam}, J.; {Armour}, W.; {Williams}, C.; {Mort},
  B.; {Dulwich}, F.; {Salvini}, S.; {Magro}, A.; {Roberts}, S.; {Serylak}, M.;
  et~al.
\newblock {Limits on fast radio bursts at 145 MHz with ARTEMIS, a real-time
  software backend}.
\newblock {\em \mnras} {\bf 2015}, {\em 452},~1254--1262,
 doi:10.1093/mnras/stv1306.

\bibitem[{Tingay} \em{et~al.}(2013){Tingay}, {Goeke}, {Bowman}, {Emrich},
  {Ord}, {Mitchell}, {Morales}, {Booler}, {Crosse}, {Wayth}, {Lonsdale},
  {Tremblay}, {Pallot}, {Colegate}, {Wicenec}, {Kudryavtseva}, {Arcus},
  {Barnes}, {Bernardi}, {Briggs}, {Burns}, {Bunton}, {Cappallo}, {Corey},
  {Deshpande}, {Desouza}, {Gaensler}, {Greenhill}, {Hall}, {Hazelton}, {Herne},
  {Hewitt}, {Johnston-Hollitt}, {Kaplan}, {Kasper}, {Kincaid}, {Koenig},
  {Kratzenberg}, {Lynch}, {Mckinley}, {Mcwhirter}, {Morgan}, {Oberoi},
  {Pathikulangara}, {Prabu}, {Remillard}, {Rogers}, {Roshi}, {Salah}, {Sault},
  {Udaya-Shankar}, {Schlagenhaufer}, {Srivani}, {Stevens}, {Subrahmanyan},
  {Waterson}, {Webster}, {Whitney}, {Williams}, {Williams}, and
  {Wyithe}]{tingay13}
{Tingay}, S.J.; {Goeke}, R.; {Bowman}, J.D.; {Emrich}, D.; {Ord}, S.M.;
  {Mitchell}, D.A.; {Morales}, M.F.; {Booler}, T.; {Crosse}, B.; {Wayth}, R.B.;
  et~al.
\newblock {The Murchison Widefield Array: The Square Kilometre Array Precursor
  at Low Radio Frequencies}.
\newblock {\em Publ. Astron. Soc. Aust.} {\bf 2013}, {\em 30},~e007,
   doi:10.1017/pasa.2012.007.

\bibitem[{Tingay} \em{et~al.}(2015){Tingay}, {Trott}, {Wayth}, {Bernardi},
  {Bowman}, {Briggs}, {Cappallo}, {Deshpande}, {Feng}, {Gaensler}, {Greenhill},
  {Hancock}, {Hazelton}, {Johnston-Hollitt}, {Kaplan}, {Lonsdale}, {McWhirter},
  {Mitchell}, {Morales}, {Morgan}, {Murphy}, {Oberoi}, {Prabu}, {Udaya
  Shankar}, {Srivani}, {Subrahmanyan}, {Webster}, {Williams}, and
  {Williams}]{tingay15}
{Tingay}, S.J.; {Trott}, C.M.; {Wayth}, R.B.; {Bernardi}, G.; {Bowman}, J.D.;
  {Briggs}, F.; {Cappallo}, R.J.; {Deshpande}, A.A.; {Feng}, L.; {Gaensler},
  B.M.; et~al.
\newblock {A Search for Fast Radio Bursts at Low Frequencies with Murchison
  Widefield Array High Time Resolution Imaging}.
\newblock {\em  Astron. J.} {\bf 2015}, {\em 150},~199,
  doi:10.1088/0004-6256/150/6/199.

\bibitem[{Rowlinson} \em{et~al.}(2016){Rowlinson}, {Bell}, {Murphy}, {Trott},
  {Hurley-Walker}, {Johnston}, {Tingay}, {Kaplan}, {Carbone}, {Hancock},
  {Feng}, {Offringa}, {Bernardi}, {Bowman}, {Briggs}, {Cappallo}, {Deshpande},
  {Gaensler}, {Greenhill}, {Hazelton}, {Johnston-Hollitt}, {Lonsdale},
  {McWhirter}, {Mitchell}, {Morales}, {Morgan}, {Oberoi}, {Ord}, {Prabu},
  {Udaya Shankar}, {Srivani}, {Subrahmanyan}, {Wayth}, {Webster}, {Williams},
  and {Williams}]{rowlinson16}
{Rowlinson}, A.; {Bell}, M.E.; {Murphy}, T.; {Trott}, C.M.; {Hurley-Walker},
  N.; {Johnston}, S.; {Tingay}, S.J.; {Kaplan}, D.L.; {Carbone}, D.; {Hancock},
  P.J.; et~al.
\newblock {Limits on Fast Radio Bursts and other transient sources at 182 MHz
  using the Murchison Widefield Array}.
\newblock {\em \mnras} {\bf 2016}, {\em 458},~3506--3522,
, doi:10.1093/mnras/stw451.

\bibitem[{Trott} \em{et~al.}(2013){Trott}, {Tingay}, and {Wayth}]{ttw13}
{Trott}, C.M.; {Tingay}, S.J.; {Wayth}, R.B.
\newblock {Prospects for the Detection of Fast Radio Bursts with the Murchison
  Widefield Array}.
\newblock {\em \apjl} {\bf 2013}, {\em 776},~L16,
 doi:10.1088/2041-8205/776/1/L16.

\bibitem[{Fedorova} and {Rodin}(2019)]{fedrod19}
{Fedorova}, V.A.; {Rodin}, A.E.
\newblock {Search for Fast Radio Bursts in the Direction of the Galaxies M31
  and M33}.
\newblock {\em Astron. Rep.} {\bf 2019}, {\em 63},~877--890,
 doi:10.1134/S1063772919110039.

\bibitem[{Fedorova} and {Rodin}(2021)]{fedrod21}
{Fedorova}, V.A.; {Rodin}, A.E.
\newblock {Comparative Analysis of the Observational Properties of Fast Radio
  Bursts at the Frequencies of 111 and 1400 MHz}.
\newblock {\em Astron. Rep.} {\bf 2021}, {\em 65},~776--804,
 doi:10.1134/S1063772921100097.

\bibitem[{Caleb} \em{et~al.}(2016){Caleb}, {Flynn}, {Bailes}, {Barr},
  {Bateman}, {Bhandari}, {Campbell-Wilson}, {Green}, {Hunstead}, {Jameson},
  {Jankowski}, {Keane}, {Ravi}, {van Straten}, and {Krishnan}]{caleb16}
{Caleb}, M.; {Flynn}, C.; {Bailes}, M.; {Barr}, E.D.; {Bateman}, T.;
  {Bhandari}, S.; {Campbell-Wilson}, D.; {Green}, A.J.; {Hunstead}, R.W.;
  {Jameson}, A.; et~al.
\newblock {Fast Radio Transient searches with UTMOST at 843 MHz}.
\newblock {\em \mnras} {\bf 2016}, {\em 458},~718--725,
 doi:10.1093/mnras/stw109.

\bibitem[{Bailes} \em{et~al.}(2017){Bailes}, {Jameson}, {Flynn}, {Bateman},
  {Barr}, {Bhandari}, {Bunton}, {Caleb}, {Campbell-Wilson}, {Farah},
  {Gaensler}, {Green}, {Hunstead}, {Jankowski}, {Keane}, {Krishnan}, {Murphy},
  {O'Neill}, {Os{\l}owski}, {Parthasarathy}, {Ravi}, {Rosado}, and
  {Temby}]{bailes17}
{Bailes}, M.; {Jameson}, A.; {Flynn}, C.; {Bateman}, T.; {Barr}, E.D.;
  {Bhandari}, S.; {Bunton}, J.D.; {Caleb}, M.; {Campbell-Wilson}, D.; {Farah},
  W.; et~al.
\newblock {The UTMOST: A Hybrid Digital Signal Processor Transforms the
  Molonglo Observatory Synthesis Telescope}.
\newblock {\em Publ. Astron. Soc. Aust.} {\bf 2017}, {\em 34},~e045,
 doi:10.1017/pasa.2017.39.

\bibitem[{Caleb} \em{et~al.}(2017){Caleb}, {Flynn}, {Bailes}, {Barr},
  {Bateman}, {Bhandari}, {Campbell-Wilson}, {Farah}, {Green}, {Hunstead},
  {Jameson}, {Jankowski}, {Keane}, {Parthasarathy}, {Ravi}, {Rosado}, {van
  Straten}, and {Venkatraman Krishnan}]{caleb17}
{Caleb}, M.; {Flynn}, C.; {Bailes}, M.; {Barr}, E.D.; {Bateman}, T.;
  {Bhandari}, S.; {Campbell-Wilson}, D.; {Farah}, W.; {Green}, A.J.;
  {Hunstead}, R.W.; et~al.
\newblock {The first interferometric detections of fast radio bursts}.
\newblock {\em \mnras} {\bf 2017}, {\em 468},~3746--3756,
 doi:10.1093/mnras/stx638.

\bibitem[{Farah} \em{et~al.}(2019){Farah}, {Flynn}, {Bailes}, {Jameson},
  {Bateman}, {Campbell-Wilson}, {Day}, {Deller}, {Green}, {Gupta}, {Hunstead},
  {Lower}, {Os{\l}owski}, {Parthasarathy}, {Price}, {Ravi}, {Shannon},
  {Sutherland}, {Temby}, {Krishnan}, {Caleb}, {Chang}, {Cruces}, {Roy},
  {Morello}, {Onken}, {Stappers}, {Webb}, and {Wolf}]{farah19}
{Farah}, W.; {Flynn}, C.; {Bailes}, M.; {Jameson}, A.; {Bateman}, T.;
  {Campbell-Wilson}, D.; {Day}, C.K.; {Deller}, A.T.; {Green}, A.J.; {Gupta},
  V.; et~al.
\newblock {Five new real-time detections of fast radio bursts with UTMOST}.
\newblock {\em \mnras} {\bf 2019}, {\em 488},~2989--3002,
 doi:10.1093/mnras/stz1748.

\bibitem[{Bhandari} \em{et~al.}(2018){Bhandari}, {Bannister}, {Murphy}, {Bell},
  {Raja}, {Marvil}, {Hancock}, {Whiting}, {Flynn}, {Collier}, {Kaplan},
  {Allison}, {Anderson}, {Heywood}, {Hotan}, {Hunstead}, {Lee-Waddell},
  {Madrid}, {McConnell}, {Popping}, {Rhee}, {Sadler}, and
  {Voronkov}]{bhandari18}
{Bhandari}, S.; {Bannister}, K.W.; {Murphy}, T.; {Bell}, M.; {Raja}, W.;
  {Marvil}, J.; {Hancock}, P.J.; {Whiting}, M.; {Flynn}, C.M.; {Collier}, J.D.;
 et~al.
\newblock {A pilot survey for transients and variables with the Australian
  Square Kilometre Array Pathfinder}.
\newblock {\em \mnras} {\bf 2018}, {\em 478},~1784--1794,
 doi:10.1093/mnras/sty1157.

\bibitem[{Macquart} \em{et~al.}(2019){Macquart}, {Shannon}, {Bannister},
  {James}, {Ekers}, and {Bunton}]{macquart19}
{Macquart}, J.P.; {Shannon}, R.M.; {Bannister}, K.W.; {James}, C.W.; {Ekers},
  R.D.; {Bunton}, J.D.
\newblock {The Spectral Properties of the Bright Fast Radio Burst Population}.
\newblock {\em \apjl} {\bf 2019}, {\em 872},~L19,
doi:10.3847/2041-8213/ab03d6.

\bibitem[{CHIME/FRB Collaboration} \em{et~al.}(2019){CHIME/FRB Collaboration},
  {Amiri}, {Bandura}, {Bhardwaj}, {Boubel}, {Boyce}, {Boyle}, {Brar},
  {Burhanpurkar}, {Chawla}, {Cliche}, {Cubranic}, {Deng}, {Denman}, {Dobbs},
  {Fandino}, {Fonseca}, {Gaensler}, {Gilbert}, {Giri}, {Good}, {Halpern},
  {Hanna}, {Hill}, {Hinshaw}, {H{\"o}fer}, {Josephy}, {Kaspi}, {Landecker},
  {Lang}, {Masui}, {Mckinven}, {Mena-Parra}, {Merryfield}, {Milutinovic},
  {Moatti}, {Naidu}, {Newburgh}, {Ng}, {Patel}, {Pen}, {Pinsonneault-Marotte},
  {Pleunis}, {Rafiei-Ravandi}, {Ransom}, {Renard}, {Scholz}, {Shaw}, {Siegel},
  {Smith}, {Stairs}, {Tendulkar}, {Tretyakov}, {Vanderlinde}, and
  {Yadav}]{chime_13_19}
{CHIME/FRB Collaboration}; {Amiri}, M.; {Bandura}, K.; {Bhardwaj}, M.;
  {Boubel}, P.; {Boyce}, M.M.; {Boyle}, P.J.; {Brar}, C.; {Burhanpurkar}, M.;
  {Chawla}, P.; {Cliche}, J.F.; et~al.
\newblock {Observations of fast radio bursts at frequencies down to 400
  megahertz}.
\newblock {\em \nat} {\bf 2019}, {\em 566},~230--234,
   doi:10.1038/s41586-018-0867-7.

\bibitem[{Fonseca} \em{et~al.}(2020){Fonseca}, {Andersen}, {Bhardwaj},
  {Chawla}, {Good}, {Josephy}, {Kaspi}, {Masui}, {Mckinven}, {Michilli},
  {Pleunis}, {Shin}, {Tendulkar}, {Bandura}, {Boyle}, {Brar}, {Cassanelli},
  {Cubranic}, {Dobbs}, {Dong}, {Gaensler}, {Hinshaw}, {Land ecker}, {Leung},
  {Li}, {Lin}, {Mena-Parra}, {Merryfield}, {Naidu}, {Ng}, {Patel}, {Pen},
  {Rafiei-Ravandi}, {Rahman}, {Ransom}, {Scholz}, {Smith}, {Stairs},
  {Vanderlinde}, {Yadav}, and {Zwaniga}]{chime9rep}
{Fonseca}, E.; {Andersen}, B.C.; {Bhardwaj}, M.; {Chawla}, P.; {Good}, D.C.;
  {Josephy}, A.; {Kaspi}, V.M.; {Masui}, K.W.; {Mckinven}, R.; {Michilli}, D.;
 et~al.
\newblock {Nine New Repeating Fast Radio Burst Sources from CHIME/FRB}.
\newblock {\em arXiv} {\bf 2020}, arXiv:2001.03595.

\bibitem[{Josephy} \em{et~al.}(2021){Josephy}, {Chawla}, {Curtin}, {Kaspi},
  {Bhardwaj}, {Boyle}, {Brar}, {Cassanelli}, {Fonseca}, {Gaensler}, {Leung},
  {Lin}, {Masui}, {McKinven}, {Mena-Parra}, {Michilli}, {Ng}, {Pleunis},
  {Rafiei-Ravandi}, {Rahman}, {Sanghavi}, {Scholz}, {Smith}, {Stairs},
  {Tendulkar}, and {Zwaniga}]{josephy21}
{Josephy}, A.; {Chawla}, P.; {Curtin}, A.P.; {Kaspi}, V.M.; {Bhardwaj}, M.;
  {Boyle}, P.J.; {Brar}, C.; {Cassanelli}, T.; {Fonseca}, E.; {Gaensler}, B.M.;
 et~al.
\newblock {No Evidence for Galactic Latitude Dependence of the Fast Radio Burst
  Sky Distribution}.
\newblock {\em arXiv} {\bf 2021}, arXiv:2106.04353.

\bibitem[{Petroff} \em{et~al.}(2014){Petroff}, {van Straten}, {Johnston},
  {Bailes}, {Barr}, {Bates}, {Bhat}, {Burgay}, {Burke-Spolaor}, {Champion},
  {Coster}, {Flynn}, {Keane}, {Keith}, {Kramer}, {Levin}, {Ng}, {Possenti},
  {Stappers}, {Tiburzi}, and {Thornton}]{petroff14}
{Petroff}, E.; {van Straten}, W.; {Johnston}, S.; {Bailes}, M.; {Barr}, E.D.;
  {Bates}, S.D.; {Bhat}, N.D.R.; {Burgay}, M.; {Burke-Spolaor}, S.; {Champion},
  D.; et~al.
\newblock {An Absence of Fast Radio Bursts at Intermediate Galactic Latitudes}.
\newblock {\em \apjl} {\bf 2014}, {\em 789},~L26,
 doi:10.1088/2041-8205/789/2/L26.

\bibitem[{Connor} \em{et~al.}(2020){Connor}, {Miller}, and
  {Gardenier}]{connor20}
{Connor}, L.; {Miller}, M.C.; {Gardenier}, D.W.
\newblock {Beaming as an explanation of the repetition/width relation in FRBs}.
\newblock {\em \mnras} {\bf 2020}, {\em 497},~3076--3082,
  doi:10.1093/mnras/staa2074.

\bibitem[{Locatelli} \em{et~al.}(2020){Locatelli}, {Bernardi}, {Bianchi},
  {Chiello}, {Magro}, {Naldi}, {Pilia}, {Pupillo}, {Ridolfi}, {Setti}, and
  {Vazza}]{locatelli20}
{Locatelli}, N.T.; {Bernardi}, G.; {Bianchi}, G.; {Chiello}, R.; {Magro}, A.;
  {Naldi}, G.; {Pilia}, M.; {Pupillo}, G.; {Ridolfi}, A.; {Setti}, G.; et~al.
\newblock {The Northern Cross fast radio burst project - I. Overview and pilot
  observations at 408 MHz}.
\newblock {\em \mnras} {\bf 2020}, {\em 494},~1229--1236,
   doi:10.1093/mnras/staa813.

\bibitem[{Bernardi} \em{et~al.}(2021){Bernardi}, {Pilia}, {Bianchi}, {Magro},
  {Naldi}, {Pupillo}, {Setti}, and {Addis}]{bernardi21}
{Bernardi}, G.; {Pilia}, M.; {Bianchi}, G.; {Magro}, A.; {Naldi}, G.;
  {Pupillo}, G.; {Setti}, G.; {Addis}, A.
\newblock {One more burst from FRB 180916.J0158+65 observed with the Medicina
  Northern Cross at 408 MHz}.
\newblock {\em  Astron. Telegr.} {\bf 2021}, {\em 14480},~1.

\bibitem[{Sand} \em{et~al.}(2020){Sand}, {Gajjar}, {Pilia}, {Kudale}, {Joshi},
  {Jagtap}, {Ray}, {Deshpande}, {Bijay}, {Dey}, {Kalita}, {Bandyopadhyay},
  {Jena}, {Bhattacharya}, {Waratkar}, {Wagle}, {Singha}, {Bagchi}, {Surnis},
  {Bhat}, {Mishra}, {Konar}, and {Maan}]{sand20}
{Sand}, K.R.; {Gajjar}, V.; {Pilia}, M.; {Kudale}, S.; {Joshi}, B.C.; {Jagtap},
  V.S.; {Ray}, A.; {Deshpande}, A.; {Bijay}, B.; {Dey}, B.; et~al.
\newblock {Low-frequency detection of FRB180916 with the uGMRT}.
\newblock {\em  Astron. Telegr.} {\bf 2020}, {\em 13781},~1.

\bibitem[{Marthi} \em{et~al.}(2020){Marthi}, {Gautam}, {Li}, {Lin}, {Main},
  {Naidu}, {Pen}, and {Wharton}]{marthi20}
{Marthi}, V.R.; {Gautam}, T.; {Li}, D.Z.; {Lin}, H.H.; {Main}, R.A.; {Naidu},
  A.; {Pen}, U.L.; {Wharton}, R.S.
\newblock {Detection of 15 bursts from the fast radio burst 180916.J0158+65
  with the upgraded Giant Metrewave Radio Telescope}.
\newblock {\em \mnras} {\bf 2020}, {\em 499},~L16--L20,
 doi:10.1093/mnrasl/slaa148.

\bibitem[{Keane} \em{et~al.}(2016){Keane}, {Johnston}, {Bhandari}, {Barr},
  {Bhat}, {Burgay}, {Caleb}, {Flynn}, {Jameson}, {Kramer}, {Petroff},
  {Possenti}, {van Straten}, {Bailes}, {Burke-Spolaor}, {Eatough}, {Stappers},
  {Totani}, {Honma}, {Furusawa}, {Hattori}, {Morokuma}, {Niino}, {Sugai},
  {Terai}, {Tominaga}, {Yamasaki}, {Yasuda}, {Allen}, {Cooke}, {Jencson},
  {Kasliwal}, {Kaplan}, {Tingay}, {Williams}, {Wayth}, {Chandra}, {Perrodin},
  {Berezina}, {Mickaliger}, and {Bassa}]{keane16}
{Keane}, E.F.; {Johnston}, S.; {Bhandari}, S.; {Barr}, E.; {Bhat}, N.D.R.;
  {Burgay}, M.; {Caleb}, M.; {Flynn}, C.; {Jameson}, A.; {Kramer}, M.;
 et~al.
\newblock {The host galaxy of a fast radio burst}.
\newblock {\em \nat} {\bf 2016}, {\em 530},~453--456,
doi:10.1038/nature17140.

\bibitem[{Sokolowski} \em{et~al.}(2018){Sokolowski}, {Bhat}, {Macquart},
  {Shannon}, {Bannister}, {Ekers}, {Scott}, {Beardsley}, {Crosse}, {Emrich},
  {Franzen}, {Gaensler}, {Horsley}, {Johnston-Hollitt}, {Kaplan}, {Kenney},
  {Morales}, {Pallot}, {Sleap}, {Steele}, {Tingay}, {Trott}, {Walker}, {Wayth},
  {Williams}, and {Wu}]{sokolowski18}
{Sokolowski}, M.; {Bhat}, N.D.R.; {Macquart}, J.P.; {Shannon}, R.M.;
  {Bannister}, K.W.; {Ekers}, R.D.; {Scott}, D.R.; {Beardsley}, A.P.; {Crosse},
  B.; {Emrich}, D.; et~al.
\newblock {No Low-frequency Emission from Extremely Bright Fast Radio Bursts}.
\newblock {\em \apjl} {\bf 2018}, {\em 867},~L12,
 doi:10.3847/2041-8213/aae58d.

\bibitem[{Houben} \em{et~al.}(2019){Houben}, {Spitler}, {ter Veen}, {Rachen},
  {Falcke}, and {Kramer}]{houben19}
{Houben}, L.J.M.; {Spitler}, L.G.; {ter Veen}, S.; {Rachen}, J.P.; {Falcke},
  H.; {Kramer}, M.
\newblock {Constraints on the low frequency spectrum of FRB 121102}.
\newblock {\em \aap} {\bf 2019}, {\em 623},~A42,
 doi:10.1051/0004-6361/201833875.

\bibitem[{Aggarwal} \em{et~al.}(2020){Aggarwal}, {Law}, {Burke-Spolaor},
  {Bower}, {Butler}, {Demorest}, {Linford}, and {Lazio}]{aggarwal20}
{Aggarwal}, K.; {Law}, C.J.; {Burke-Spolaor}, S.; {Bower}, G.; {Butler}, B.J.;
  {Demorest}, P.; {Linford}, J.; {Lazio}, T.J.W.
\newblock {VLA/Realfast Detection of a Burst from FRB 180916.J0158+65 and Tests
  for Periodic Activity}.
\newblock {\em Res. Notes Am. Astron. Soc.} {\bf 2020},
  {\em 4},~94,  doi:10.3847/2515-5172/ab9f33.

\bibitem[{James} \em{et~al.}(2019){James}, {Ekers}, {Macquart}, {Bannister},
  and {Shannon}]{james19}
{James}, C.W.; {Ekers}, R.D.; {Macquart}, J.P.; {Bannister}, K.W.; {Shannon},
  R.M.
\newblock {The slope of the source-count distribution for fast radio bursts}.
\newblock {\em \mnras} {\bf 2019}, {\em 483},~1342--1353,
 doi:10.1093/mnras/sty3031.

\bibitem[{Scholz} \em{et~al.}(2020){Scholz}, {Cook}, {Cruces}, {Hessels},
  {Kaspi}, {Majid}, {Naidu}, {Pearlman}, {Spitler}, {Bandura}, {Bhardwaj},
  {Cassanelli}, {Chawla}, {Gaensler}, {Good}, {Josephy}, {Karuppusamy},
  {Keimpema}, {Kirichenko}, {Kirsten}, {Kocz}, {Leung}, {Marcote}, {Masui},
  {Mena-Parra}, {Merryfield}, {Michilli}, {Naudet}, {Nimmo}, {Pleunis},
  {Prince}, {Rafiei-Ravandi}, {Rahman}, {Shin}, {Smith}, {Stairs}, {Tendulkar},
  and {Vanderlinde}]{scholz20}
{Scholz}, P.; {Cook}, A.; {Cruces}, M.; {Hessels}, J.W.T.; {Kaspi}, V.M.;
  {Majid}, W.A.; {Naidu}, A.; {Pearlman}, A.B.; {Spitler}, L.; {Bandura}, K.M.;
 et~al.
\newblock {Simultaneous X-ray and Radio Observations of the Repeating Fast
  Radio Burst FRB 180916.J0158+65}.
\newblock {\em arXiv} {\bf 2020}, arXiv:2004.06082.

\bibitem[{Nicastro} \em{et~al.}(2021){Nicastro}, {Guidorzi}, {Palazzi},
  {Zampieri}, {Turatto}, and {Gardini}]{nicastro21}
{Nicastro}, L.; {Guidorzi}, C.; {Palazzi}, E.; {Zampieri}, L.; {Turatto}, M.;
  {Gardini}, A.
\newblock {Multiwavelength Observations of Fast Radio Bursts}.
\newblock {\em Universe} {\bf 2021}, {\em 7},~76,
 doi:10.3390/universe7030076.

\end{thebibliography}


%


\end{document}